\documentclass[10pt,journal,compsoc, twocolumn]{IEEEtran}
\emergencystretch=\maxdimen
\usepackage{cite}
\usepackage{pifont}
\usepackage{subcaption}
\usepackage{amsmath,amssymb,amsfonts}
\usepackage{algorithmic}
\usepackage{algorithm}
\usepackage{graphicx}
\usepackage{textcomp}
\usepackage{xcolor}
\usepackage{comment}
\usepackage{array}
\usepackage{booktabs}
\usepackage{makecell}
\usepackage{multirow}
\usepackage{mathtools}
\usepackage{threeparttable}

\IEEEoverridecommandlockouts

\begin{document}
\newtheorem{definition}{\it Definition}
\newtheorem{theorem}{\bf Theorem}
\newtheorem{lemma}{\it Lemma}
\newtheorem{corollary}{\it Corollary}
\newtheorem{remark}{\it Remark}
\newtheorem{example}{\it Example}
\newtheorem{case}{\bf Case Study}
\newtheorem{assumption}{\it Assumption}
\newtheorem{property}{\it Property}
\newtheorem{proposition}{\it Proposition}

\newcommand{\hP}[1]{{\boldsymbol h}_{{#1}{\bullet}}}
\newcommand{\hS}[1]{{\boldsymbol h}_{{\bullet}{#1}}}

\newcommand{\ba}{\boldsymbol{a}}
\newcommand{\baq}{\overline{q}}
\newcommand{\bA}{\boldsymbol{A}}
\newcommand{\bb}{\boldsymbol{b}}
\newcommand{\bB}{\boldsymbol{B}}
\newcommand{\bc}{\boldsymbol{c}}
\newcommand{\bcO}{\boldsymbol{\cal O}}
\newcommand{\bh}{\boldsymbol{h}}
\newcommand{\bH}{\boldsymbol{H}}
\newcommand{\bl}{\boldsymbol{l}}
\newcommand{\bm}{\boldsymbol{m}}
\newcommand{\bn}{\boldsymbol{n}}
\newcommand{\bo}{\boldsymbol{o}}
\newcommand{\bO}{\boldsymbol{O}}
\newcommand{\bp}{\boldsymbol{p}}
\newcommand{\bq}{\boldsymbol{q}}
\newcommand{\bR}{\boldsymbol{R}}
\newcommand{\bs}{\boldsymbol{s}}
\newcommand{\bS}{\boldsymbol{S}}
\newcommand{\bT}{\boldsymbol{T}}
\newcommand{\bu}{\boldsymbol{u}}
\newcommand{\bv}{\boldsymbol{v}}
\newcommand{\bw}{\boldsymbol{w}}

\newcommand{\balpha}{\boldsymbol{\alpha}}
\newcommand{\bbeta}{\boldsymbol{\beta}}
\newcommand{\bOmega}{\boldsymbol{\Omega}}
\newcommand{\bTheta}{\boldsymbol{\Theta}}
\newcommand{\bphi}{\boldsymbol{\phi}}
\newcommand{\btheta}{\boldsymbol{\theta}}
\newcommand{\bvarpi}{\boldsymbol{\varpi}}
\newcommand{\bpi}{\boldsymbol{\pi}}
\newcommand{\bpsi}{\boldsymbol{\psi}}
\newcommand{\bxi}{\boldsymbol{\xi}}
\newcommand{\bx}{\boldsymbol{x}}
\newcommand{\by}{\boldsymbol{y}}

\newcommand{\cA}{{\cal A}}
\newcommand{\bcA}{\boldsymbol{\cal A}}
\newcommand{\cB}{{\cal B}}
\newcommand{\cE}{{\cal E}}
\newcommand{\cG}{{\cal G}}
\newcommand{\cH}{{\cal H}}
\newcommand{\bcH}{\boldsymbol {\cal H}}
\newcommand{\cK}{{\cal K}}
\newcommand{\cO}{{\cal O}}
\newcommand{\cR}{{\cal R}}
\newcommand{\cS}{{\cal S}}
\newcommand{\dcS}{\ddot{{\cal S}}}
\newcommand{\ds}{\ddot{{s}}}
\newcommand{\cT}{{\cal T}}
\newcommand{\cU}{{\cal U}}
\newcommand{\wt}[1]{\widetilde{#1}}

\newcommand{\mA}{\mathbb{A}}
\newcommand{\mE}{\mathbb{E}}
\newcommand{\mG}{\mathbb{G}}
\newcommand{\mR}{\mathbb{R}}
\newcommand{\mS}{\mathbb{S}}
\newcommand{\mU}{\mathbb{U}}
\newcommand{\mV}{\mathbb{V}}
\newcommand{\mW}{\mathbb{W}}

\newcommand{\uq}{\underline{q}}
\newcommand{\ubq}{\underline{\boldsymbol q}}

\newcommand{\gre}[1]{\textcolor[rgb]{0,1,0}{#1}}
\newcommand{\blu}[1]{\textcolor[rgb]{0,0,1}{#1}}
\newcommand{\higl}[1] {\textcolor{black}{#1}}

\title{
SANSee: A Physical-layer Semantic-aware
Networking Framework for Distributed Wireless
Sensing}

\author{
Huixiang~Zhu, Yong~Xiao, \IEEEmembership{Senior~Member, IEEE}, Yingyu Li, Guangming~Shi, \IEEEmembership{Fellow, IEEE}, \\and Marwan Krunz, \IEEEmembership{Fellow, IEEE}

\thanks{*This work is accepted at IEEE Transactions on Mobile Computing. Copyright may be transferred without notice, after which this version may no longer be accessible.

A preliminary version of this paper was presented at the IEEE International Conference on Network Protocols (ICNP), Reykjavik, Iceland, October 2023 \cite{ZhuHuixiang2023ICNPWorkshop}. 

H. Zhu is with the School of Electronic Information and Communications at the Huazhong University of Science and Technology, Wuhan, China 430074 (e-mail: zhuhuixiang@hust.edu.cn).

Y. Xiao is affiliated with the School of Electronic Information and Communications at the Huazhong University of Science and Technology, Wuhan 430074, China, the Peng Cheng Laboratory, Shenzhen, Guangdong 518055, China, and the Pazhou Laboratory (Huangpu), Guangzhou, Guangdong 510555, China (e-mail: yongxiao@hust.edu.cn).

Y. Li is with the School of Mechanical Engineering and Electronic Information, China University of Geosciences, Wuhan, China 430074 (e-mail: liyingyu29@cug.edu.cn).

G. Shi is with the Peng Cheng Laboratory, Shenzhen, Guangdong 518055, China, the School of Artificial Intelligence, the Xidian University, Xi’an, Shaanxi 710071, China, and the Pazhou Laboratory (Huangpu), Guangzhou, Guangdong 510555, China (e-mail: gmshi@xidian.edu.cn).


M. Krunz is with the Department of Electrical and Computer Engineering, the University of Arizona, Tucson, AZ 85721 (e-mail: krunz@arizona.edu).


}
}
\maketitle

\begin{abstract}
Contactless device-free wireless sensing has recently attracted significant interest due to its potential to support a wide range of immersive human-machine interactive applications using ubiquitously available radio frequency (RF) signals. Traditional approaches focus on developing a single global model based on a combined dataset collected from different locations. However, wireless signals are known to be location and environment specific. Thus, a global model results in inconsistent and unreliable sensing results. It is also unrealistic to construct individual models for all the possible locations and environmental scenarios. Motivated by the observation that signals recorded at different locations are closely related to a set of physical-layer semantic features, in this paper we propose SANSee, a semantic-aware networking-based framework for distributed wireless sensing. SANSee allows models constructed in one or a limited number of locations to be transferred to new locations without requiring any locally labeled data or model training. SANSee is built on the concept of physical-layer semantic-aware network (pSAN), which characterizes the semantic similarity and the correlations of sensed data across different locations. A pSAN-based zero-shot transfer learning solution is introduced to allow receivers in new locations to obtain location-specific models by directly aggregating the models trained by other receivers. We theoretically prove that models obtained by SANSee can approach the locally optimal models. Experimental results based on real-world datasets are used to verify that the accuracy of the transferred models obtained by SANSee matches that of the models trained by the locally labeled data based on supervised learning approaches.    
\end{abstract}

\begin{IEEEkeywords}
Semantic-aware network, physical-layer semantics, distributed wireless sensing, zero-shot transfer learning.
\end{IEEEkeywords}

\section{Introduction}
Wireless sensing has recently attracted significant interest due to its potential to achieve device-free movement detection and tracking in a wide range of applications, including smart healthcare, urban sensing, and unmanned surveillance systems. It is a key enabler of emerging applications that require immersive contact-free human-machine interactions, including augmented reality/virtual reality (AR/VR) and Tactile Internet\cite{xiao2020toward, XY2018TactileInternet}. Recent results show that by detecting changes in the RF signal propagation and reflection patterns caused by the human body, it is possible to recognize a wide range of human actions and gestures, such as falling, walking, sitting, etc. Furthermore, if wireless sensing data collected by multiple receivers can be jointly analyzed, more fine-grained human gestures, such as hand gestures and finger movement, can be detected \cite{liu2019wireless}.

\begin{figure}[!ht]
     \centering
     \begin{subfigure}[b]{0.24\textwidth}
         \centering
        \includegraphics[width=1\textwidth]{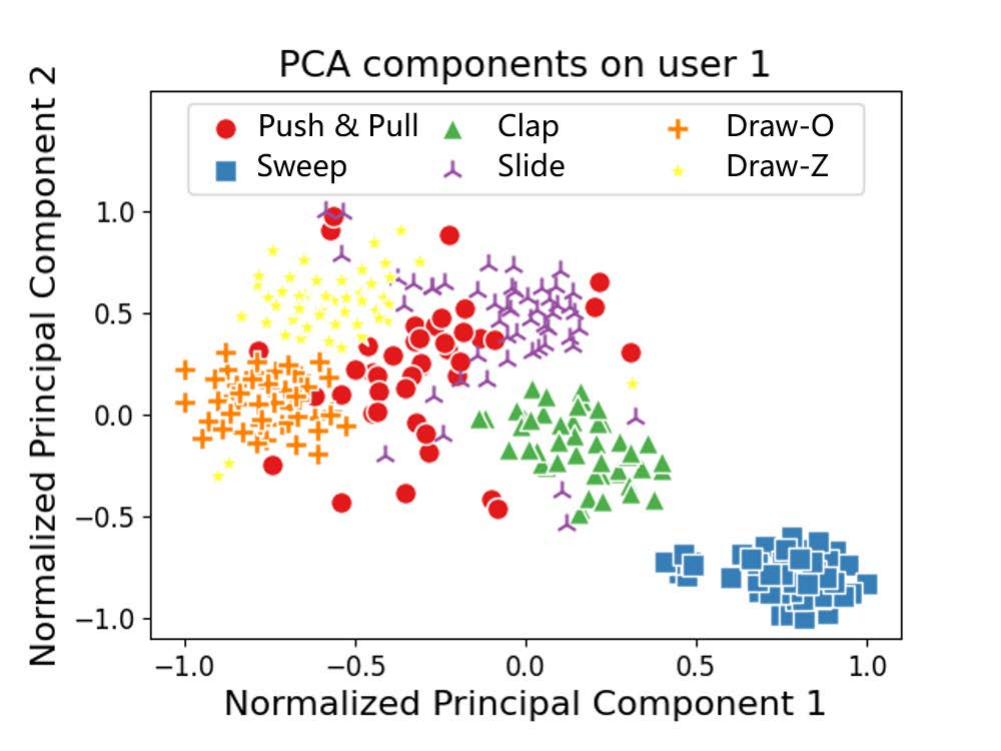}
         \caption{\bf }
         \label{dif_gesture}
     \end{subfigure}
     \hfill
     \begin{subfigure}[b]{0.24\textwidth}
     \centering
     \includegraphics[width=1\textwidth]{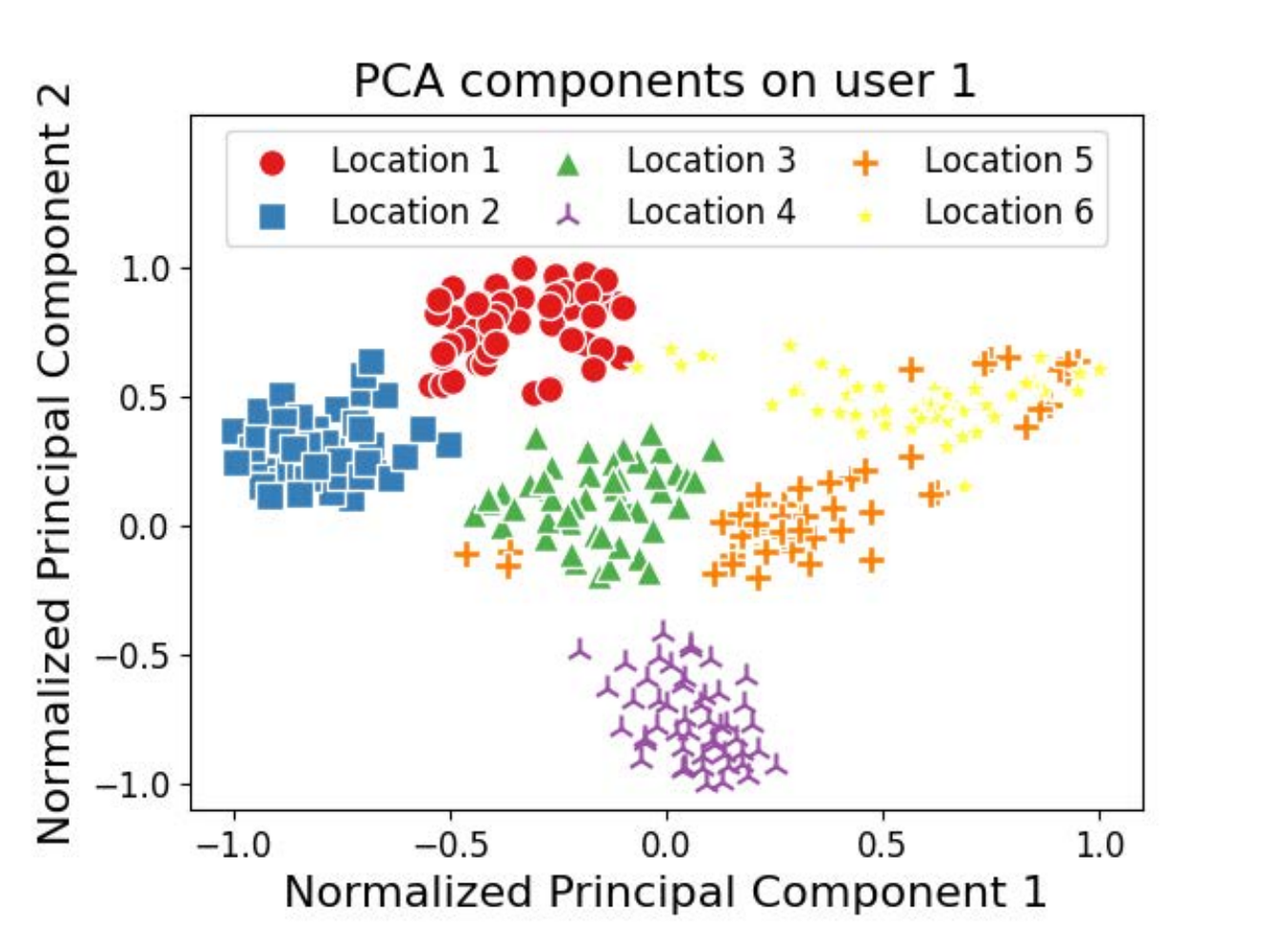}
         \caption{\bf }
         \label{dif_location}
     \end{subfigure}
    \caption{
    (a) Visualization of the statistical features of wireless signals recorded at the same location when different human gestures are performed, and 
    (b) visualization of statistical diversity of wireless signals recorded by receivers at different locations when the same gesture is performed.}
    \label{motivation_cluster}
\end{figure}

Most existing works on wireless sensing adopt a
one-fits-all approach, in which a centralized model is trained based on wireless sensing data recorded from a few locations and applied to a much wider range of locations and environments. 
However, wireless signals are known to exhibit highly temporal and spatial heterogeneity. Specifically, wireless signals are highly dependent on location, environment, and human-related factors. For example, different locations of transmitters, receivers, and objects, as well as room layouts will result in drastically different signal characteristics and data distributions. Furthermore, different body movement patterns (e.g., human gestures) and orientations will also result in different spatial and temporal variations of wireless sensing data. To shed more light on this observation, in Fig. \ref{motivation_cluster} we
use principal components analysis (PCA) to reduce data dimension and then visualize the resulting 2-dimension statistical features of wireless sensing signals \cite{van2008visualizing}, i.e., channel state information (CSI), recorded when the same person performs different gestures at the same location (Fig. \ref{motivation_cluster}(a)) and when the person performs the same gesture at different locations (Fig. \ref{motivation_cluster}(b)). We can observe that the statistical features vary significantly when different gestures are performed or when receivers are deployed at different locations. Accordingly, training a single global model by combining sensing data collected at different locations, while ignoring the unique features of each individual location, environment, and gesture profile, will significantly reduce the wireless sensing accuracy and will result in highly unreliable sensing performance across different locations and gestures.

One possible solution is to train separate models for different locations and environments. Unfortunately, this approach incurs too much overload and relies on a large number of high-quality labeled data samples. Also, due to physical space and cost limitations, it is generally unrealistic to have a highly dense deployment of sensors and receivers to collect data that covers all spatial and temporal resolutions of different users and their gestures.
\higl{To summarize, due to the heterogeneity of wireless signals and the scarcity of lablled samples, it is difficult for conventional distributed wireless sensing solutions to achieve a desired gesture recognition accuracy, especially when most receivers cannot collect labeled data samples or construct local models due to their limited computational and storage capabilities.}

To overcome the above challenges, we propose SANSee, a distributed wireless sensing framework that transfers the gesture recognition models trained for one or a few locations to new locations without training new models or collecting new data samples. Our proposed model is motivated by the observation that the statistics of the wireless signals recorded in a given location are closely related to a set of physical-layer semantic features, such as the spatial layout, environmental features, and gesture profiles. These physical-layer semantic features can be utilized to infer the statistical correlations between wireless sensing signals across different locations and environments for location-specific model construction and transfer. More specifically, we develop a novel {\it physical-layer semantic-aware networking} (pSAN) framework to characterize the similarity between physical-layer semantic features and correlations of wireless signal distributions at different locations and environmental scenarios. 
We then propose a {\it pSAN-based zero-shot transfer learning solution}, in which receivers at new locations and environments obtain location-specific gesture recognition models by directly aggregating the already trained models of other receivers. In our solution, the aggregation coefficients of the model transfer are calculated based on the correlations between semantic features of different locations. We theoretically prove that the aggregated model obtained by SANSee approaches the locally optimal model without requiring any locally labeled data or local model training. Extensive experiments conducted based on real-world datasets are presented to corroborate our theoretical results.

The key contributions of this paper are as follows:
\begin{itemize}
\item We identify the physical-layer semantic features, including environment-related and gesture-related semantics, called E- and G-semantics, respectively, that determine the distributions of wireless sensing signals under different physical environments and gesture profiles. We then introduce the pSAN framework, which captures similarity between physical-layer semantics of different locations at different physical environments.

\item We develop a zero-shot transfer learning solution based on pSAN, which allows receivers in new locations to obtain location-specific models by linearly aggregating the models trained by a few receivers. 

\item We present theoretical bounds on model training error and transfer errors of SANSee. We prove that the localized models obtained by SANSee approaches the locally optimal model in each specific location even without locally labeled data or local model training.

\item Extensive experiments are conducted based on real-world wireless sensing datasets consisting of multiple types of human gestures recorded at 18 different locations. Our results show that our proposed model aggregation solutions can match models trained by real labeled data, obtained through supervised learning.
\end{itemize}

The remainder of this paper is organized as follows. Related works are reviewed in Section \ref{Section_RelatedWork}. We introduce the system model and problem formulation in Section \ref{Section_SystemModel}. An overview of SANSee framework is provided in Section \ref{SANSee Overview}. The detailed procedures of physical-layer semantics estimation are discussed in Section \ref{Section_PhysicallayerSemanticEstimation}. \higl{The concept of semantic similarity and pSAN-based model correlation network are introduced in Section \ref{Section_SemanticSimilarity}. 
Model training and transfer algorithms are proposed in Sections 6.2 and 6.3, respectively. Theoretical results about model training error and transfer error are derived in Section \ref{Theoretical_Results}. Experimental results are presented in Section \ref{Performance_Evaluation}, and we conclude the paper in Section \ref{Conclusion}.}

\section{Related Work}
\label{Section_RelatedWork}
\noindent
{\bf RF-based Wireless Sensing:}
Distributed wireless sensing has emerged as a promising area of research, leveraging ubiquitous wireless signals to enable contactless and device-free localization, tracking, and activity recognition\cite{ma2019wifi, tan2022commodity}. 
Most existing works focus on capturing the spatial and temporal dynamics of a few parameters, such as Doppler frequency shift (DFS), Time-of-Flight (ToF), and Angle-of-Arrival (AoA)\cite{yousefi2017survey, wang2019survey}.  In \cite{kotaru2015spotfi} the authors proposed SpotFi for decimeter-level human localization based on the AoA and relative ToF information of dominant incident signals from the target to multiple receivers. In \cite{li2017indotrack} the authors designed a human trajectory tracking system named IndoTrack to achieve successive tracking in an indoor environment. The main idea behind IndoTrack is to first extract accurate DFS from noisy channel state information samples and then jointly estimate target velocity and location via probabilistic co-modeling of DFS and AoA information from wireless receivers. In \cite{zheng2019Widar} the authors proposed Widar3.0 to achieve cross-domain gesture recognition by feeding the domain-independent Body Coordinate Velocity Profile (BVP), extracted from CSIs into a hybrid deep learning model, which consists of a convolutional neural network (CNN) for spatial feature extraction and a recurrent neural network (RNN) for temporal modeling.


\noindent
{\bf Semantic-Aware Networking:} Utilizing semantic knowledge to enhance communication and networking performance has recently attracted significant interest\cite{XY2021SemanticCommMagazine, BeyondTransmittingBitsPaper}. Most existing works focus on extracting human language-inspired semantic information to compress various forms of human generated signals, and improve communication efficiency and reliability\cite{weng2021semantic,  9830752, xie2021deep}. For example, in \cite{weng2021semantic} the authors adopted an attention mechanism-based solution to
compress speech signals in which essential speech information is identified by providing higher weights to them when training the neural network. In \cite{xie2021deep}, the authors considered a Transformer-based language text compression for maximizing the system capacity and minimizing the semantic errors by recovering the meaning of sentences. Multi-modal data compression was also investigated in \cite{9830752}, where a task-oriented semantic communications framework was proposed to unify the structure of transmitters for different tasks. In addition to compressing and recovering data bits, recent studies suggested that semantic information has a higher efficiency in recovering signals with high human-oriented perception quality. The so-called rate-distortion-perception tradeoff has been investigated in semantic communication\cite{chai2023rate,xiao2022rate}, where studies show that in some cases the receiver can directly infer the semantic information source satisfying certain distortion and perception constraints without requiring any data communication from the transmitter.
Recently, semantic information has also been utilized to enable high-level reasoning and inference in communication networks\cite{xiao2022imitation,xiao2023reasoning}. More specifically, the so-called implicit semantic-aware communication network was proposed in \cite{xiao2022imitation} in which the semantic correlations have been exploited to infer implicit information, such as clue information or background knowledge that are closely related to the data information sent over the network.
In addition to communication networks,
semantic knowledge has recently been extended to 
other fields, such as mmWave beam tracking\cite{FeiFeiGao2023EnvironmentSemantics}, 
image and video segmentation\cite{feng2020taplab}, emotional analysis\cite{liu2023emotion}, and affective computing\cite{deng2022multimodal}.
In contrast to all these existing works, in this paper, we introduce the concept of physical-layer semantics to capture the impact of environmental and human-related features that influence the distribution of wireless sensing data.  To the best of our knowledge, this is the first work that utilizes the semantic similarity of physical-layer features to transfer models between different locations and environmental scenarios.

\noindent
{\bf Transfer Learning-based Wireless Sensing:}
To reduce the cost of model training, transfer learning methods have been recently applied to wireless sensing, with the goal to transfer knowledge obtained from a source domain to a target domain, so as to support a variety of wireless sensing tasks \cite{yang2023sensefi}. A straightforward idea is to extract domain-independent features from labeled samples in the source domain. For example, in \cite{jiang2018towards, zou2018robust, wang2022airfi}, the authors show that adversarial architectures such as generative adversarial networks (GANs) can be used to learn the hidden relationships between the source inputs and the target outputs by combining a CNN feature extraction and a domain discriminator. Although integrating GANs into distributed wireless sensing solutions is a promising direction, it demands numerous ad-hoc “tricks” to achieve model convergence \cite{li2021deep}. In \cite{zhang2018crosssense}, CrossSense was introduced as the state-of-the-art wireless transfer technique on WiFi-based gait identification and gesture recognition applications.
\higl{To enable cross-domain sensing, CrossSense employs an artificial neural network (ANN) based mixture-of-experts strategy, where multiple specialized sensing models, or experts, are used to capture the mapping from diverse sourcing inputs to the targeting outputs.}

\noindent
{\bf Federated Learning-based Wireless Sensing:}
Federated learning (FL) is an emerging solution that enables distributed model training by utilizing model parameter sets instead of private data samples for sharing \cite{mcmahan2017communication}. FL-based wireless sensing solutions have recently attracted significant interest due to their unique advantages, including decentralization, low communication overload, and privacy protection \cite{nguyen2021federated, yang2023sensefi}. For instance, the authors in \cite{hernandez2021wifederated} designed WiFederated for WiFi-based human activity recognition, which was the first FL-based wireless sensing framework proposed to overcome the challenge posed by the centralized model training paradigm. In \cite{zhang2022Cross} the authors introduced a cross-domain federated learning framework called CDFL, which aims at addressing the scarcity of labeled wireless data by generating simulated training data using a
physical model guided by public datasets in other
domains. Recent works\cite{liu2019floc, nagia2022federated} also investigated distributed indoor localization by combining FL and wireless sensing 
based on receivers deployed across different locations.

\section{System Model and Problem Formulation}
\label{Section_SystemModel}
\subsection{System Model}

We consider human gesture recognition based on a distributed wireless sensing system consisting of one or more Wi-Fi transmitters and a set ${\cal K}$ of $K$ receivers deployed at different locations across the considered area. Each receiver records wireless signals (e.g., CSI data) that are reflected and scattered by human users when performing a set of gestures. We focus on the decentralized sensing scenario in which each receiver stores its recorded wireless signals locally which, due to the constraints in data privacy, cannot be exposed to others. We assume that only a subset of receivers ${\cal K}^L$ for ${\cal K}^L \subseteq {\cal K}$ can have labeled wireless sensing data samples. Each receiver in ${\cal K}^L$ can then construct a location-specific model to recognize different gestures of the human users based on its local dataset. There are some other receivers, denoted as subset ${\cal K}^N = {\cal K}\setminus {\cal K}^L$ that cannot have any labeled data and therefore cannot construct any local models using traditional supervised learning approaches. As mentioned earlier, due to the spatial heterogeneity of wireless sensing signals, receivers at different locations require different models to recognize the same gestures. In other words,
receivers in ${\cal K}^N$ cannot directly utilize the gesture recognition models of receivers in ${\cal K}^L$ for their local gesture recognition tasks.

\subsection{Physical-layer Semantics}

We observe that the statistics of received CSI signals are closely related to the semantic information of the physical environment,
such as the size and layout of rooms, the location of transmitters and receivers, and the human users' gesture profiles, such as the speed of movement of different body parts when performing different gestures, etc. Motivated by this observation, we investigate whether it is possible to develop a model transferring solution that allows one or a limited number of receivers with labeled data to transfer their locally trained models to other receivers, especially receivers without any labeled dataset, based on the correlations of environmental and gesture-related semantic features.

Let us first identify the key semantic features in wireless sensing systems that may influence the distribution of the CSI data recorded at each receiver. It is known that the CSI signal recorded by a receiver is mainly characterized by the wireless links connecting the transmitter and receiver, influenced by the gesture-performing human users as well as the physical objects located along side of the channels. More specifically, the CSI signal recorded by receiver $k$ at arrival time $\alpha$, subcarrier frequency $\theta$, and antenna $\beta$ can be written as \cite{zheng2019Widar}:
\begin{eqnarray}\label{eq_ReceivedSignal}
\lefteqn{ H_k\left(\alpha, \theta, \beta \right)= \left( \sum_{n\in {\cal L}_S} A_{k,n} e^{ -j 2 \pi \theta \tau_{k,n}(\theta, \beta)} \right.} \nonumber \\
&&\;\;\; \left. +\sum_{m\in {\cal L}_M} A_{k,m}(\alpha) e^{-j 2 \pi \theta \tau_{k,m}(\alpha, \theta, \beta)} \right) e^{j\epsilon(\alpha, \theta, \beta)},
\end{eqnarray}
where ${\cal L}_S$ and ${\cal L}_M$ are sets of stationary and dynamic path components, respectively, and $e^{j\epsilon(\alpha, \theta, \beta)}$ is the phase error caused by asynchronization between transceivers and hardware imperfection. For each propagation path $l$ for $l\in {\cal L}_S \cup {\cal L}_M$, $A_{k, l}$ and $\tau_{k, l}$ are the channel attenuation factor and time delay, respectively. Here dynamic path components correspond to the received signals reflected by the moving targets, while the stationary path components correspond to the signals received from the direct paths and the reflection signals from static objects such as walls and furniture. Since the CSI can only be sampled as discrete signals in time (packet), frequency (subcarrier), and space (antenna) \cite{qian2018widar2}, the time delay of static and dynamic signal paths, respectively, in (\ref{eq_ReceivedSignal}) can be written as follows:
\begin{eqnarray}
\tau_{k,n}\left(\theta, \beta \right)&=& \tau_{k,0} + \Delta \beta_{k,n} \cdot \varpi_{k,0}, \mbox{for } n \in {\cal L}_S \label{eq_signalphase_1} \\
 \tau_{k,m}\left(\alpha, \theta, \beta \right)&=& \tau_{k,0} -\frac{\rho_{k,0}}{\Delta \theta_{k,m}}\Delta \alpha_{k,m} + \Delta \beta_{k,m} \cdot \varpi_{k,0}, \nonumber \\
 &&\;\;\;\;\;\;\;\;\;\;\;\;\;\;\;\;\;\;\;\;\;\;\;\;\;\;\;\;\;\;\; \mbox{for } m \in {\cal L}_M  \label{eq_signalphase_2}
\end{eqnarray}
where $\Delta \alpha_{k,l}$, $\Delta \theta_{k,l}$, $\Delta \beta_{k,l}$ for $l\in {\cal L}_S \cup {\cal L}_M$ are differences of packets, subcarriers, and spatial positions, respectively, between two consecutive CSI samples of
$H_k\left(\alpha, \theta, \beta \right)$ in (\ref{eq_ReceivedSignal}). $H_k\left(0,0,0 \right)$ is defined as the CSI reference signal with the time delay $\tau_{k,0}$, DFS $\rho_{k,0}$ and AoA $\varpi_{k,0}$.

From (\ref{eq_ReceivedSignal}), we can observe that the CSI signals recorded by receiver $k \in  {\cal K}$ are closely related to the following two types of physical-layer semantics:

\noindent
{\bf Environment-related semantics (E-semantics)}: include the semantic information related to the physical environment such as environmental layout and the relative locations and orientations of transmitters, receivers, and human users. We therefore can write the feature vector of E-semantics of receiver $k$ as $\bu_k = \langle A_{k,n}, \tau_{k,n}, \varpi_{k,n} \rangle_{n\in {\cal L}_S}$.

\noindent
{\bf Gesture-related semantics (G-semantics)}: include the semantic information associated with gestures such as the users' body coordinates and movement patterns of gestures.
\higl{We can write the feature vector of G-semantics of receiver $k$ as $\bv_k =\langle A_{k,m}, \tau_{k,m}, \varpi_{k,m}, \rho_{k,m} \rangle_{m\in {\cal L}_M}$. }

We can then rewrite (\ref{eq_ReceivedSignal}) into the following form:
\begin{eqnarray}\label{eq_RS}
 H_k\left(\alpha, \theta, \beta \right)&=&  \sum_{n\in {\cal L}_S} p_{k,n}(\theta, \beta; \bu_k) \nonumber \\
&& +\sum_{m\in {\cal L}_M} q_{k,m}(\alpha, \theta, \beta; \bv_k),
\end{eqnarray}
where $p_{k,n}(\theta, \beta; \bu_k)=A_{k,n} e^{ -j 2 \pi \theta \tau_{k,n}(\theta, \beta)+j\epsilon(\alpha, \theta, \beta)}$ and $q_{k,m}(\alpha, \theta, \beta; \bv_k)=A_{k,m}(\alpha) e^{-j 2 \pi \theta \tau_{k,m}(\alpha, \theta, \beta)+j\epsilon(\alpha, \theta, \beta)}$ are stationary and dynamic path component signals, respectively.

We combine both E- and G-semantics and write the physical-layer semantic feature vector of wireless signals recorded by receiver $k$ as $\bphi_k = \langle \bu_k, \bv_k \rangle$. We can observe that the physical-layer semantics are location-specific and therefore each receiver $k$ has a unique semantic feature vector $\bphi_k$ which plays a key role in determining the probability distribution of the locally received CSI signals.

\subsection{Physical-Layer Semantic-Aware Network}

Let us now formally introduce the concept of physical-layer semantic-aware network (pSAN) as follows:
\begin{definition}
A {\em physical-layer semantic-aware network} (pSAN) is a wireless sensing network in which the physical-layer semantics, including both E- and G-semantics, can be aware, known,  and utilized, by each receiver.
\end{definition}

In pSAN, the similarity of physical-layer semantics between different receivers
can be used to infer correlations between different location-specific models trained by these receivers. Recall that only a subset ${\cal K}^L$ of $K^L$ receivers can have labeled CSI signals. To simplify our description, we use $k'$ for $k'\in {\cal K}^L$ to denote the $k$th receiver with labeled CSI data. Let ${\cal D}_{k'}$ be the set of labeled CSI data at receiver $k'$. We assume the labeled data samples at different receivers in ${\cal K}^L$ are associated with the same set of gesture classes. Similarly, let $k''$ for $k'' \in {\cal K}^N$ be the $k''$th receiver that does not have any labeled data.

The key idea is to establish a mapping function that converts different high-dimensional physical-layer semantics into the same low-dimensional semantic space to capture the similarity between the key statistic features of physical-layer semantics that determine the gesture recognition models trained by different receivers. Specifically, let $\bar\bphi_k$ be the low-dimensional semantic vectors converted from $\bphi_k$ to the semantic space for $k\in {\cal K}$.
\higl{Common metrics for measuring semantic similarity include energy-based and statistic-based metrics. In this paper, we mainly focus on energy-based semantic similarity. We will present a formal definition and give a more detailed discussion in Section 6. Without loss of generality, in this paper, we use $S \left( \bar\bphi_{j}, \bar\bphi_{k} \right)$ to denote the semantic similarity between two semantic features $\bphi_{j}$ and $ \bphi_{k}$.}

\subsection{Problem Formulation}

Each labeled CSI data $\zeta_{k',i} = \langle x_{k',i}, y_{k',i} \rangle$ recorded by receiver $k'$ for $k'\in{\cal K}^L$ consists of a CSI signal $x_{k',i}$, e.g., an instance of CSI signal recorded by receiver $k'$, and a class label $y_{k',i}$ that belongs to one of a set of gesture classes $\cal Y$. Let ${\cal D}_{k'}$ be the set of local training data samples at receiver $k'$. Each receiver $k'\in {\cal K}^L$ can then construct a local model $\pmb{\omega}_{k'}$ by minimizing its local objective function,
\begin{eqnarray}\label{local_objective_function}
\min_{\pmb{\omega}_{k'}} F_{k'} \left(\pmb{\omega}_{k'} \right) = {\frac{1}{|{\cal D}_{k'}|}} \sum_{\zeta_{k',i} \in {\cal D}_{k'}}\left[f_{k'}\left(\pmb{\omega}_{k'}; \zeta_{k',i} \right)\right],
\end{eqnarray}
where $\pmb{\omega}_{k'}$ is the model parameters of receiver $k'$.

We also need to learn a semantic-aware model transfer function to transfer models learned by receivers with labeled data to those receivers without any labeled data according to their semantic similarity. In our considered decentralized wireless sensing scenario, the CSI data recorded by each receiver cannot be exposed to others. It is however possible for the receivers to expose their locally trained models to other receivers. In the rest of this paper, we will develop a pSAN-based model aggregation and transfer approach in which each receiver $k''\in {\cal K}^N$ can directly obtain a location-specific model by aggregating models that are already trained by receivers in ${\cal K}^L$.

The main objective is to design an appropriate model transfer approach, so the transferred model at receiver $k''$ can approach the locally optimal model $\pmb{\omega}^{*}_{k''}$, i.e., we write the problem as follows:
\begin{eqnarray}\label{main_objective}
\min_{\pmb{\omega}_{k''}} \left[F_{k''} \left(\pmb{\omega}_{k''} \right) - F_{k''} \left(\pmb{\omega}^*_{k''} \right)\right], \quad\forall k'' \in {\cal K}^N,
\end{eqnarray}
where $\pmb{\omega}_{k''}$ is the transferred model obtained by receiver $k''$ which, if we consider a linear model transfer framework, can be obtained as follows:
\begin{eqnarray}\label{aggregate_euqa}
   \higl{\pmb{\omega}_{k''} = \sum_{k' \in {\cal K}'} \xi \left( S ( \bar\bphi_{k'}, \bar\bphi_{k''} ) \right) \pmb{\omega}_{k'},}
\end{eqnarray}
\higl{where $S ( \bar\bphi_{k'}, \bar\bphi_{k''} )$ denotes the semantic similarity between semantics $\bar\bphi_{k'}$ of receiver $k'$ and semantics $\bar\bphi_{k''}$ of receiver $k''$,}
$\xi (\cdot)$ is a semantic-aware model transfer function that maps the semantic similarity between receivers $k'$ and $k''$ to a normalized model aggregation coefficient value.
\higl{We will give a more detailed discussion on how to obtain $S ( \cdot,\cdot )$ and $\xi (\cdot)$ in Section 6 and prove the convergence result of our proposed solutions later in Section 7.}

\begin{figure}[!ht]
     \centering
     \begin{subfigure}[b]{0.45\textwidth}
        \centering
        \includegraphics[width=1\textwidth]{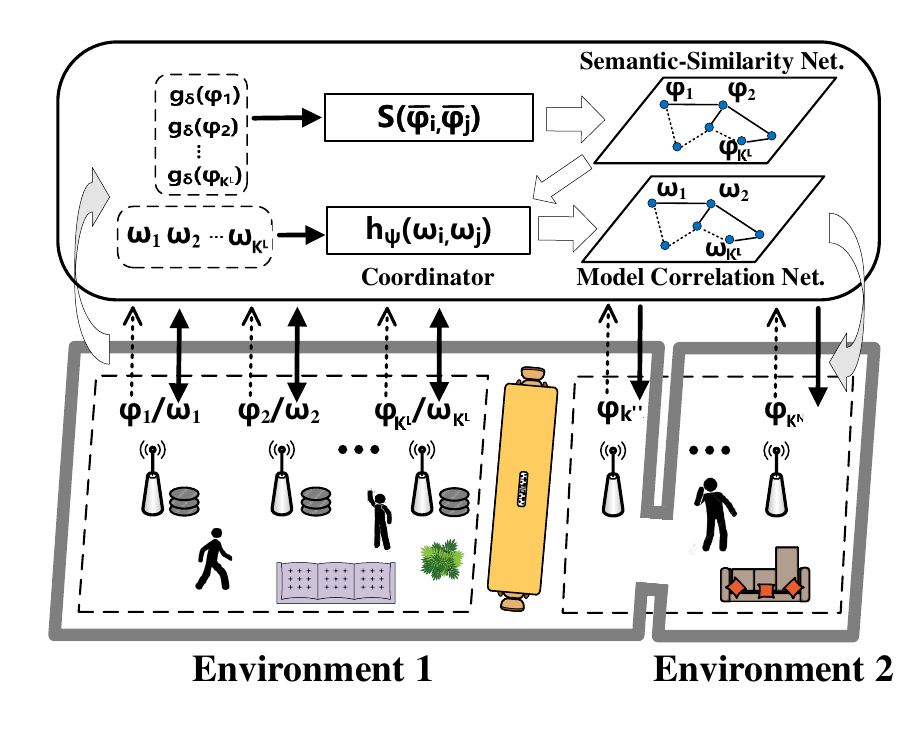}
         \caption{}
         \label{overview_v9}
     \end{subfigure}
    \begin{subfigure}[b]{0.45\textwidth}
        \centering
        \includegraphics[width=1\textwidth]{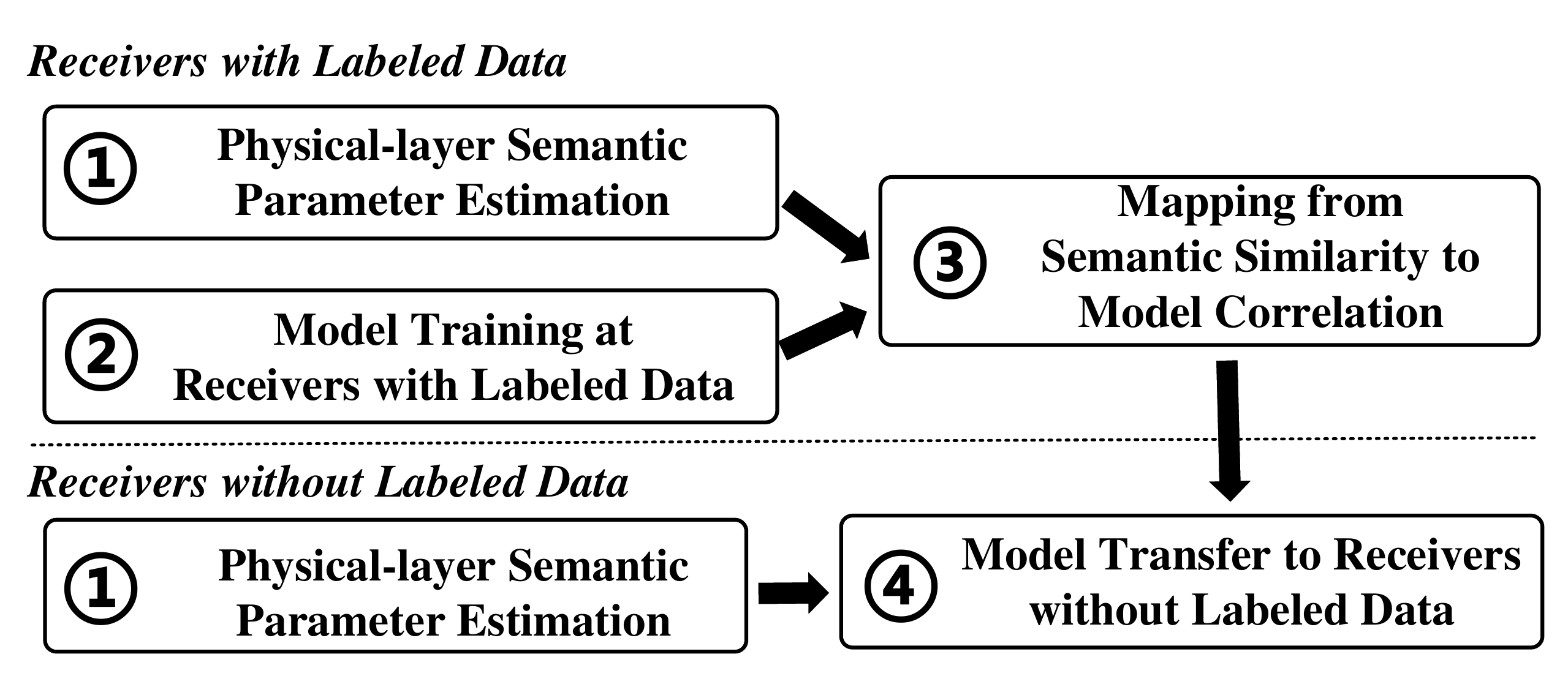}
        \caption{}
         \label{System_Flowchart}
     \end{subfigure}
    \caption{
    (a) SANSee framework and (b) key
    training procedures.}
    \label{overview}
    \vspace{-0.5cm}
\end{figure}

\section{SANSee Overview}\label{SANSee Overview}

\higl{The architectural framework and key training procedures of SANSee are illustrated in Fig. \ref{overview_v9} and \ref{System_Flowchart}, respectively. The detailed operations are described as follows: 
}

\noindent
{\bf Physical-layer Semantics Estimation:}
Each receiver needs to first estimate key physical-layer semantic parameters that influence its local CSI data. Note that estimating semantic parameters does not require any labeled CSI data. To estimate E- and G- semantics separately, each receiver needs to first separate its CSI signals by applying the high-pass and low-pass filters, respectively, and then apply the maximum likelihood estimation (MLE) approach to estimate the combination of different semantic parameters.

\noindent
{\bf Mapping from Semantic Similarity to Model Correlation:} After each receiver has successfully estimated its physical-layer semantics, we then need to construct a mapping function that can convert the semantic similarity to the model correlation between different receivers. To characterize the semantic similarity between different receivers, we introduce a low-dimensional semantic space in which the distance between any two physical-layer semantics is proportional to their semantic similarity. We then construct a mapping function to map the high-dimensional semantic feature vector into the semantic space. We also introduce a correlation coefficient to characterize the model correlation between local models trained by different receivers. Finally, we design a novel loss function to simultaneously optimize parameters of the semantic mapping function and the calculation function of the model correlation coefficient to match semantic similarity with model correlations.

\noindent
{\bf Model Training at Receivers with Labeled Data:} All the receivers with labeled data will jointly construct their location-specific models. We adopt a personalized federated learning-based solution for receivers to collaboratively train their location-specific models without exposing their local datasets. After successfully training their models, all the receivers with labeled data will link their models with their physical-layer semantics and establish a mapping function to convert semantic similarity to model correlation coefficients.

\noindent
{\bf Model Transfer at Receivers without Labeled Data:} Each receiver without labeled data will rely on the coordinator to construct its location-specific model based on the correlated model trained by receivers with labeled data. More specifically, each receiver without labeled data will submit its locally estimated semantic features to the coordinator. The coordinator will then apply the previously constructed mapping function to calculate the model correlation coefficients for all the correlated models obtained by receivers with labeled data, and finally send the aggregated model to each corresponding receiver.


\section{Physical-layer Semantics Estimation}
\label{Section_PhysicallayerSemanticEstimation}
The first step in pSAN is to quantify the impact of physical-layer semantics on the CSI data recorded by each receiver. From (\ref{eq_ReceivedSignal}), we can observe that, the raw CSI signal $H_k(\alpha, \theta, \beta)$ obtained by each receiver consists of phase error term $e^{j\epsilon(\alpha, \theta, \beta)}$ which may result in inaccurate estimation of physical-layer semantics. This issue can be addressed when the receiver has two or more antennas, in which the phase error term can be canceled by performing conjugate multiplication and amplitude adjustment on CSI signals received by two antennas \cite{li2017indotrack}. Let ${\hat H}_k(\alpha, \theta, \beta)$ be the phase error-canceled version of the CSI signal of receiver $k$. We also use $\hat{q}_{k,m}$ and $\hat{p}_{k,n}$ to denote dynamic and stationary path components in ${\hat H}_k(\alpha, \theta, \beta)$, respectively.

\higl{By applying a high-pass filter, we can separate the sum of dynamic components related to G-semantics $\bv_k$ from the raw CSI of receiver $k \in {\cal K}$, denoted as ${\hat H}^{M}_k\left(\alpha, \theta, \beta\right)=$ $\sum_{m\in {\cal L}_M} {\hat q}_{k,m}(\alpha, \theta, \beta; \bv_k)$.} We can then adopt the maximum likelihood estimation (MLE) to estimate the G-semantics parameters consisting of a collection of parameters of all dynamic signal components, i.e., $\bv_k=\{\bv_{k,m} \}_{m\in {\cal L}_M}$ with $\bv_{k,m}=\langle A_{k,m}, \tau_{k,m}, \varpi_{k,m}, \rho_{k,m} \rangle$. More specifically, the G-semantics $\bv^{*}_k$ of receiver $k$ can be estimated by solving the following problem:
\begin{eqnarray}\label{MLE_G}
 \bv^{*}_k=\arg\max_{\bv_k} \{-\sum_{\alpha \in {\cal A}, \theta \in {\bf \Theta}, \beta \in {\cal B}}| {\hat H}^{M}_k\left(\alpha, \theta, \beta\right) \nonumber\\
 -\sum_{m\in {\cal L}_M}q_{k,m}(\alpha, \theta, \beta; \bv_{k,m})|^2\},
\end{eqnarray}
where ${\hat H}^M_k\left(\alpha, \theta, \beta\right)$ is the obtained from real-measured CSI signal and $q_{k,m}(\alpha, \theta, \beta; \bv_{k,m})$ is the estimated components. ${\cal A}$, ${\bf \Theta}$, ${\cal B}$ are the sets of possible packets, subcarriers, and antennas, i.e., $\alpha \in {\cal A}, \theta \in {\bf \Theta}, \beta \in {\cal B}$.

Similarly, we can extract the sum of stationary components ${\hat H}^{S}_k(\theta, \beta)=\sum_{n\in {\cal L}_S} p_{k,n}(\theta, \beta; \bu_k)$ related to E-semantics by applying a low-pass filter and estimate parameters in E-semantics, i.e., $\bu_k=\{\bu_{k,n} \}_{n\in {\cal L}_S}$ with $\bu_{k,n}=\langle A_{k,n}, \tau_{k,n}, \varpi_{k,n}\rangle$ as follows:
\begin{eqnarray}\label{MLE_P}
 \bu^{*}_k=\arg\max_{\bu_k} \{-\sum_{ \theta \in {\bf \Theta}, \beta \in {\cal B}}| {\hat H}^{S}_k(\theta, \beta) \nonumber \\
 -\sum_{n\in {\cal L}_S}p_{k,n}(\theta, \beta; \bu_{k,n})|^2\},
\end{eqnarray}
where ${\hat H}^S_k\left(\alpha, \theta, \beta\right)$ is obtained from the real-measured CSI signal and $p_{k,n}( \theta, \beta; \bu_{k,n})$ is the estimated components.

We can observe that, it is generally difficult to derive closed-form solutions of $\bv^{*}_k$ and $\bu^{*}_k$ in (\ref{MLE_G}) and (\ref{MLE_P}). We can however adopt a modified Space Alternating Generalized Expectation Maximization (mSAGE) algorithm to estimate the values of $\bv^{*}_k$ and $\bu^{*}_k$ using an iteration-based  approach\cite{Fleury1999Channel}. We use superscript $t$ to denote the operation in the $t$th iteration. The $m$th dynamic signal path component can be calculated by first performing the expectation step as follows:
\begin{eqnarray}\label{SAGE}
\lefteqn{ q^{t+1}_{k,m} (\alpha, \theta, \beta;  \bv^t_{k,m})=q^t_{k,m}(\alpha, \theta, \beta; \bv^t_{k,m})}\\
&&+\pi_H\left({\hat H}^{M}_k\left(\alpha, \theta, \beta\right)-\sum_{m'\in {\cal L}_M}q^t_{k,m}\left(\alpha, \theta, \beta; \bv^t_{k, m'}\right)\right),\nonumber
\end{eqnarray}
where $\bv^t_{k, m'}$ is the G-semantics of the $m$-th path estimated in the $t$th iteration of receiver $k$, and $\pi_H$ is the non-negative step size and its default value can be set as 1. We then obtain the optimal value of parameter $\bv^*_{k,m}$
by maximizing the magnitude of the signal received at the $m$th signal path component $z_{k,m} (\tau, \varpi, \rho; q^{t+1}_{k,m}) = \sum_{\alpha \in {\cal A}, \theta \in {\bf \Theta}, \beta \in {\cal B}}$ $|e^{2\pi ( \Delta\theta_{k,m}\tau_{k,m}+f_c \Delta \beta_{k,m} \varpi_{k,m}-\Delta\alpha_{k,m} \rho_{k,m}) }q^{t+1}_{k,m}\left(\alpha, \theta, \beta; \bv^t_{k,m} \right)|^2$, i.e. $\bv^*_{k,m}$ is given by,
\begin{eqnarray}\label{SAGE_objective}
\bv^*_{k,m} = \arg \max\limits_{\bv_{k,m}} z_{k,m} (\tau, \varpi, \rho; q^{t+1}_{k,m}),
\end{eqnarray}
where $f_c$ is the carrier frequency of the wireless channel, and $\Delta \alpha_{k,m}$, $\Delta \theta_{k,m}$, $\Delta \beta_{k,m}$ are defined previously in (\ref{eq_signalphase_2}). To solve (\ref{SAGE_objective}), we apply the following steps to sequentially estimate each individual parameter $\tau^{t+1}_{k,m}$, $\varpi^{t+1}_{k,m}$, $\rho^{t+1}_{k,m}$, and $A^{t+1}_{k,m}$ in $\bv_{k,m}$ as follows:
\begin{eqnarray}\label{SAGE_steps}
\tau^{t+1}_{k,m}=\arg\max_{\tau}|z_{k,m}(\tau, \varpi^t_{k,m}, \rho^t_{k,m}; q^{t+1}_{k,m} (\alpha, \theta, \beta;  \bv^t_{k,m}))|^2,&&\\
\varpi^{t+1}_{k,m}=\arg\max_{\varpi}|z_{k,m}(\tau^{t+1}_{k,m}, \varpi, \rho^t_{k,m}; q^{t+1}_{k,m} (\alpha, \theta, \beta;  \bv^t_{k,m}))|^2,&&\\
\rho^{t+1}_{k,m}=\arg\max_{\rho}|z_{k,m}(\tau^{t+1}_{k,m}, \varpi^{t+1}_{k,m}, \rho; q^{t+1}_{k,m} (\alpha, \theta, \beta;  \bv^t_{k,m}))|^2,&&\\
A^{t+1}_{k,m}=\frac{z_{k,m}(\tau^{t+1}_{k,m}, \varpi^{t+1}_{k,m}, \rho^{t+1}_{k,m}; q^{t+1}_{k,m} (\alpha, \theta, \beta;  \bv^t_{k,m}))}{|{\cal A}|\cdot|{\bf \Theta}|\cdot|{\cal B}|}.&&
\end{eqnarray}
The above iteration process ends when the difference between two successive estimations of $\bv_{k,m}$ is within a pre-defined threshold $\varsigma$. For stationary component signal estimation, we can follow a similar approach to estimate the parameters of $\bu_k$ for receiver $k$. The detailed procedures of the physical-layer semantics estimation process are summarized in Algorithm \ref{Algorithm1}.

\begin{algorithm}[H]
\caption{Physical-layer Semantics Estimation Algorithm of Receiver $k$}\label{Algorithm1}\footnotesize
	{\bf Input}: CSI $ H_k(\alpha, \theta, \beta)$; Numbers of estimated paths $L_M$ and $L_N$; Pre-defined threshold $\varsigma$; Initial iteration $t=0$; Initial values $\bu_k=0$, $\bv_k=0$.\\
	{\bf Output}: Physical-layer semantics $\phi_k=\langle \bu^{*}_k,\bv^{*}_k \rangle$.
    \begin{itemize}
    \item[ 1:] Cancel $ H_k(\alpha, \theta, \beta)$ by denoising and obtain ${\hat H}_k (\alpha, \theta, \beta)$ ;
    \item[ 2:] {\bf While} $\|\bv^t_{k,m}-\bv^{t+1}_{k,m} \|\le\varsigma$ {\bf do}
    \item[ 3:] \quad {\bf For} $m$ = 1, $\cdots$, $L_M$ {\bf do}
    \item[ 4:] \quad \quad Apply a high-pass filter to obtain ${\hat H}^{M}_k\left(\alpha, \theta, \beta\right)$;
    \item[ 5:] \quad \quad Calculate  $q^{t+1}_{k,m}(\alpha, \theta, \beta; \bv^t_{k,m})$ by using (\ref{SAGE});
    \item[ 6:] \quad \quad Estimate parameters of $\bv^{t+1}_{k,m}$ by using (\ref{SAGE_steps})-(15);
    \item[ 7:] \quad {\bf End for}
    \item[ 8:] \quad $t=t+1$;
    \item[ 9:] {\bf End while}
    \item[ 10:] {\bf While} $\|\bu^t_{k,n}-\bu^{t+1}_{k,n} \|\le\varsigma$ {\bf do}
    \item[ 11:] \quad {\bf For} $n$ = 1, $\cdots$, $L_N$ {\bf do}
    \item[ 12:] \quad \quad Apply a low-pass filter to obtain ${\hat H}^{S}_k\left( \theta, \beta\right)$;
    \item[ 13:] \quad \quad Calculate  $p^{t+1}_{k,n}(\theta, \beta; \bu^t_{k,n})$ by substituting ${\hat H}^{S}_k$ into (\ref{SAGE});
    \item[ 14:] \quad \quad Estimate  $\bu^{t+1}_{k,n}$ by substituting $p^{t+1}_{k,n}$ into  (\ref{SAGE_steps})-(15);
    \item[ 15:] \quad {\bf End for}
    \item[ 16:] \quad $t=t+1$;
    \item[ 17:] {\bf End while}
    \end{itemize}
\end{algorithm}

\section{\higl{Model Training and Transfer}}
\label{Section_ModelTrainandTransfer}

\subsection{\higl{Semantic Similarity and Model Correlations}}
\label{Section_SemanticSimilarity}

\higl{From the previous discussion, we can observe that the physical-layer semantics directly affect the distributions of the CSI data at each receiver. It is known that, for a given algorithmic framework, the distribution of training dataset and the resulting model are in one-to-one correspondence. Thus, in this section, we aim to develop a mapping function that converts the semantic similarity to the correlations of models.}

 Motivated by the fact that physical-layer semantics of each receiver consist of multiple key parameters that have different impacts on the performance of different gesture-recognition tasks, we need to first convert the high-dimensional physical-layer semantics of different receivers into a low-dimensional space referred to as the (physical-layer) semantic space. In the semantic space, the distance between different semantics of different receivers is proportional to the correlations of their local gesture recognition models, e.g., the larger the distance (similarity) between receivers' semantics, the higher the correlations between different local models of different receivers. In this way, we can use the semantic similarity to transfer models from some receivers, e.g., receivers with labelled data, to other receivers, e.g., receivers without labeled data, without requiring any extra model training.

In this paper, we consider a neural network-based mapping function to convert the high-dimensional physical-layer semantics $\boldsymbol{\phi}_k$ into the low-dimensional version $\bar{\boldsymbol{\phi}}_k$ in the semantic space. We can write the mapping function that outputs the low-dimensional semantic representation as $\bar{\boldsymbol{\phi}}_k=g_{\delta}(\boldsymbol{\phi}_k)$, where $\delta$ is the parameters of the mapping function.

Let $S\left(\bar{\boldsymbol{\phi}}_k, \bar{\boldsymbol{\phi}}_j\right)$ be the semantic similarity between receivers $k$ and $j$ in the semantic space. We consider a general framework in which semantic similarity can be measured using different metrics. For example, if the Euclidean distance has been adopted to measure similarity between two semantics $\bar{\boldsymbol{\phi}}_k$ and $\bar{\boldsymbol{\phi}}_j$ in semantic space, we can write:
\begin{eqnarray}\label{MSE}
S\left(\bar{\boldsymbol{\phi}}_k, \bar{\boldsymbol{\phi}}_j\right)=S\left(g_{\delta}(\boldsymbol{\phi}_k), g_{\delta}(\boldsymbol{\phi}_j)\right)=|\bar{\boldsymbol{\phi}}_k- \bar{\boldsymbol{\phi}}_j|^2.
\end{eqnarray}
We can also use other types of metrics such as statistic-based similarity metrics, including cross-entropy (CE) and Jensen–Shannon divergence (JSD), by following the same line in \cite{xiao2022imitation}.

Next, we need to define the correlation between gesture-recognition models trained based on datasets available at different receivers. In this paper, we adopt a linear correlation in which the correlation between different models $\pmb{\omega}_j$ and $\pmb{\omega}_k$ is characterized by a linear coefficient $\xi_{j,k}$. If suppose model $\pmb{\omega}_j$ is correlated with a set of models, e.g., $\{\pmb{\omega}_k\}_{k\in{\cal K}^L}$ for $j\notin {\cal K}^L$, we then can write model $\pmb{\omega}_{j}$ as a linear combination of all the correlated models with normalized coefficients given by $\pmb{\omega}_j=\sum_{k\in{\cal K}^L}\xi_{j,k}\pmb{\omega}_k$, where $\xi_{j,k}$ satisfies $0\le\xi_{j,k}\le 1 $ and $\sum_{k\in{\cal K}^L}\xi_{j,k}=1$. Suppose the model correlation coefficient $\xi_{j,k}$ can also be learned by a neural network $h_{\psi}$ with parameter $\psi$, i.e., we can write $\xi_{j,k}=h_{\psi}(\pmb{\omega}_j, \pmb{\omega}_k)$.

Finally, we can use the following loss function to train parameters $\delta$ and $\psi$ to match the semantic similarity with the model correlation:
\begin{eqnarray}\label{transfer_loss}
{\cal L}(\delta, \psi) = \sum_{j,k\in{\cal K}^L}|h_{\psi}(\pmb{\omega}_j, \pmb{\omega}_k)-S\left(\bar{\boldsymbol{\phi}}_k, \bar{\boldsymbol{\phi}}_j\right)|^2.
\end{eqnarray}
The models $\delta$ and $\psi$ can be trained at the same time by minimizing the above loss function using the standard SGD approach.

In SANSee, $\delta$ and $\psi$ are first trained based on the set of receivers with labeled data ${\cal K}^L$. The receivers without labeled data in ${\cal K}^N$ can then directly obtain their local models by performing a linear combination operation on the set of models $\{\pmb{\omega}_k\}_{k\in{\cal K}^L}$. We will give a more detailed discussion on the model construction process at receivers in ${\cal K}^L$ as well as the model transfer process from receivers in ${\cal K}^L$ to receivers ${\cal K}^N$ in the next section.


\subsection{Model Training at Receivers with labelled data}



\higl{In this paper, we follow a commonly adopted FL setting in which receivers optimize their model parameters to minimize the loss functions based on their local data distributions. In other words, for a given model, the optimal model parameters obtained based on the local data minimize the loss function and maximize the output accuracy of the trained model. The optimal parameters of the trained models directly reflect the correlation between the data distributions of different receivers and therefore can be used to decide the set of receivers with similar data distributions. The above results have been verified both theoretically and practically in many FL-based applications and have already served as the foundation of many well-developed personalized FL solutions \cite{t2020personalized, pmlr-v139-li21h, li_federated_2020, pmlr-v119-karimireddy20a, Li_2021_CVPR, dinh2022new}.}

\higl{In fact, the difference between model parameters learned by different receivers due to different distributions of the local datasets is commonly referred to as the client drift problem. This problem results in slow convergence and even divergence of the model training process in model-aggregation-based FL approaches \cite{tan_towards_2022}. To address the client drift problem, an attention-inducing function $\lambda \sum_{{k'} < {j'}} R\left(\| \pmb{\omega}_{k'}- \pmb{\omega}_{{j'}}\|^2\right)$ is introduced in the regularized loss function, which improves the collaboration between the personalized models trained by different receivers. The attention-inducing function enhances the convergence and performance of personalized models through an attentive message-passing mechanism, which is model agnostic and can coordinate various intermediate results, with proven convergence for both convex and non-convex models \cite{huang2021personalized}.}  Specifically, we consider an attention-inducing function-based personalized federated learning solution in which all receivers in ${\cal K}^L$ collaborate in training a set of ${\cal K}^L$ location-specific models, denoted as $\pmb{\Omega}= \langle\pmb{\omega}_k\rangle_{k\in {\cal K}^L}$ by minimizing the following objective functions:
\begin{eqnarray}\label{global_obj}
 \mathcal{J}({\boldsymbol \Omega}) \coloneqq
 \sum_{k'\in \mathcal{K}^L}\left( F_{k'}({\boldsymbol \omega}_{k'})+\lambda \sum_{{k'} < {j'}} R\left(\| \pmb{\omega}_{k'}- \pmb{\omega}_{{j'}}\|^2\right)\right),
\end{eqnarray}
where $\lambda >0$ is a non-negative collaboration parameter, $R(\|\pmb{\omega}_{k'}- \pmb{\omega}_{{j'}}\|^2)$ is a regularizer which is an attention-inducing function included here to encourage collaborations between receivers with correlated models. In particular, we follow a commonly adopted setting \cite{huang2021personalized}
and use the negative exponential function to characterize the difference between models $\pmb{\omega}_{k'}$ and $\pmb{\omega}_{{j'}}$, defined as follows:
\begin{eqnarray}\label{neg_exp}
R(\|\pmb{\omega}_{k'}- \pmb{\omega}_{{j'}}\|^2)=1-e^{-\|\pmb{\omega}_{k'}- \pmb{\omega}_{{j'}}\|^2/{\sigma_R}},
\end{eqnarray}
where ${\sigma_R}$ is the difference parameter that controls the relative difference between models.
\higl{The added attention-inducing function in the objective function in (18) is an increasing function of the difference between model parameters of receivers. Thus, when minimizing the objective function at a receiver, other models learned by receivers with higher (lower) similarity in the local data distributions will have higher (lower) weights. Moreover, the regularization also smooths the difference between the model parameters at different receivers. This further reduces the variations of the model parameter differences, especially at the beginning of the model training process, which further improves the convergence and robustness of the personalized model aggregation. We then describe the detailed personalized model training process.}

\higl{In this paper, we adopt a standard SGD-based FL setting as introduced in \cite{nguyen2021federated} to iteratively construct personalized models for receivers in the set ${\cal K}^L$. Specifically, a coordinator is pre-assigned and announced to all the receivers, which would periodically upload their local model parameters to the coordinator for model aggregation and download the updated models for the next round of local model training. The proposed model is flexible; the coordinator is a logical entity deployed at any receiver, e.g., a coordination receiver, or a physical entity installed at a dedicated central server. In the former case, all other receivers periodically upload their intermediate local models to the coordinating receiver, which in turn aggregates the received models with its own model. In the latter case, all receivers upload their intermediate local models to the central server for model aggregation once in a while. Both scenarios have already been widely applied in many FL applications.} In the rest of this section, we use the superscript $(\cdot)^{t,e}$ to denote the parameters in $e$th local iteration of the $t$th global coordination round, i.e., $\pmb{\omega}^{t,e}$ is the model downloaded from the coordinator at the beginning of the $t$th round. In the $t$th coordination round, each receiver $k' \in {\cal K}^L$ updates its local model as follows:
\begin{equation}
\begin{aligned}
\pmb{\omega}^{t,e+1}_{k'} = \pmb{\omega}^{t,e}_{k'}-\eta \nabla \widetilde{F}_{k'}(\pmb{\omega}^{t,e}_{k'}), \; \mbox{for}\; e = 0, \ldots, E-1
\label{Local_SGD}
\end{aligned}
\end{equation}
where $\pmb{\omega}^{t,e}_{k'}$ denotes the local model of receiver $k'$ in the $e$th iteration in the $t$th coordination round,
$\eta$ is the local learning rate, and $\nabla \widetilde{F}_{k'}(\pmb{\omega}^{t,e}_{k'})$ is the unbiased stochastic gradient. At the end of the $E$th local iteration, receivers will upload models
$\{\pmb{\omega}^{t, E}_{1},\cdots,\pmb{\omega}^{t, E}_{K^L}\}$ to the coordinator for global model updating. At the coordinator, the following step will be performed for each receiver $k'$ to obtain the next-round model $\pmb{\omega}^{t+1}_{k'}$ for each receiver $k'$ as follows, for ${k',j'\in{\cal K}^L}$:
\begin{eqnarray}
 \pmb{\omega}^{t+1}_{k'} = \pmb{\omega}^{t,E}_{k'}- \sum_{{{k'} \neq j'}} \widetilde{\eta}\lambda\nabla R(\| \pmb{\omega}^{t,E}_{k'}- \pmb{\omega}^{t,E}_{j'}\|^2),
 \label{Server_gradient}
\end{eqnarray}
where $\widetilde{\eta}=\eta E$ is the step size. Repeat the above processes until the preset target loss $\epsilon_J$ is reached.

In fact, the step in (\ref{Server_gradient}) at the coordinator is in essence to update the model for each receiver $k'$ by performing a linear combination given by
\begin{equation}
\begin{aligned}
 \pmb{\omega}^{t+1}_{k'} = \sum_{{j'} \in {\cal K}^L} \xi^t_{{k'},{j'}} \pmb{\omega}^{t,E}_{{j'}}
 \label{convex_combination}
\end{aligned}
\vspace{-0.2cm}
\end{equation}
where $\xi^t_{{k'},{j'}}$ is given by
\begin{equation}
\xi^t_{{k'},{j'}}=\left\{
\begin{aligned}
&\widetilde{\eta} \lambda R^{'}(\|\pmb{\omega}^{t,E}_{k'}-
 \pmb{\omega}^{t,E}_{{j'}}\|^2) , & {k'} \neq {j'}, \\
&1- \widetilde{\eta} \lambda \sum^{K^L}_{{j'} \neq {k'} } R^{'}(\|\pmb{\omega}^{t,E}_{k'}-
 \pmb{\omega}^{t,E}_{{j'}}\|^2), & {k'} = {j'},
\end{aligned}
\right.
\label{333}
\end{equation}
where $R^{'}(\|\pmb{\omega}^{t,E}_{k'}- \pmb{\omega}^{t,E}_{{j'}}\|^2)=\frac{e^{-\|\pmb{\omega}^{t,E}_{k'}- \pmb{\omega}^{t,E}_{{j'}}\|^2/{\sigma_R}}}{\sigma_R}$.
Note that, the value of $R(\|\pmb{\omega}^{t,E}_{k'}- \pmb{\omega}^{t,E}_{{j'}}\|^2)$ decreases as the model correlation between models $\pmb{\omega}^{t,E}_{k'}$ and $\pmb{\omega}^{t,E}_{{j'}}$ increase. Also, since $0\le R^{'}(\|\pmb{\omega}^{t,E}_{k'}- \pmb{\omega}^{t,E}_{{j'}}\|^2) \le \frac{1}{\sigma_R}$, we can ensure $0\le \xi^t_{{k'},{j'}}\le 1$ by choosing a proper local learning rate, e.g., $\eta \le \frac{1}{\lambda E(K^L-1)}$. \higl{The coordinator needs to perform only simple linear combining operations based on the models uploaded by a limited number of receivers with labeled data. Thus its computational load is negligible compared to that of the local model training process at each receiver. More specifically, in SANSee, we follow similar model aggregation operations as the existing personalized FL solutions in which the coordinator performs linear combining of model parameters received from the receivers for personalized model coordination. The overhead of such a model aggregation approach is generally considered negligible by many existing works in FL \cite{Ala_2024_FLsurvey}. }

\begin{algorithm}[H]
	\caption{Model Training Algorithm}\label{Algorithm2}\footnotesize
	{\bf Input}: 
    Target loss $\epsilon_J$; Local SGD steps $E$; Set of receivers with labelled data ${\cal K}^L$; labelled data $\{\mathcal{D}_1$, $\dots$, $\mathcal{D}_{K^L}\}$; \\
	{\bf Output}: Personalized models of labeled receivers $\{\pmb{\omega}^T_0$, $\dots$, $\pmb{\omega}^T_{K^L-1}\}$.

    \begin{itemize}
    \item[ 1:] Server broadcasts an initial model $\pmb{\omega}_0$ to all receivers in ${\cal K}^L$;
    \item[ 2:] {\bf While} $\mathcal{J}({\boldsymbol \Omega}) \ge \epsilon_J$ {\bf do}
    \item[ 3:] \quad {\bf For} receiver $k' \in {\cal K}^L$ {\bf in parallel do}
    \item[ 4:] \quad \quad {\bf For} $e$ = 0, $\cdots$, $E-1$ {\bf do}
    \item[ 5:] \quad \quad \quad Uniformly sample a mini-batch $\zeta^{t,e}_{k'}$ from $\mathcal{D}_{k'}$;
    \item[ 6:] \quad \quad \quad Perform SGD iterations on $\pmb{\omega}^{t,e}_{k'}$ by using (\ref{Local_SGD});
    \item[ 7:] \quad \quad {\bf End for}
    \item[ 8:] \quad {\bf End parallel for}
    \item[ 9:] \quad {\bf For} $k' \in {\cal K}^L$ {\bf do on coordinator}
    \item[ 10:] \quad \quad Obtain coefficient $\xi^t_{{k'},{j'}}$ by using (\ref{333});
    \item[ 11:] \quad \quad Update next-round model $\pmb{\omega}^{t+1}_{k'}$ by using (\ref{convex_combination});
    \item[ 12:] \quad {\bf End for on coordinator}
    \item[ 13:] {\bf End for}
    \end{itemize}
\end{algorithm}

\subsection{Model Transfer to Receivers without labelled data}

\higl{Let us now develop a model transfer solution that maps the personalized models constructed by receivers with labelled data to receivers without any labelled data. Specifically, each receiver ${k'}\in{\cal K}^L$ with a labeled dataset first establishes a semantic mapping pair $\langle {\boldsymbol{\phi}_{k'}}, \pmb{\omega}_{k'} \rangle$ consisting of its location-specific semantics ${\boldsymbol{\phi}_{k'}}$ obtained in Section 5 and its local model $\pmb{\omega}_{k'}$ constructed in Section 6.2.} We can then follow the same line as Section 6.1 to jointly develop two modules: a semantic mapping functional module $g_{\delta}(\cdot)$ with parameter $\delta$ and a model correlation functional module $h_{\psi}(\cdot, \cdot)$ with parameter $\psi$.

\higl{The detailed procedures for implementing model transfer in SANSee are illustrated in Fig. 3. The semantic mapping functional module $g_{\delta}(\cdot)$ with parameter $\delta$ is
implemented based on a 4-convolutional block-based CNN architecture in which each block consists of a 3$\times$3 convolutional layer followed by a batch normalization and a ReLU layer. Two max-pool layers are then inserted after the first two blocks to extract important features while simultaneously reducing the data dimensions. After that, the resulting low-dimensional semantics $\{\bar{\boldsymbol{\phi}}_1, \cdots, \bar{\boldsymbol{\phi}}_{k'}\}$ are concatenated and fed into a feature concatenation layer, followed by two convolutional blocks, a fully connected ReLU layer and a fully connected sigmoid layer that outputs the semantic similarity between any pairs of input physical-layer semantics. The model correlation functional module $h_{\psi}(\cdot, \cdot)$ with parameter $\psi$ is implemented using the convolutional block concatenated with two fully connected layers. Finally, the objective loss function $\mathcal{L}(\delta, \psi)$ given in (17) is used to establish the mapping relationship between semantic similarity and model correlations. To minimize the loss function $\mathcal{L}(\delta, \psi)$, we jointly optimize both functional modules by solving the following problem:
\begin{eqnarray}\label{transfer_loss_v2}
\langle \delta^{*}, \psi^{*} \rangle = \arg \min\limits_{\langle \delta, \psi \rangle} \mathcal{L}(\delta, \psi).
\end{eqnarray}}

\begin{figure}[!ht]
    \centering
    \includegraphics[width=0.48\textwidth]{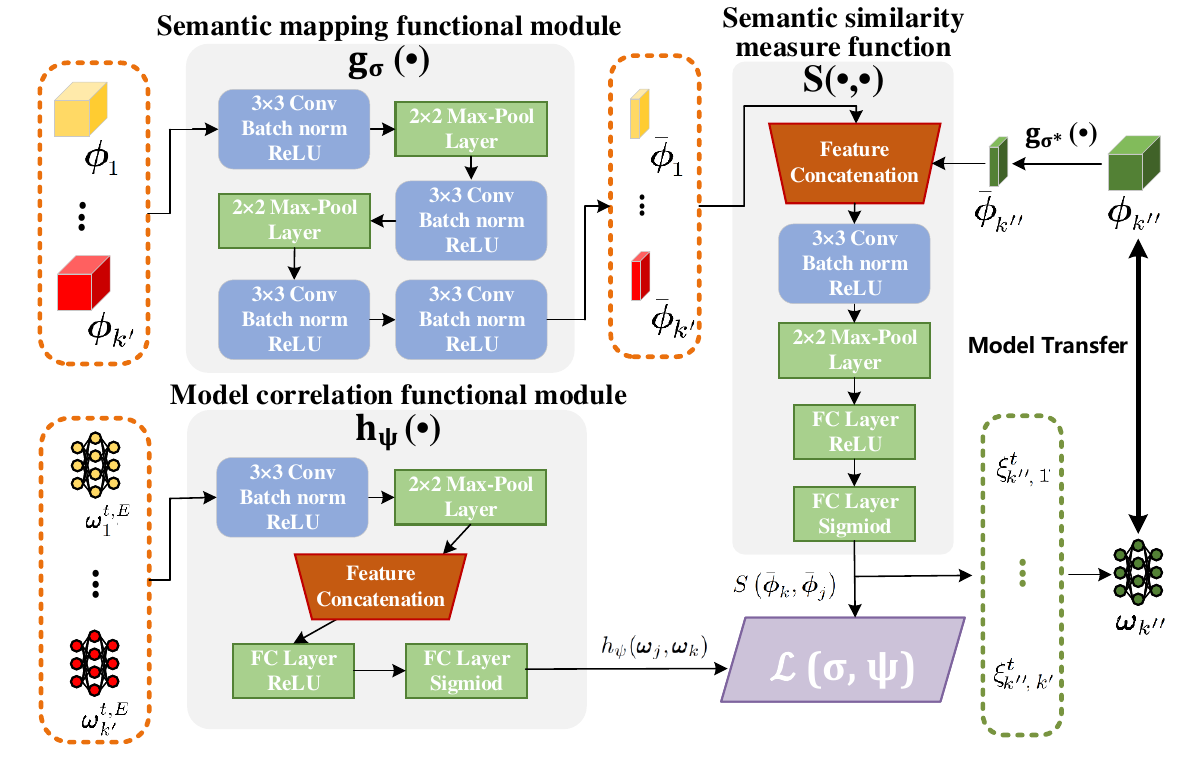}
    \caption{Detailed architectural components of the proposed model transfer solution.}
    \label{overview_v2}
\end{figure}

\higl{In this case, receivers with no any labeled data can obtain a location-specific model by performing linear combinations of all personalized models at receivers with labeled data, i.e., the location-specific model $\pmb{\omega}_{k''}$ of receiver $k''\in{\cal K}^N$ can be calculated as $\pmb{\omega}_{k''} = \sum_{k' \in {\cal K}'} \xi_{ k'', k'} \pmb{\omega}_{k'}$, where $\xi_{k'', k'}$ is the model aggregation coefficient predicted by the optimized semantic mapping functional module, i.e., $\xi_{k'', k'}=S(g_{\delta^{*}}(\boldsymbol{\phi}_{k''}), g_{\delta^{*}}(\boldsymbol{\phi}_{k'}))$.} \higl{SANSee does not require any labelled data at the target receivers. Furthermore, the model transfer process involves only linear operations summation and therefore, compared to existing transfer learning solutions \cite{li2021deep,zhang2018crosssense,wang2022airfi}. SANSee significantly reduces the data labelling overhead as well as the required computational cost at the target receivers.} We illustrate the detailed procedures of model transfer in Algorithm 3. As will be proved in the next section, the model obtained by each receiver $k''\in{\cal K}^N$ without labeled data can approach to the real local model $\pmb{\omega}^{*}_{k''}$.

\begin{algorithm}[H]
	\caption{pSAN-based Model Transfer  Algorithm}\label{Algorithm4}\footnotesize
	{\bf Input}: Raw CSI samples of all receivers. \\
	{\bf Output}: Transfer models $\{\pmb{\omega}_{k''}\}_{k''\in{\cal K}^N}$ of receivers in ${\cal K}^N$.
    \begin{itemize}
    \item[ 1:] Estimate physical-layer semantics $\{\boldsymbol{\phi}_1$, $\dots$, $\boldsymbol{\phi}_{K}\}$ of all receivers by using Alg. \ref{Algorithm1};
    \item[ 2:] Obtain local models $\{\pmb{\omega}_1$, $\dots$, $\pmb{\omega}_{K^L}\}$ of receivers in  ${\cal K}^L$ by using Alg. \ref{Algorithm2};
    \item[ 3:] Construct a set of semantic mapping pairs $\{\boldsymbol{\phi}_{k'}, \pmb{\omega}_{k'} \}_{k'\in {\cal K}^L}$;
    \item[ 4:] Train the two functional modules by minimizing (\ref{transfer_loss}) ;
    \item[ 5:] {\bf For} $k''\in{\cal K}^N$ {\bf do on coordinator}
    \item[ 6:] \quad Calculate aggregation coefficients $\{\xi_{k'', 1}$, $\cdots$, $\xi_{k'', K^L}\}$ by using the optimized modules;
    \item[ 7:] \quad Obtain the transfer model $\pmb{\omega}_{k''}$ by performing a linear combination;
    \item[ 8:] {\bf End for on coordinator}
    \end{itemize}
\end{algorithm}

\section{Theoretical Results} \label{Theoretical_Results}
In this section, we present the theoretical results related to our proposed SANSee architecture. As mentioned earlier, SANSee is a distributed personalized model construction framework that involves two major steps: (1) Local model training: it first trains a set of models at the receivers with labelled data and, (2) model transfer: these trained models will be transferred to new receivers at novel locations without requiring any labelled data. In the rest of this section,
we derive theoretical bounds of 
the following two types of errors:
\begin{itemize}
    \item[(1)] {\bf (Local) Model Training Error}: corresponds to the performance gap between the models $\pmb{\Omega} = \langle \pmb{\omega}_k \rangle_{k\in {\cal K}^L}$ trained by receivers based on their locally recorded datasets and the ground truth model $\pmb{\Omega}^{*}=\langle \pmb{\omega}^*_k \rangle_{k\in {\cal K}^L}$, given by $\mathbb{E}[\mathcal{J}(\pmb{\Omega})-\mathcal{J}(\pmb{\Omega}^{*})]$ where $\mathcal{J}(\pmb{\Omega})$ is defined previously in (\ref{global_obj}).
   \higl{We will a present detailed discussion in Section \ref{Section_TheorecticalBoundModelTraining}.}
    \item[(2)] {\bf Model Transfer Error}:  corresponds to the error of the transferred models obtained by the receivers without labelled data and the ground truth model, defined previously in (\ref{main_objective}).  
    We will present a detailed discussion in Section \ref{Section_TheorecticalBoundModelTransfer}. 
\end{itemize}

\subsection{Model Training Error}
\label{Section_TheorecticalBoundModelTraining}
We use superscript $T$ to denote the models trained in the $T$th coordination round, e.g., we use $\pmb{\Omega}^0$ and $\pmb{\Omega}^T$ to denote the models in the $0$th (initial model vector) and $T$th coordination round. We can then
prove the following result about the model training error. 

\begin{theorem}
\label{Theorem_TrainError}
Suppose the following assumptions hold:
\begin{itemize}
    \item[] {\it Assumption 1}: \higl{(Strong Convexity) $F_1, \cdots, F_K$ are all $\mu$-convex: i.e., $\frac{\mu}{2}\|\pmb{\nu}-\pmb{\omega}\|^2 \le F_k(\pmb{\nu})-F_k(\pmb{\omega})$ $-\langle \nabla F_k(\pmb{\omega}), \pmb{\nu}-\pmb{\omega}\rangle$, for all $\pmb{\nu}, \pmb{\omega} \in \mathbb{R}^d$ and $k\in{\cal K}$,}

    \item[] {\it Assumption 2}: \higl{(Lipschitz Smoothness) $F_1, \cdots, F_K$ are all $L$-smooth: i.e., $\frac{L}{2}\|\pmb{\nu}-\pmb{\omega}\|^2\ge F_k(\pmb{\nu})-F_k(\pmb{\omega})$ $-\langle \nabla F_k(\pmb{\omega}), \pmb{\nu}-\pmb{\omega}\rangle $, for all $\pmb{\nu}, \pmb{\omega} \in \mathbb{R}^d$ and $k\in{\cal K}$,}

    \item[] {\it Assumption 3}: (Bounded Variance) The variance of stochastic gradients on all local objective functions is bounded:  $\mathbb{E}\|\nabla \widetilde{F}_k(\pmb{\omega}, \zeta_k)-\nabla F_k(\pmb{\omega})\|^2 \le \sigma^2_F$, for all $\pmb{\omega} \in \mathbb{R}^d$ and $k\in{\cal K}$,

    \item[] {\it Assumption 4}: (Bounded Gradient) The gradient of the attention-inducing
function is bounded:  $\nabla R(\|\pmb{\nu}- \pmb{\omega}\|^2) \le \kappa_R$, for all $\pmb{\nu}, \pmb{\omega} \in \mathbb{R}^d$.
\end{itemize}

Then, there exist $\lambda >2L$, $T\ge \frac{4}{\tilde{\eta}_1 \mu}$, and learning rate $\eta\le\frac{\tilde{\eta}_1}{E}$ such that 
\begin{equation}
\begin{aligned}
\lefteqn{\mathbb{E}[\mathcal{J}(\pmb{\Omega}^{T})-\mathcal{J}(\pmb{\Omega}^{*})] \le  {\mathcal{O}}\left( \|\pmb{\Omega}^0-\pmb{\Omega}^*\|^2 e^{\frac{-\tilde{\eta}_1 \mu T}{4} } \right.} \\
&&\;\;\;\;\;\;\;\;\;\;\;\;\;\;\;\;\;\;\;\;\; +\left. \frac{(1+\mu T)( E \Gamma^L +\sigma^2_{F}/B)}{\mu^3T^2EK^L}\right) ,
\label{theorem1_a}
\end{aligned}
\end{equation}
where  $\tilde{\eta}_1\coloneqq\left(12(L+\lambda\kappa_R)+\frac{128\lambda \kappa_R L^2}{\mu^2}+\frac{96 L^2}{\mu}\right)^{-1}$, $\Gamma^L=\sum_{k'\in \mathcal{K}^L}\| \nabla F_{k'}({\pmb{\omega}}^{*}_{k'})\|^2$, and $B$ is mini-batch size. ${\mathcal{O}}(\cdot)$ is the big-O notation which ignores poly-logarithmic and constant numerical factors.
\end{theorem}

\begin{IEEEproof}
See Appendix \ref{Proof_Theorem_TrainError}.
\end{IEEEproof}

We can observe that the assumptions introduced in the above theorem are reasonable in many practical scenarios. More specifically, as discussed in \cite{li_federated_2020, tan_towards_2022}, Assumptions 1-3 are satisfied by many commonly adopted loss functions such as cross-entropy, L2 regularization, etc. Assumption 4 can also be achieved by choosing many commonly used regularizers such as the negative exponential function\cite{t2020personalized}.


We can observe from Theorem \ref{Theorem_TrainError} that the model training error is closely related to the initial model selection, mini-batch size $B$, the number of local iterations between consecutive coordination rounds $E$, and the total number of receivers participating in the model training $K^L$. More specifically, $\|\pmb{\Omega}^0-\pmb{\Omega}^*\|^2$ term in (\ref{theorem1_a}) quantifies the error caused by the selection of the initial model. Since this term is multiplied with term $e^{\frac{-\tilde{\eta}_1 \mu T}{4} }$, we can reduce the impact of incorrect selection of the initial model by increasing the value of $\tilde{\eta}_1$ which can be achieved by choosing a smaller value $\lambda$. We can also observe that the model training error always increases with the values of $B$, $E$, and $K^L$. Increasing these parameters however will result in higher computation of complexity and longer coordination delay during each coordination round.

It is known that, in most existing FL-based solutions, the convergence rate is always adversely affected by the heterogeneity level of datasets at the model training participating receivers\cite{li_federated_2020, t2020personalized}. Our result in Theorem \ref{Theorem_TrainError} can also capture this issue. More specifically, the term $\Gamma^L$ is a commonly used metric to measure the heterogeneity level, i.e., level of non-iid, among datasets at the receivers. We can observe that model training error increases with the value of $\Gamma^L$. We can however observe that the impact of the non-iid decreases when the number of coordination rounds $T$ becomes large.

\subsection{Model Transfer Error}
\label{Section_TheorecticalBoundModelTransfer}
In the model transfer step, receiver $k''\in {\cal K}^N$ can directly obtain its model by aggregating the already trained models of others, e.g., receiver $k'$ for $k'\in {\cal K}^L$. Therefore, the model transfer error is closely related to the data distribution difference between receivers in sets ${\cal K}^L$ and ${\cal K}^N$. In this paper, we use a commonly used metric, total variation distance, to quantify the data distribution difference which is defined as follows:

\begin{definition}\label{TV distance definition}
  For the data
  distributions $\mathcal{P}_{k'}$ and $\mathcal{P}_{k''}$ of receivers $k'$ and $k''$
  over the dataset $\mathcal{D}$, the total variation distance between them is defined as $\|\mathcal{P}_{k'}-\mathcal{P}_{k''}\|_{TV} \coloneqq \sup_{\zeta \in \mathcal{D}} |\mathcal{P}_{k'}(\zeta)-\mathcal{P}_{k''}(\zeta)|$, where $\zeta$ is data uniformly sampled from dataset $\mathcal{D}$.
\end{definition}

We can then prove the following result about the model transfer error.
\begin{theorem}
\label{Theorem_TransferError}
Suppose Assumptions 1-4 and the following assumptions hold:
\begin{itemize}
    \item[] {\it Assumption 5}: The local objective function is $M$-bounded: $F_k(\cdot, \zeta_k)\le M$, for all $ k \in {\cal K}$.
\end{itemize}

Then, we have
\begin{eqnarray}
\lefteqn{ \mathbb{E}[F_{k''}(\pmb{\omega}_{k''})-F_{k''}(\pmb{\omega}^{*}_{k''})] \le \frac{\epsilon+(\lambda + \Gamma^N)K^L}{K^L}} \label{eq_theorem2_3} \\
&&\;\;\;\;\;\;\;\;\;\;\;\;\; +  M\sum_{k'\in \cal{K}^L}\|\xi_{k'', k'}\mathcal{P}_{k''}-\frac{\mathcal{P}_{k'}}{K^L}\|_{TV},\nonumber \label{eq_theorem2_1}
\end{eqnarray}
\higl{where $\epsilon$ is the model training error derived in (\ref{theorem1_a}), $\boldsymbol{\phi}_{k''}$ are defined in Sections \ref{Section_SemanticSimilarity} and \ref{Section_ModelTrainandTransfer}, $\pmb{\omega}_{k''}=$ $\sum_{k'\in {\cal K}^L} \xi_{k'', k'} \pmb{\omega}_{k'}$,  $\xi_{k'', k'}=S(g_{\delta^{*}}(\boldsymbol{\phi}_{k''}), g_{\delta^{*}}(\boldsymbol{\phi}_{k'}))$, and
$\Gamma^N=\frac{1}{K^L}\sum_{k'\in {\cal K}^L}F_{k'}(\pmb{\omega}^{*}_{k'})-F_{k''}(\pmb{\omega}^{*}_{k''})$.}
\end{theorem}
\begin{IEEEproof}
See Appendix \ref{Proof_Theorem_TransferError}.
\end{IEEEproof}

Assumptions 5 and 6 in Theorem \ref{Theorem_TransferError} are reasonable in many transfer learning application scenarios\cite{fallah_personalized_2020, fallah2021generalization}. This is because receivers without labeled data rely on the transferred model to recognize specific human gestures when observing new testing data points. In the ideal scenario in which the transferred model can perfectly recognize the label of any new testing data point, the local objective function $F_k(\pmb{\Omega}^*_{k''}, \zeta_k)$ will be minimized and also the gradient value $\nabla F_k(\cdot, \zeta_k)$ will approach zero. Even in most non-ideal scenarios, the local objective function and its gradient based on any new testing data point need to be assumed to be bounded to derive any valid theoretical bounds.

We can observe that the theoretical bound of model transfer error depends mainly on the TV distance between data distributions of receivers $k'$ and $k''$ as shown in (\ref{eq_theorem2_1}). More specifically, as the TV distance between $\mathcal{P}_{k'}$ and $\mathcal{P}_{k''}$ reduces, the term in (\ref{eq_theorem2_1}) approaches zero. We can therefore apply solutions developed in Section \ref{Section_SemanticSimilarity} to learn the optimal model correlation coefficients $\xi_{k'', k'}$ to minimize (\ref{eq_theorem2_1}).
We can also observe that the model transfer error is also related to the model training error. However, as the number of model training receivers becomes large, the impact of the model training error on the transfer error decreases, as shown in (\ref{eq_theorem2_3}).

\section{Performance Evaluation} \label{Performance_Evaluation}
\subsection{Experimental Setup}


{\bf Dataset:} To evaluate the performance of SANSee, we conduct extensive experiments based on a public available wireless sensing dataset, Widar\cite{zheng2019Widar}, consisting of 6 types of human gestures (e.g., Push-Pull, Sweep, Clap, Slide, Draw-O, and Draw-zigzag) recorded at three different environments: classroom, hall, and office. In each environment, an off-the-shelf Wi-Fi transmitter with one activated antenna and 6 receivers, each has three activated antennas, have been deployed at different locations with the same relative distances in a 2m$\times$2m sensing area. The transmitter is set to broadcast data packets at a rate of 1,000 packets per second at 5.825 GHz Wi-Fi band. The dataset consists of 12,000 labeled gesture data samples in total.

{\bf Model:} For gesture recognition model construction, each gesture is assumed to last around 1.5 seconds and the CSI signals recorded by each receiver will be equally divided into a set of 1.5 second time segments. 
We adopt ResNet-8 model to extract the spatial and temporal features of the training data samples associated with different gestures, trained based on a cross-entropy loss function using the standard SGD algorithm. For the semantic-based transfer learning model, we design a 4-convolutional block-based CNN architecture to convert high-dimensional semantic features into low-dimensional semantic space in which each block consists of a 3$\times$3 convolutional layer followed by a batch normalization and ReLU layer. To map semantic similarity to model correlation, the low-dimensional semantic representations are fed into a feature concatenation layer, followed by 2 convolutional blocks, a fully connected ReLU layer and a fully connected sigmod layer to output the model correlation coefficient.

{\bf Platform:} We conduct our experiments on a workstation with an Intel(R) Core(TM) i9-13900K CPU@5.8GHz, 128.0GB RAM@4000.0MHz, 1 TB SSD, 4 TB HDD, and two NVIDIA GeForce RTX 4090 GPUs. The CSI data samples are processed using MatLab and gesture recognition models and pSAN-based transfer learning models are trained using Python 3.8, CUDA 12.2 and Pytorch 2.1.0 running on Ubuntu 22.04.

\begin{figure*}[!ht]
    \centering
     \begin{subfigure}[b]{0.22\textwidth}
         \centering
           \includegraphics[width=1.1\textwidth]{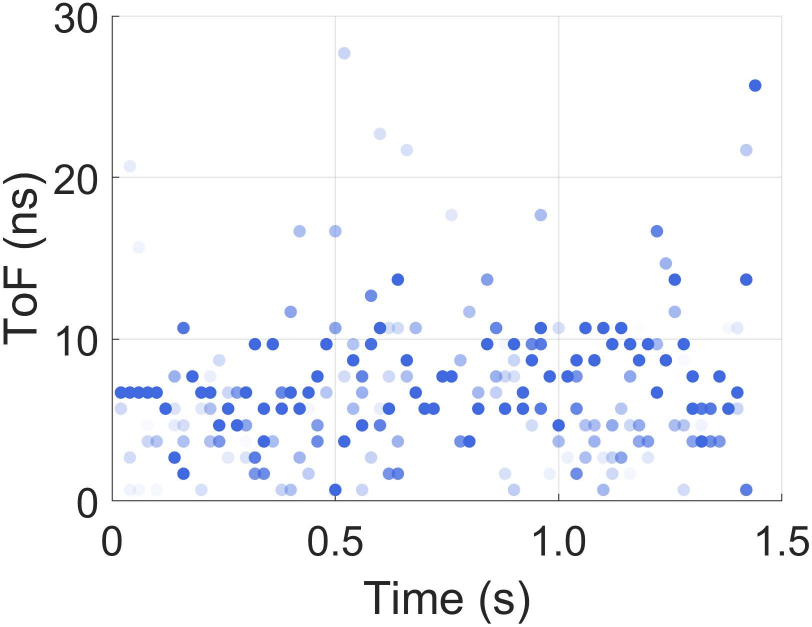}
         \caption{}
     \label{G_tof_gesture1_env3}
     \end{subfigure}
     \hfill
    \centering
     \begin{subfigure}[b]{0.22\textwidth}
         \centering
           \includegraphics[width=1.1\textwidth]{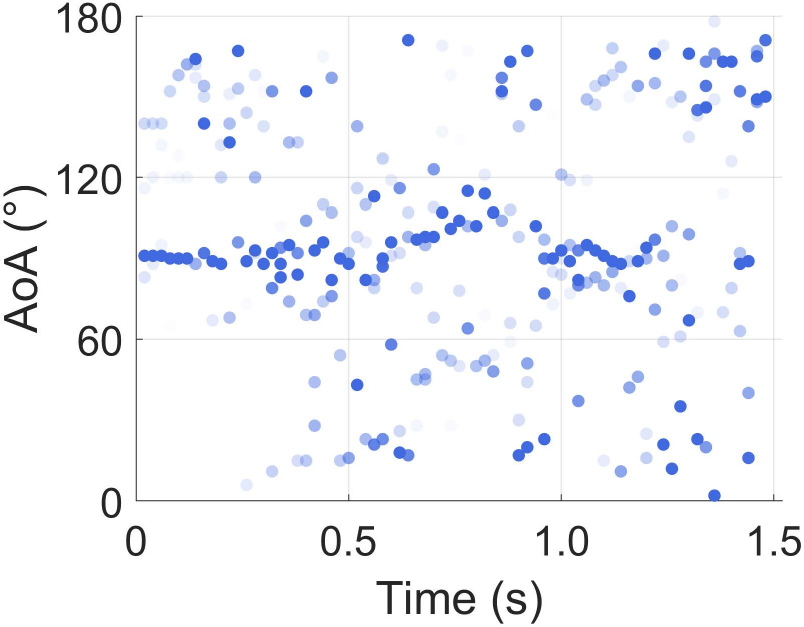}
         \caption{}
     \label{G_aoa_gesture1_env3}
     \end{subfigure}
     \hfill
    \centering
     \begin{subfigure}[b]{0.22\textwidth}
         \centering
           \includegraphics[width=1.1\textwidth]{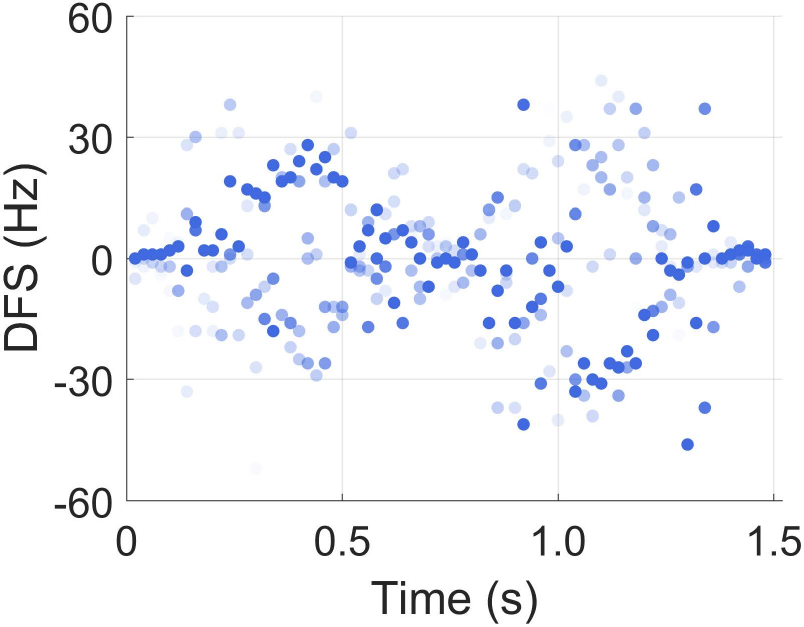}
         \caption{}
     \label{G_dop_gesture1_env3}
     \end{subfigure}
     \hfill
    \centering
     \begin{subfigure}[b]{0.22\textwidth}
         \centering
           \includegraphics[width=1.1\textwidth]{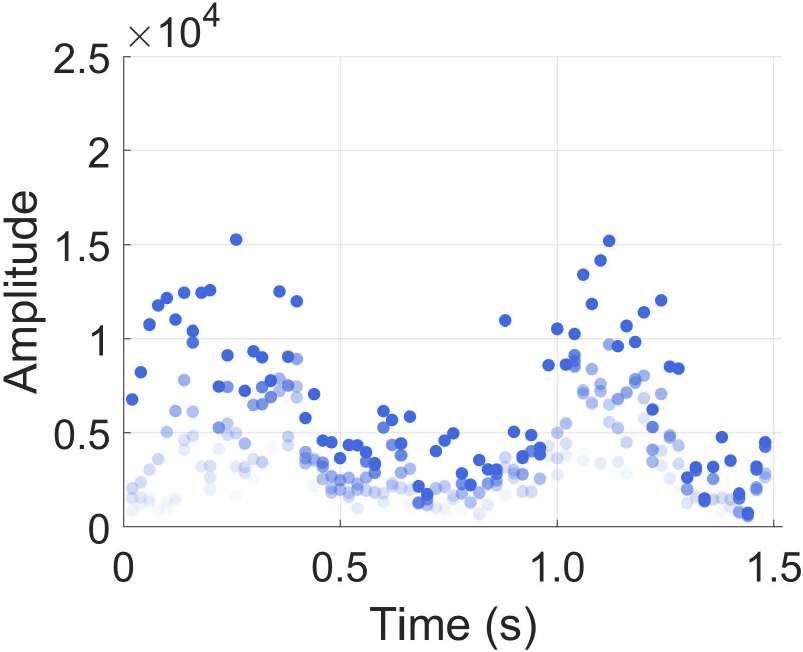}
         \caption{}
     \label{G_amp_gesture1_env3}
     \end{subfigure}
     \hfill
    \centering
     \begin{subfigure}[b]{0.22\textwidth}
         \centering
           \includegraphics[width=1.1\textwidth]{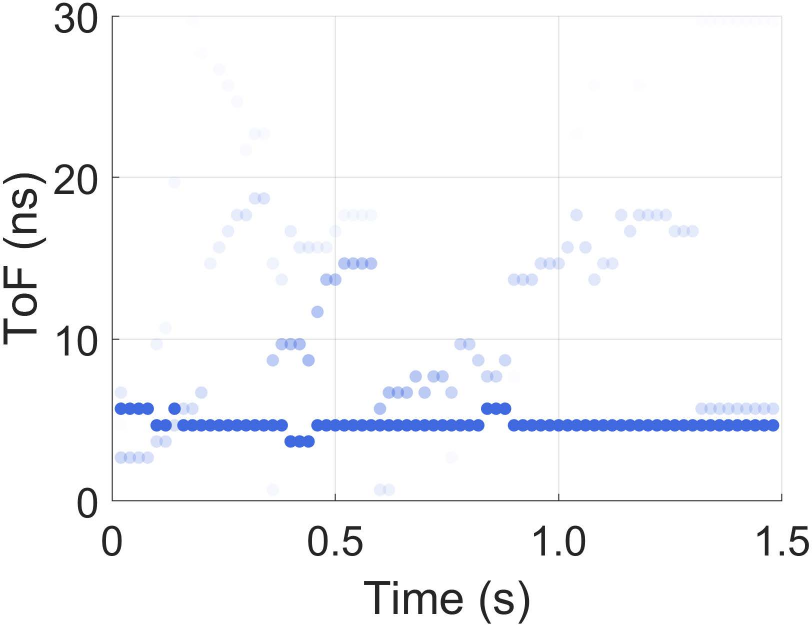}
         \caption{}
     \label{E_tof_gesture1_env3}
     \end{subfigure}
     \hfill
    \centering
     \begin{subfigure}[b]{0.22\textwidth}
         \centering
           \includegraphics[width=1.1\textwidth]{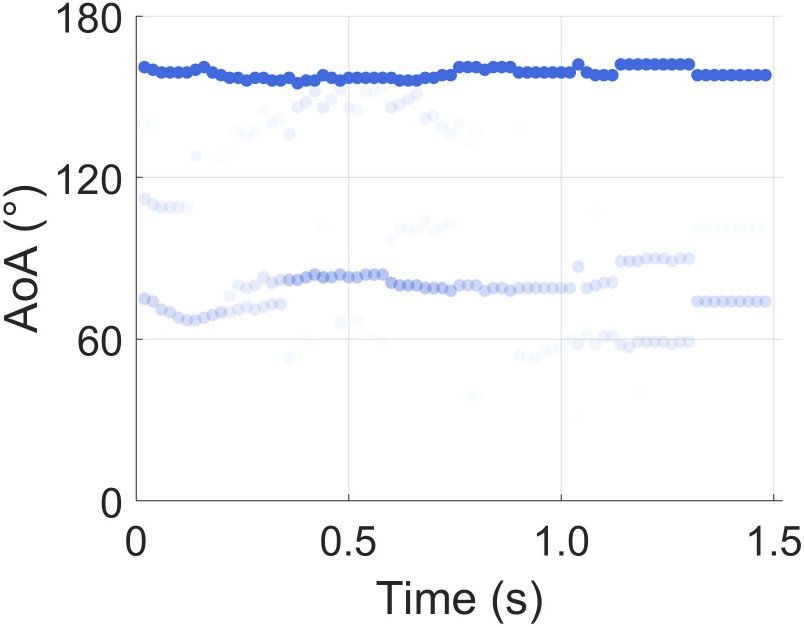}
         \caption{}
     \label{E_aoa_gesture1_env3}
     \end{subfigure}
     \hfill
    \centering
     \begin{subfigure}[b]{0.22\textwidth}
         \centering
           \includegraphics[width=1.1\textwidth]{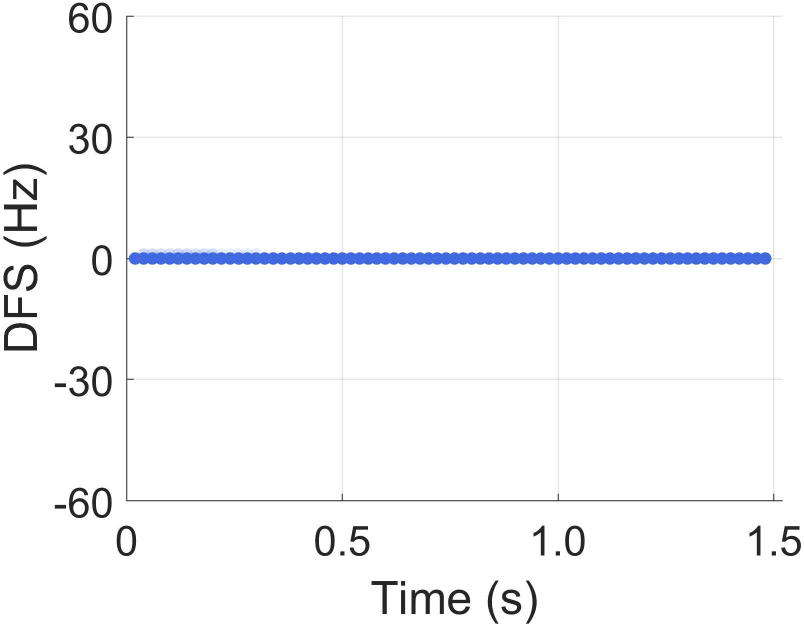}
         \caption{}
     \label{E_dop_gesture1_env3}
     \end{subfigure}
     \hfill
    \centering
     \begin{subfigure}[b]{0.22\textwidth}
         \centering
           \includegraphics[width=1.1\textwidth]{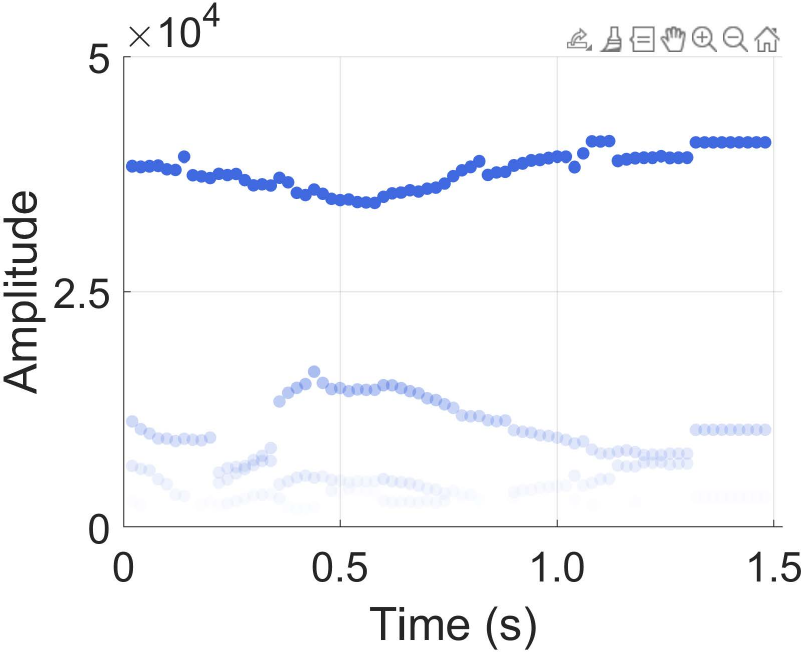}
         \caption{}
     \label{E_amp_gesture1_env3}
     \end{subfigure}

     \caption{G-semantic features (b)-(e) and E-semantic features (g)-(j) of gesture "Push \& Pull" (a) performed in the office environment (f),  based on $L=5$ estimated paths.}
     \label{Fig_AmplitudeToFAoADFS_gesture1}
\end{figure*}

\begin{figure*}[!ht]
    \centering
     \begin{subfigure}[b]{0.22\textwidth}
         \centering
           \includegraphics[width=1.1\textwidth]{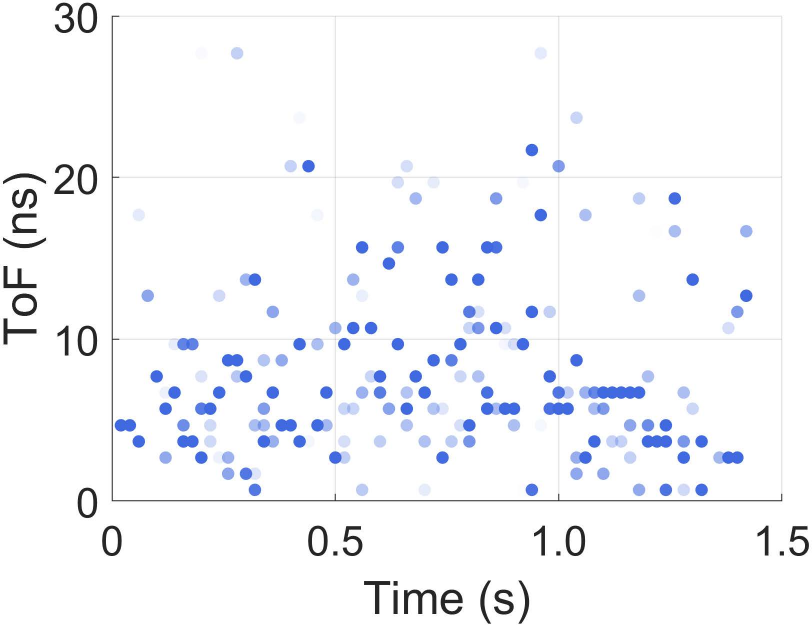}
         \caption{}
     \label{G_tof_gesture2_env3}
     \end{subfigure}
     \hfill
    \centering
     \begin{subfigure}[b]{0.22\textwidth}
         \centering
           \includegraphics[width=1.1\textwidth]{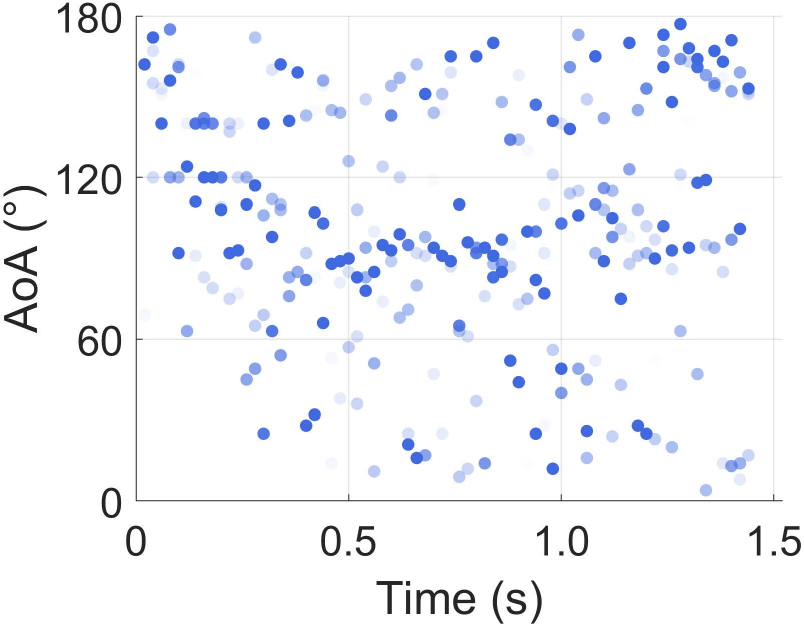}
         \caption{}
     \label{G_aoa_gesture2_env3}
     \end{subfigure}
     \hfill
    \centering
     \begin{subfigure}[b]{0.22\textwidth}
         \centering
           \includegraphics[width=1.1\textwidth]{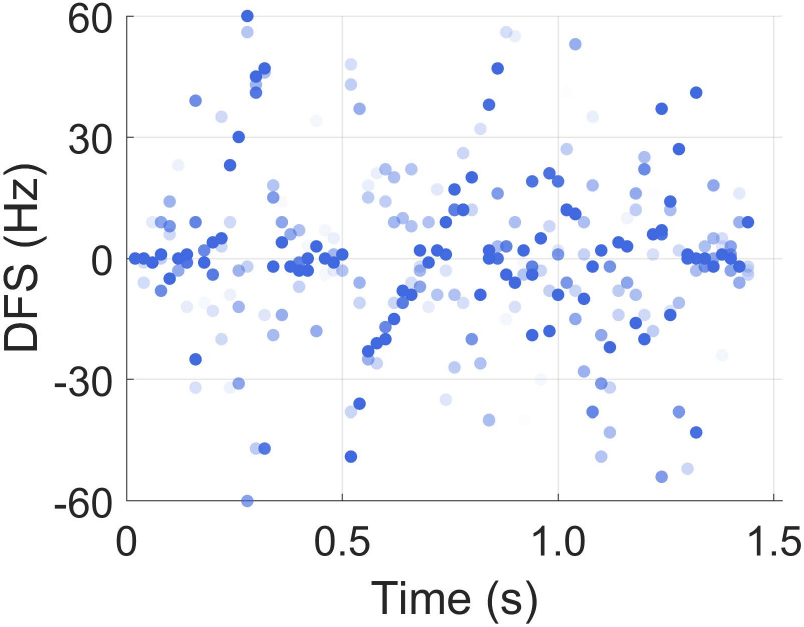}
         \caption{}
     \label{G_dop_gesture2_env3}
     \end{subfigure}
     \hfill
    \centering
     \begin{subfigure}[b]{0.22\textwidth}
         \centering
           \includegraphics[width=1.1\textwidth]{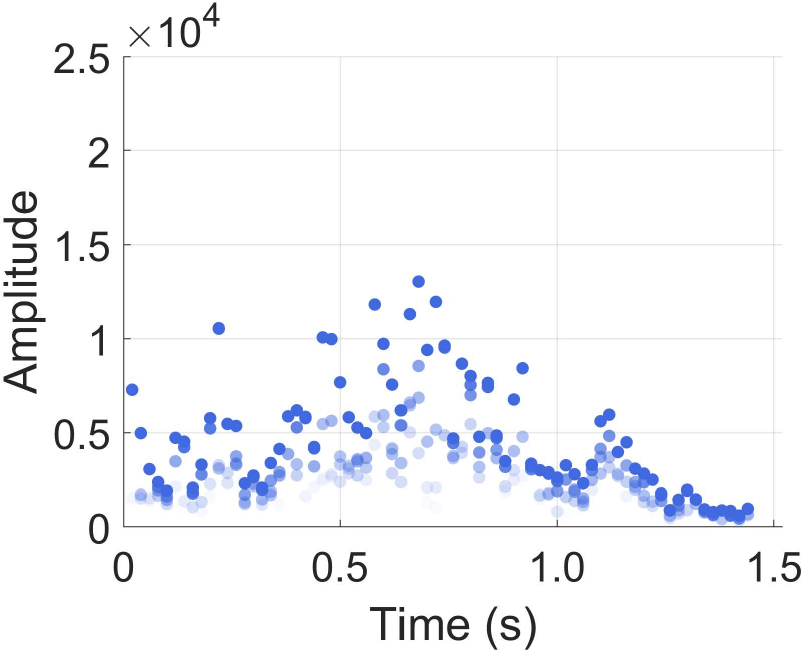}
         \caption{}
     \label{G_amp_gesture2_env3}
     \end{subfigure}
     \hfill
    \centering
     \begin{subfigure}[b]{0.22\textwidth}
         \centering
           \includegraphics[width=1.1\textwidth]{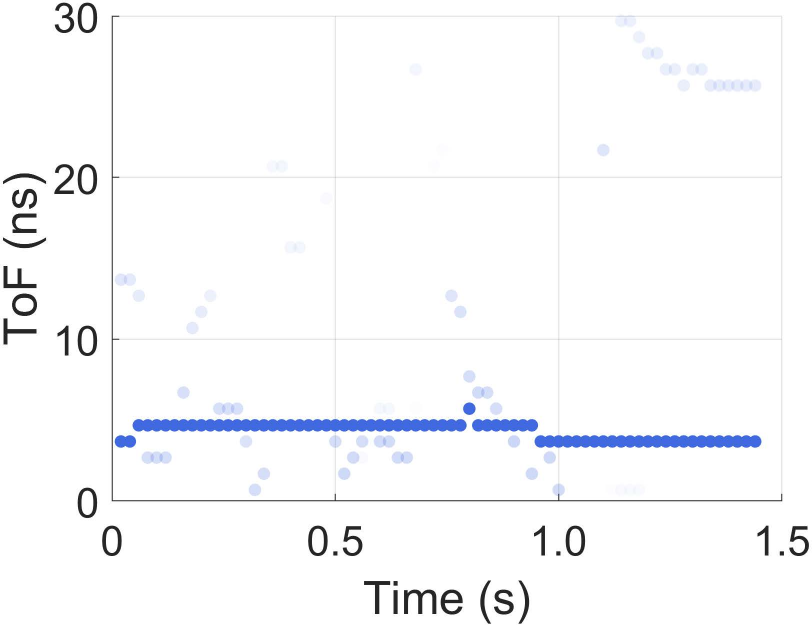}
         \caption{}
     \label{E_tof_gesture2_env3}
     \end{subfigure}
     \hfill
    \centering
     \begin{subfigure}[b]{0.22\textwidth}
         \centering
           \includegraphics[width=1.1\textwidth]{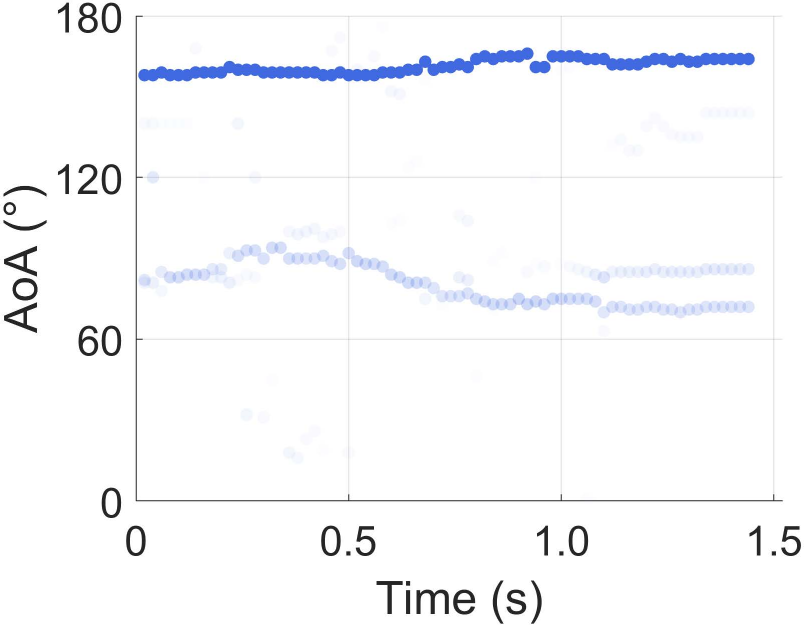}
         \caption{}
     \label{E_aoa_gesture2_env3}
     \end{subfigure}
     \hfill
    \centering
     \begin{subfigure}[b]{0.22\textwidth}
         \centering
           \includegraphics[width=1.1\textwidth]{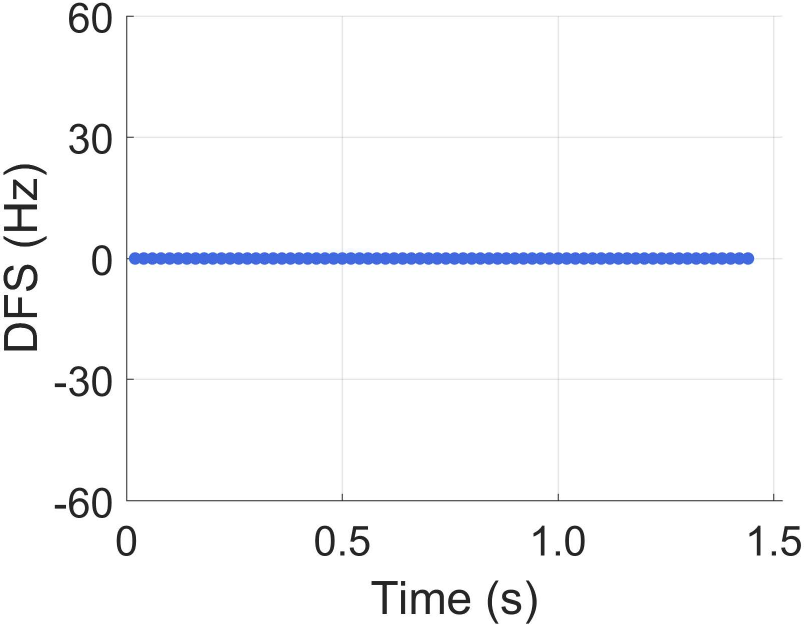}
         \caption{}
     \label{E_dop_gesture2_env3}
     \end{subfigure}
     \hfill
    \centering
     \begin{subfigure}[b]{0.22\textwidth}
         \centering
           \includegraphics[width=1.1\textwidth]{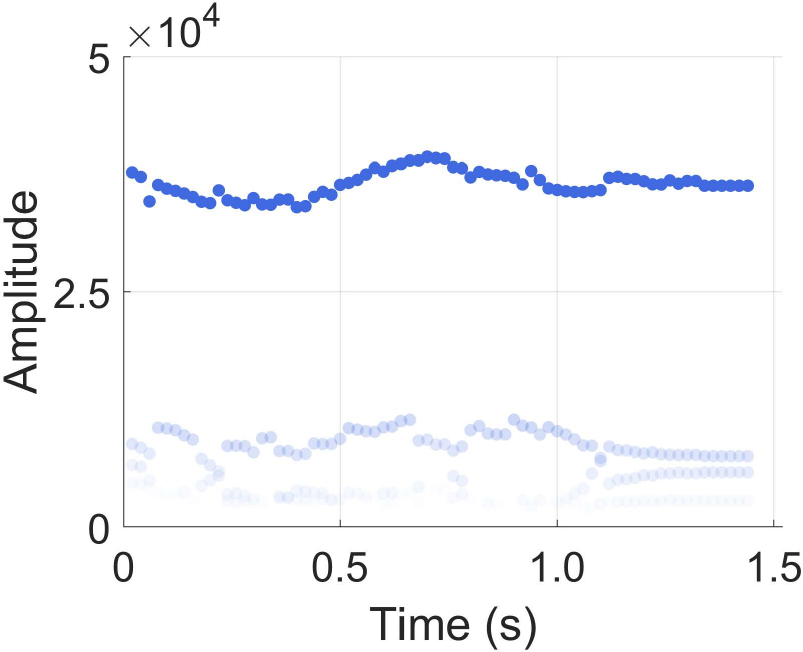}
         \caption{}
     \label{E_amp_gesture2_env3}
     \end{subfigure}

     \caption{G-semantic features (b)-(e) and E-semantic features (g)-(j) of gesture "Sweep" (a) performed in the office environment (f),  based on $L=5$ estimated paths.}
     \label{Fig_AmplitudeToFAoADFS_gesture2}
\end{figure*}

\begin{figure*}[!ht]
    \centering
     \begin{subfigure}[b]{0.22\textwidth}
         \centering
           \includegraphics[width=1.1\textwidth]{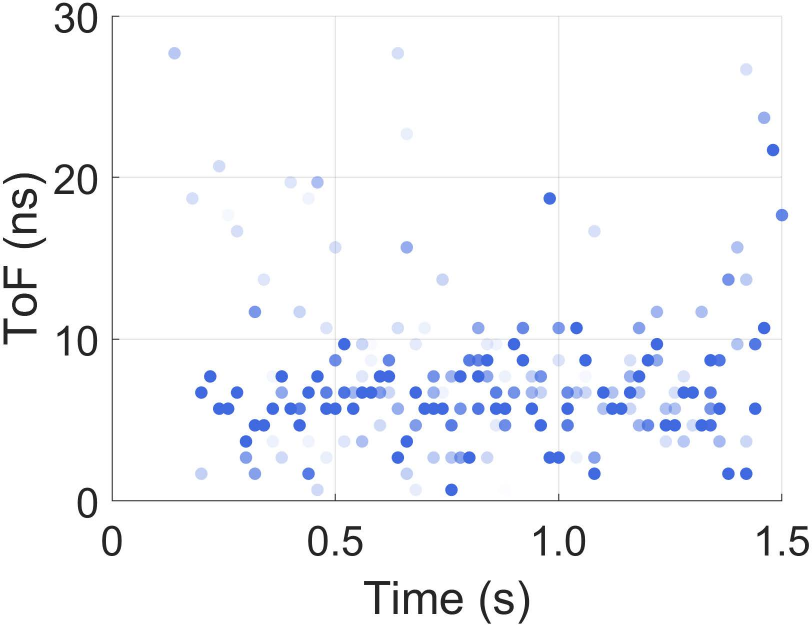}
         \caption{}
     \label{G_tof_gesture2_env1}
     \end{subfigure}
     \hfill
    \centering
     \begin{subfigure}[b]{0.22\textwidth}
         \centering
           \includegraphics[width=1.1\textwidth]{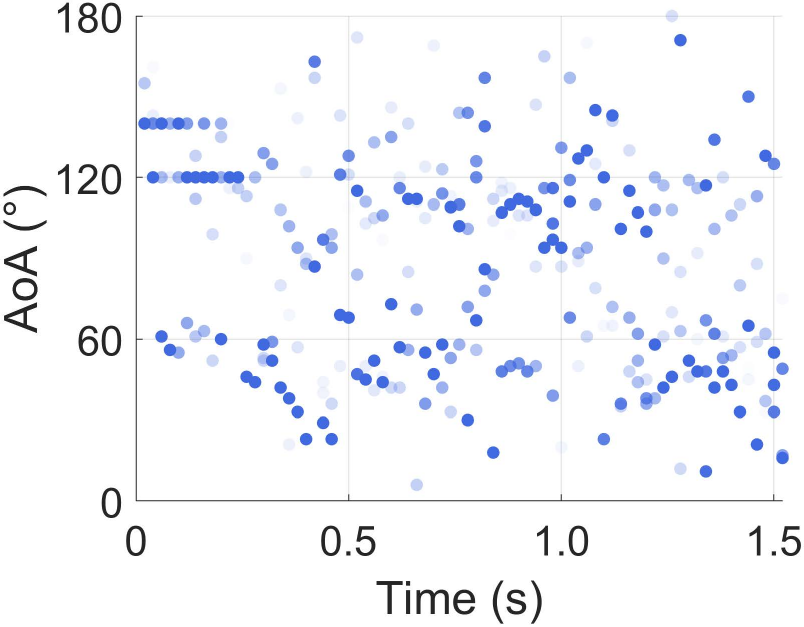}
         \caption{}
     \label{G_aoa_gesture2_env1}
     \end{subfigure}
     \hfill
    \centering
     \begin{subfigure}[b]{0.22\textwidth}
         \centering
           \includegraphics[width=1.1\textwidth]{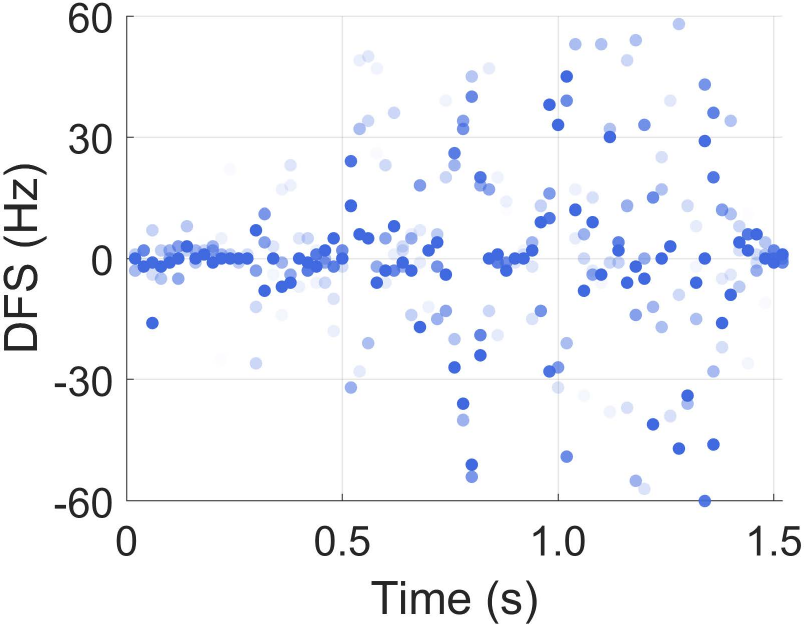}
         \caption{}
     \label{G_dop_gesture2_env1}
     \end{subfigure}
     \hfill
    \centering
     \begin{subfigure}[b]{0.22\textwidth}
         \centering
           \includegraphics[width=1.1\textwidth]{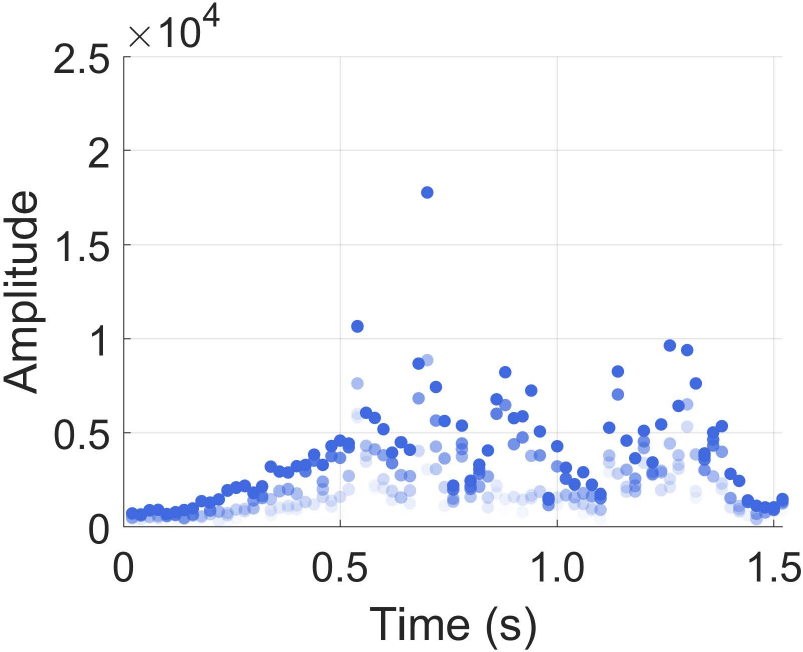}
         \caption{}
     \label{G_amp_gesture2_env1}
     \end{subfigure}
     \hfill
    \centering
     \begin{subfigure}[b]{0.22\textwidth}
         \centering
           \includegraphics[width=1.1\textwidth]{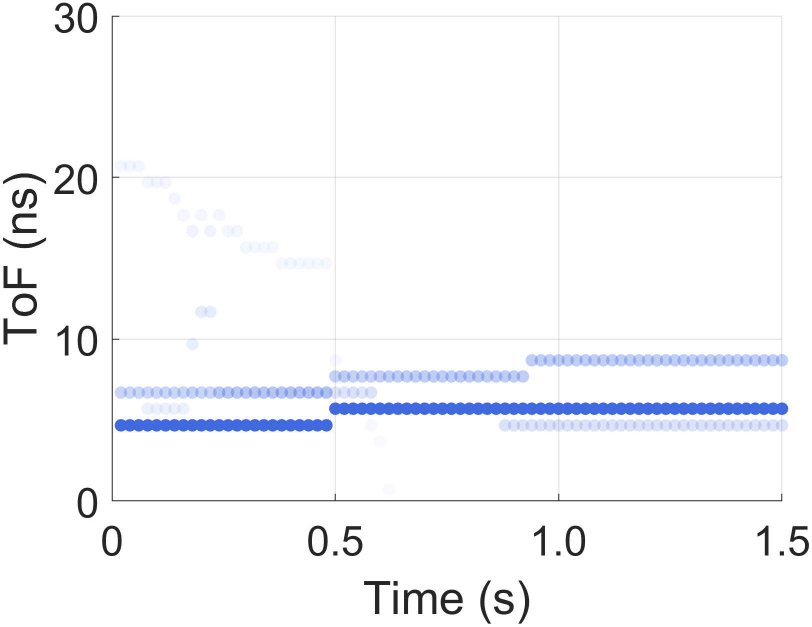}
         \caption{}
     \label{E_tof_gesture2_env1}
     \end{subfigure}
     \hfill
    \centering
     \begin{subfigure}[b]{0.22\textwidth}
         \centering
           \includegraphics[width=1.1\textwidth]{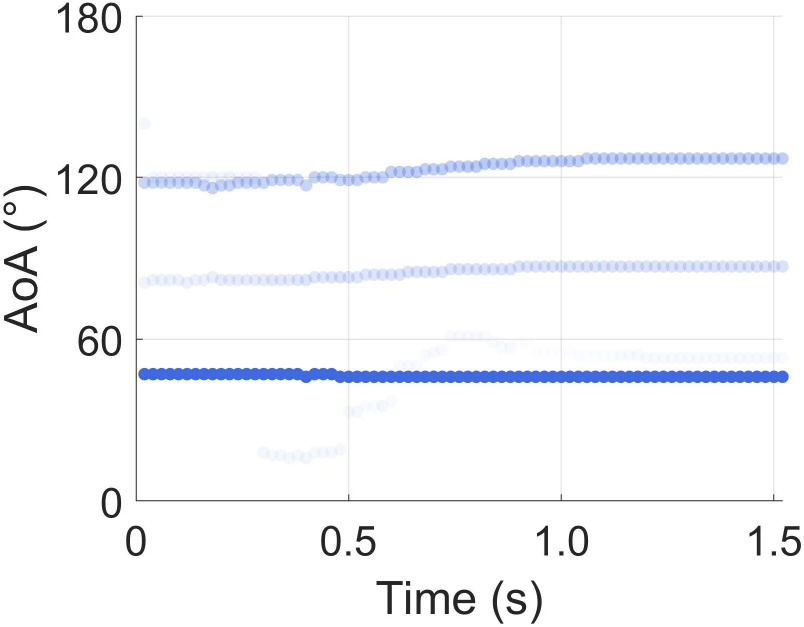}
         \caption{}
     \label{E_aoa_gesture2_env1}
     \end{subfigure}
     \hfill
    \centering
     \begin{subfigure}[b]{0.22\textwidth}
         \centering
           \includegraphics[width=1.1\textwidth]{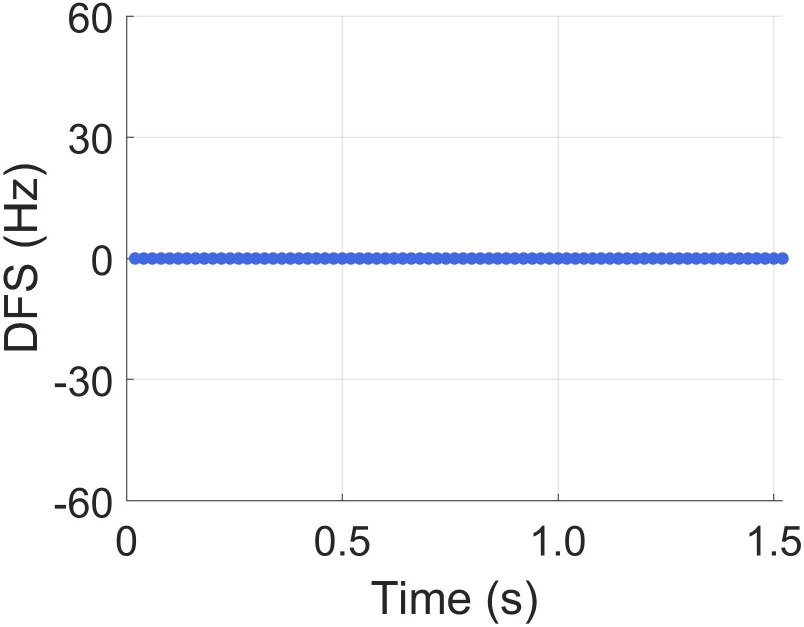}
         \caption{}
     \label{E_dop_gesture2_env1}
     \end{subfigure}
     \hfill
    \centering
     \begin{subfigure}[b]{0.22\textwidth}
         \centering
           \includegraphics[width=1.1\textwidth]{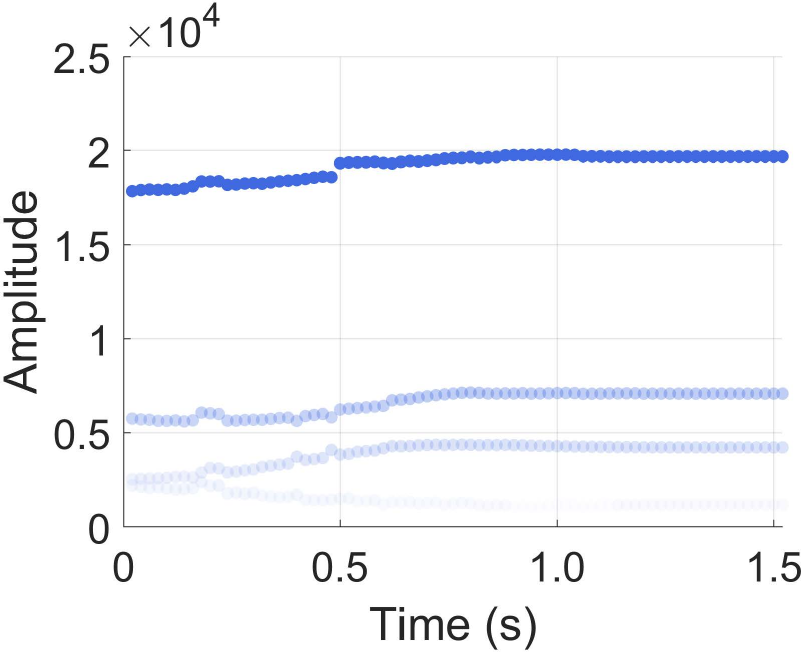}
         \caption{}
     \label{E_amp_gesture2_env1}
     \end{subfigure}

     \caption{G-semantic features (b)-(e) and E-semantic features (g)-(j) of gesture "Sweep" (a) performed in the classroom environment (f),  based on $L=5$ estimated paths.}
     \label{Fig_AmplitudeToFAoADFS_gesture1_env1}
     \vspace{-0.5cm}
\end{figure*}

\subsection{Physical-layer Semantics Estimation}
SANSee is built based on the basic idea that the physical-layer semantics play a key role in determining the distributions of wireless sensing signals as well as the models to recognize different gestures. We therefore need to first evaluate the E- and G-semantics estimated by our proposed physical-layer semantics estimation algorithm under different settings.\higl{The main idea of human gesture recognition is to detect the impact of the Doppler shift caused by human body movements on the wireless signal, particularly its higher frequency content. In this case, the magnitude of the Doppler shift of wireless signals detected by the receiver mainly depends on the gesture-performing speed as well as the signal frequency for gesture detection. It is known that body movement speeds for most human gestures, including Sweep, Clap, and Slide considered in this paper,  are between 0.25 m/sec and 4 m/sec \cite{Gupta_2012_SoundWave}, which correspond to the Doppler frequency shift between 8 Hz and 134 Hz at 5 GHz band \cite{Pu_2013_Gesture_Sensing}. We therefore set the threshold for separating the high-pass and low-pass filters to 2 Hz.} In Fig. \ref{Fig_AmplitudeToFAoADFS_gesture1}-\ref{Fig_AmplitudeToFAoADFS_gesture1_env1}, we present the physical-layer semantic features, including amplitude, ToF, AoA, and DFS of both E- and G-semantics, estimated based on Algorithm 1 proposed in Section \ref{Section_PhysicallayerSemanticEstimation}. We also show estimated results of the primary path, which responds to the reflected signal with the highest amplitude (navy blue points), with (red solid lines) and without (black dash lines) the Gaussian smoother (GS). These different features can be influenced by different semantic features of the environmental layout and gestures. For example, signal amplitudes are mainly affected by the transmit signal power as well as various power losses caused by environmental reflections, blockages, transmission distance between the transmitter and receivers. AoAs of the received signals are mainly affected by the relative orientations of the transmitter, receivers, and the gesture performing human user. In Fig. \ref{Fig_AmplitudeToFAoADFS_gesture1}-\ref{Fig_AmplitudeToFAoADFS_gesture1_env1}, we can observe that the impact of these semantic features can be perfectly captured by the stationary and dynamic path components estimated by our proposed algorithm. For example, in Fig. \ref{Fig_AmplitudeToFAoADFS_gesture1}-\ref{Fig_AmplitudeToFAoADFS_gesture2}, due to the differences in movement patterns, we can observe that all estimated G-semantics parameters of gestures "Push \& Pull"  and "Sweep" are significantly dissimilar to each other. E-semantics of  "Push \& Pull"  and "Sweep" gestures look very similar as they are recorded in the same environment office and also the three main components observed in the amplitudes of E-semantics correspond to the signals received from the direct path and two paths reflected from the walls.
Moreover, in Fig. \ref{Fig_AmplitudeToFAoADFS_gesture2}-\ref{Fig_AmplitudeToFAoADFS_gesture1_env1}, we can observe that the fluctuation patterns of the G-semantic parameters of gesture ``Sweep" look very similar to each other even they are performed in the different environments, but the G-semantics are different due to the change in the physical environments.
This also suggests that neither E-semantic nor G-semantic alone will not be able to capture the full picture of the impact of human gestures on wireless signals. Generally speaking, taking into consideration more physical-layer semantic features will result in a higher gesture recognition accuracy. It may however result in a higher computational complexity as will be discussed next.

\begin{figure}[htbp]
     \centering
     \begin{subfigure}[b]{0.32\textwidth}
         \centering
           \includegraphics[width=1\textwidth]{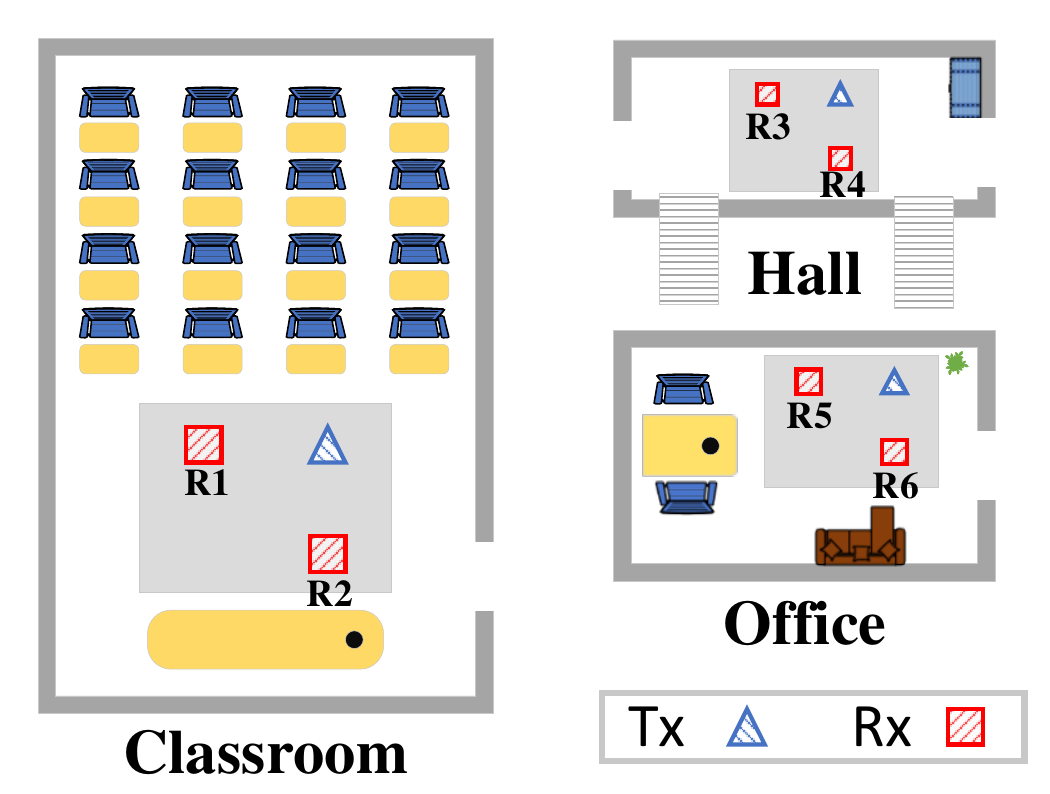}
         \caption{}
         \label{Deployment_Settings_v2}
     \end{subfigure}
     \hfill
     \begin{subfigure}[b]{0.32\textwidth}
         \centering
           \includegraphics[width=1\textwidth]{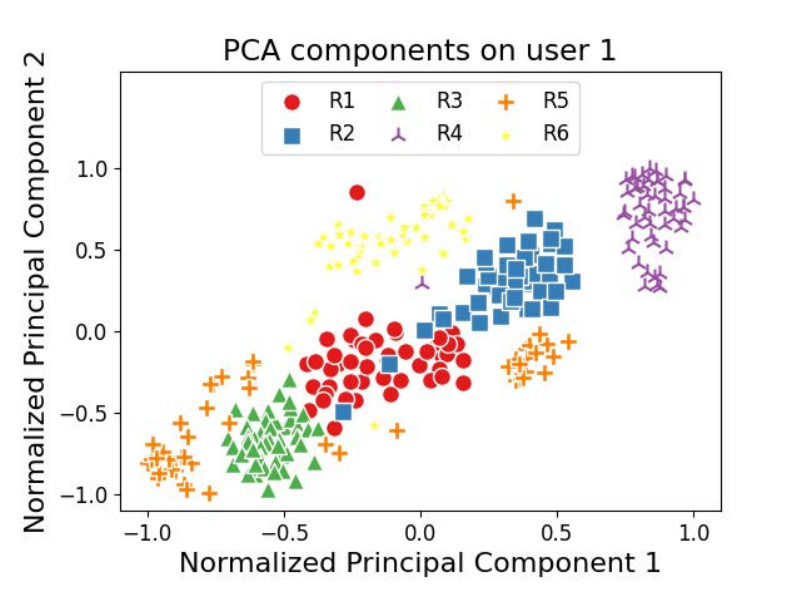}
         \caption{}         \label{tsne}
     \end{subfigure}
    \caption{(a) Locations of 6 receivers (labeled as R1-R6) deployed in 3 different environments, and (b) t-SNE-based visualization of statistical diversity of the CSIs of the same gesture recorded by
    different receivers.}
    \label{t_SNE_Deployment}
\end{figure}

Based on the above observations, we then evaluate the impact of different physical-layer semantic features on the recorded CSI signals at the receivers. In Fig. \ref{t_SNE_Deployment}, we present the t-SNE-based visualizations of the statistical features of CSI signals of the same gesture recorded at different locations in different environments. We observe that, even the relative locations and orientations of the transmitter, receivers, and the human user remain the same at different environments, the recorded CSI signals may vary significantly. This further justifies our observations that 
the traditional centralized modeling approaches, in which wireless sensing data samples recorded at different locations are combined at a centralized server to train a single global model for recognizing gestures performed at different locations, cannot provide accurate and consistent wireless sensing results at different receivers, especially in complex environments.  

\subsection{Model Training}
To evaluate the extra computational complexity introduced by considering more semantic features in the model training,  
we use the time consumption of training a given model with a fixed number of iterations as the main metric to evaluate the complexity of model training, and compare the required model training time and the resulting model accuracy when different combinations of semantic features are fed into the model during training in Table I. We can observe that the model training time almost doubles when a new semantic feature, E- or G-semantics, is added in the model training. Despite the increase in model training complexity, the accuracy of gesture recognition improves significantly, e.g., the gesture recognition accuracy improves over 50\%, increasing from 61.67\% with only amplitudes being considered to 96.34\% with all the features of both G- and E-semantics being included in the model training. We also evaluate the impact of estimating different numbers of path components in Algorithm 1 on the model complexity and accuracy in Table I. We can observe that, when the number of estimated path components increases from $L=5$ to $L=20$, the overall time consumption increases only at around 8\% and the resulting model accuracy improves around 18\%.  

\begin{table*}[!ht]
\scalebox{0.85}{
\begin{tabular}{c|ccccccc}\hline
 & \begin{tabular}[c]{@{}c@{}}Amplitude \\ Only (L=10)\end{tabular} & \begin{tabular}[c]{@{}c@{}}DFS Only\\ (L=10)\end{tabular} & \multicolumn{1}{c}{\begin{tabular}[c]{@{}c@{}}G-semantics \\ Only (L=10)\end{tabular}} & \begin{tabular}[c]{@{}c@{}}E-semantics \\ Only (L=10)\end{tabular} & \begin{tabular}[c]{@{}c@{}}Both E- and G- \\ Semantics (L=5)\end{tabular} & \begin{tabular}[c]{@{}c@{}}Both E- and G- \\ Semantics (L=10)\end{tabular} & \begin{tabular}[c]{@{}c@{}}Both E- and G- \\ Semantics (L=20)\end{tabular} \\ \hline
Dimensional Size of Semantics & $1\times100\times300$ & $1\times120\times300$ & $4\times120\times300$ & $4\times120\times300$ & $8\times120\times300$ & $8\times120\times300$ & $8\times120\times300$ \\
 Model Training Time & 1.78 h & 2.23 h & 6.04 h & 6.04 h & 13.37 h & 13.90 h & 14.44 h \\
Model Accuracy & 61.67\% & 73.87\% & 89.98\% & 16.67\% & 81.11\% & 92.90\% & 96.34\% \\ \hline
\end{tabular}
}
\caption{Comparison of model training time and accuracy when considering different combinations of semantic features and the numbers of estimated path components.}
\end{table*}

\begin{figure*}[!ht]
    \centering
       \includegraphics[width=1\textwidth]{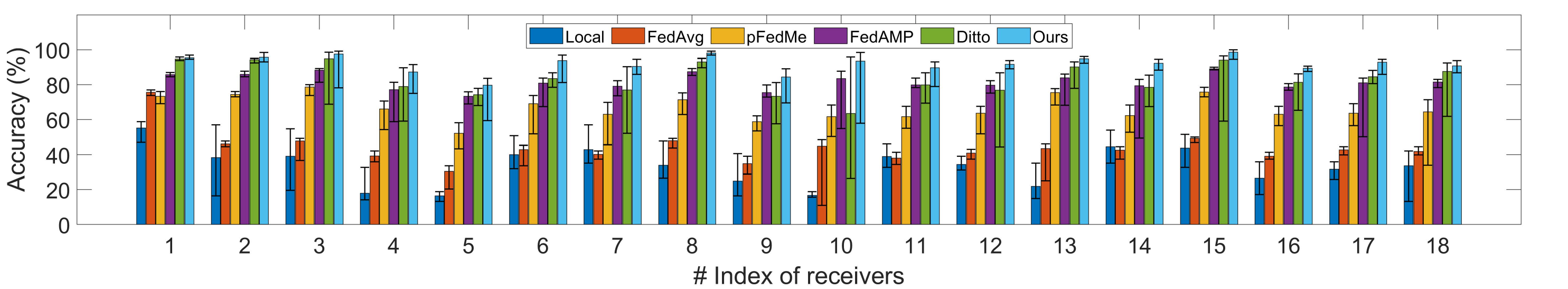}
     \caption{Wireless sensing accuracy at 18 receivers achieved by models trained by different algorithms, including local training, FedAvg, pFedMe, FedAMP, Ditto, and the proposed SANSee.}         \label{different_algorithm_errorbar}
\end{figure*}

Let us now evaluate the model training performance of SANSee for the receivers with labelled data. We compare the model accuracy of all 18 receivers at three different environments achieved by SANSee to the state-of-the-art algorithms in Fig. \ref{different_algorithm_errorbar}.
\higl{More specifically, in addition to comparing SANSee with the local training ({\it Local}) in which each receiver trains a local model based only on its local dataset and {\it FedAvg} \cite{mcmahan2017communication} in which all receivers train a single global model by periodically aggregating their local model parameters, we also consider three state-of-art personalized federated learning algorithms: {\it pFedMe}\cite{t2020personalized}, 
{\it FedAMP}\cite{huang2021personalized}, 
and {\it Ditto}\cite{pmlr-v139-li21h}.}
\higl{ Moreover, Fig. \ref{different_algorithm_errorbar} includes the average accuracy calculated based on models trained at 10 experiments. We also highlight the highest and lowest bounds on accuracy for models trained at different receivers. In Fig. \ref{different_algorithm_errorbar}, we can observe that SANSee outperforms all these existing
personalized model training algorithms and can achieve model accuracy improvements between 9.44 $\%$ and 27.64 $\%$ on average. Furthermore, the model performance at different receivers is more consistent in SANSee compared to other algorithms. More specifically, in local training, FedAvg, pFedMe, FedAMP, and Ditto algorithms, the gap between the highest and lowest model accuracy when implementing the trained models at different receivers are 18.75\%, 11.63\%, 14.02\%, 12.20\%, 22.76\%, respectively, all of which is larger than the 9.16\% gap achieved by our proposed SANSee.}


\begin{figure*}[htbp]
     \centering
     \begin{subfigure}[b]{0.32\textwidth}
         \centering
           \includegraphics[width=1\textwidth]{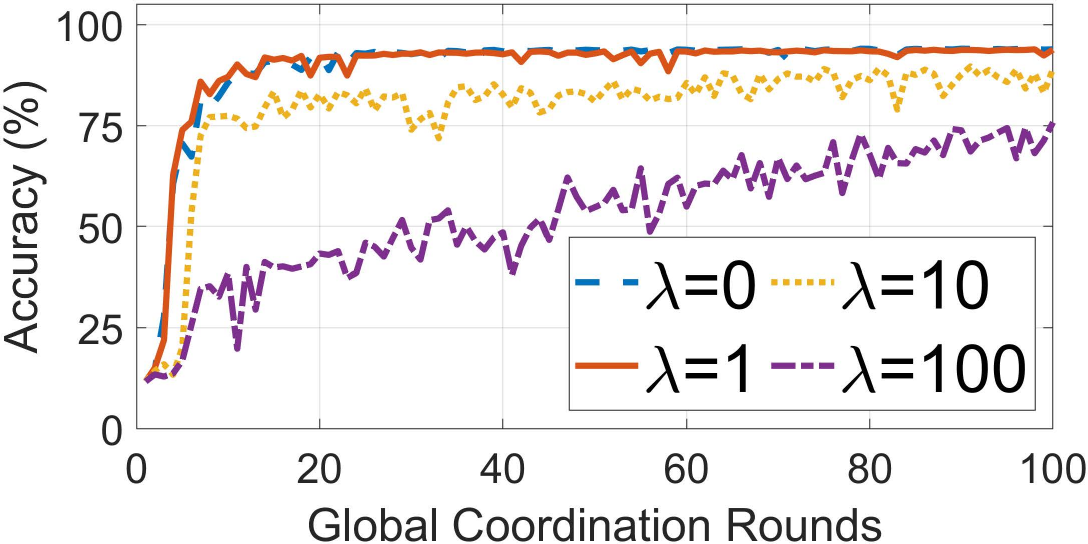}
         \caption{}
         \label{Test_accuracy_convergence_different_lambda}
     \end{subfigure}
     \hfill
     \begin{subfigure}[b]{0.32\textwidth}
         \centering
           \includegraphics[width=1\textwidth]{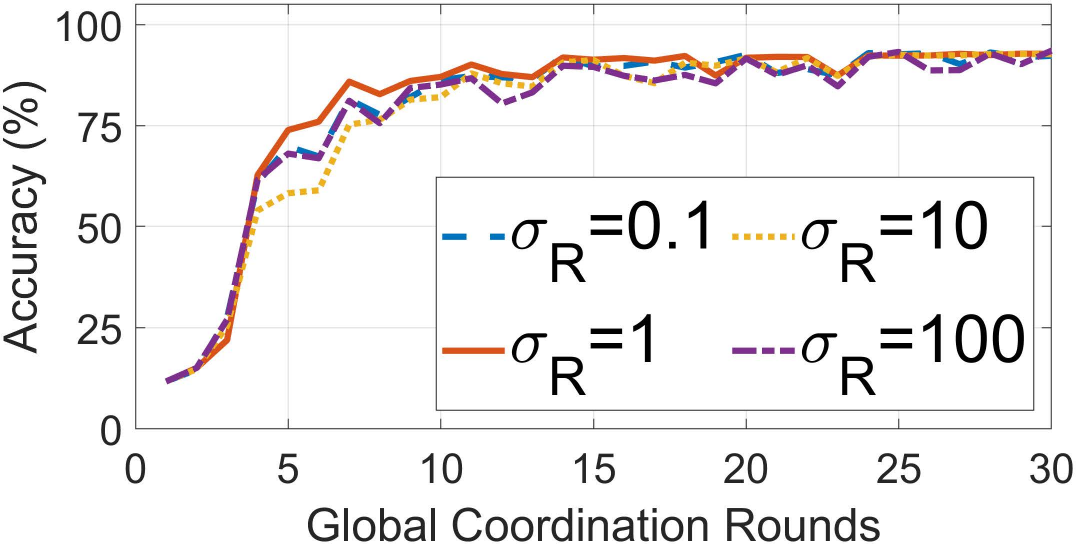}
         \caption{}         \label{Test_accuracy_convergence_different_sigma}
     \end{subfigure}
    \hfill
    \begin{subfigure}[b]{0.32\textwidth}
         \centering
           \includegraphics[width=1\textwidth]{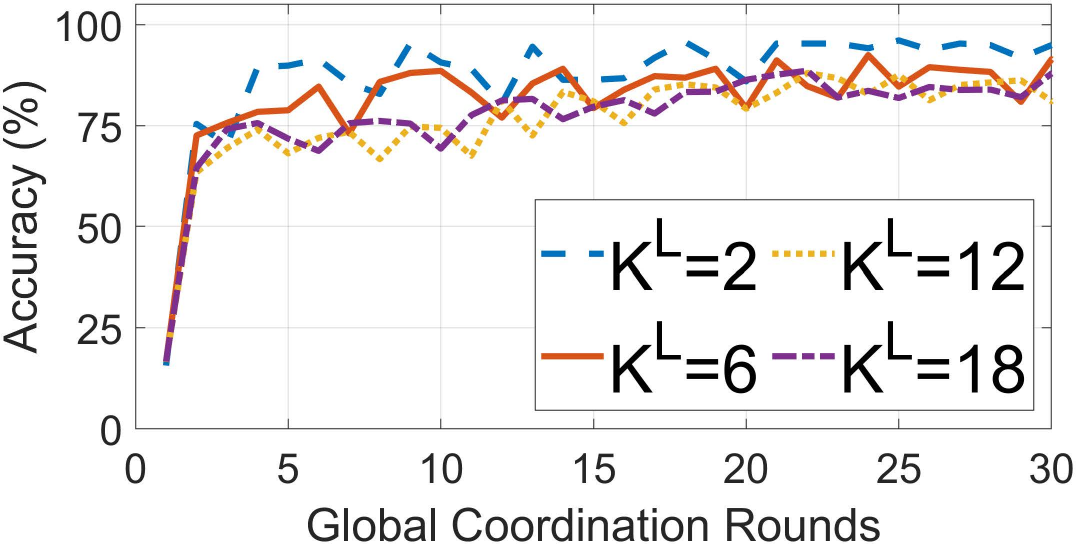}
         \caption{}         \label{Test_accuracy_convergence_different_number}
     \end{subfigure}
    \caption{Comparison of convergence rates under different model training parameters, including (a) $\lambda$ ($\sigma_R=1, K^L=18$); (b) $\sigma_R$ ($\lambda=1, K^L=18$); and (c)  $K^L$ ($\lambda=1, \sigma_R=1$).}     \label{Fig_convergence_comparison_across_different_setups_lambda}
\end{figure*}

\begin{figure*}[!ht]
     \centering
     \begin{subfigure}[b]{0.25\textwidth}
         \centering
           \includegraphics[width=1\textwidth]{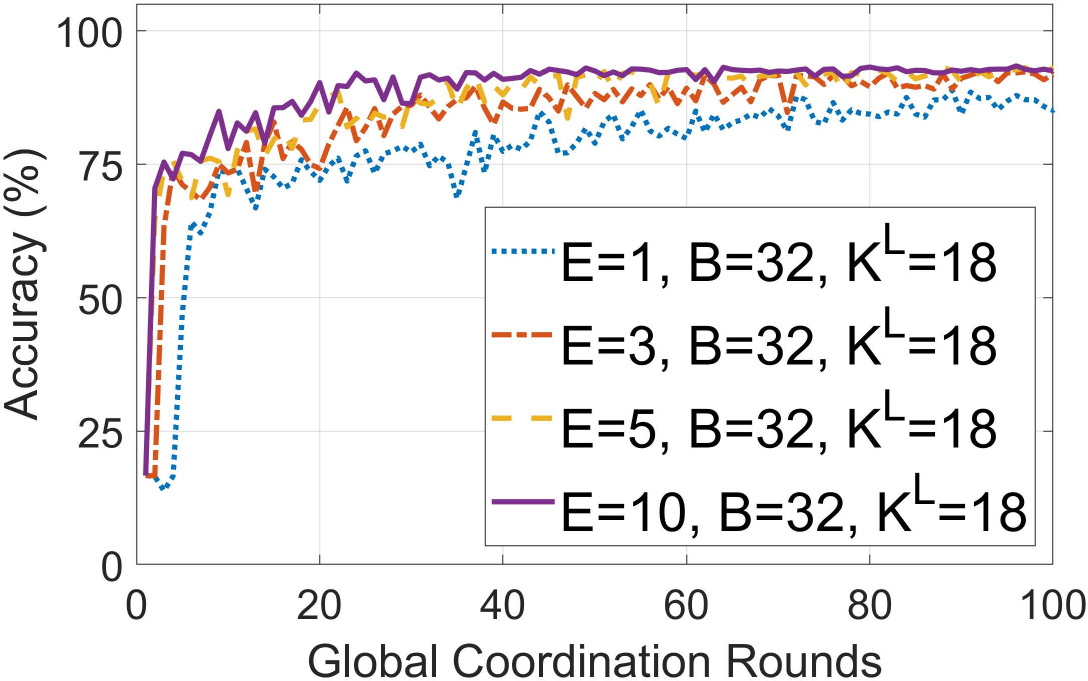}
         \caption{}
         \label{accuracy_convergence_different_E}
     \end{subfigure}
     \hfill
     \begin{subfigure}[b]{0.24\textwidth}
         \centering
           \includegraphics[width=1\textwidth]{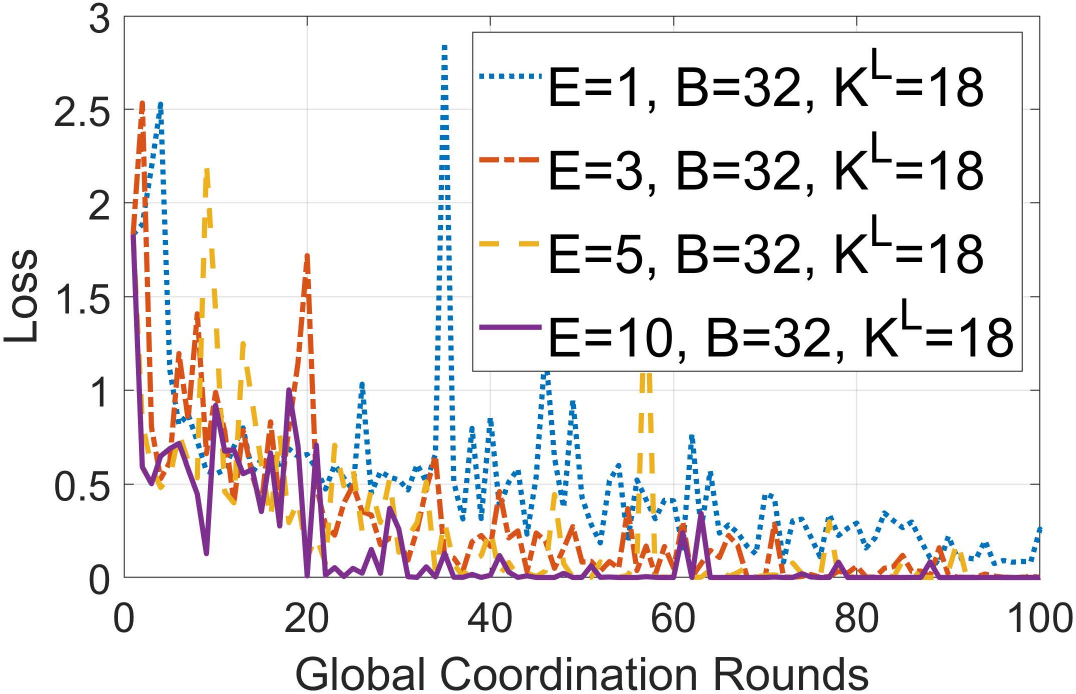}
         \caption{}
         \label{loss_convergence_different_E}
     \end{subfigure}
     \hfill
     \begin{subfigure}[b]{0.25\textwidth}
         \centering
           \includegraphics[width=1\textwidth]{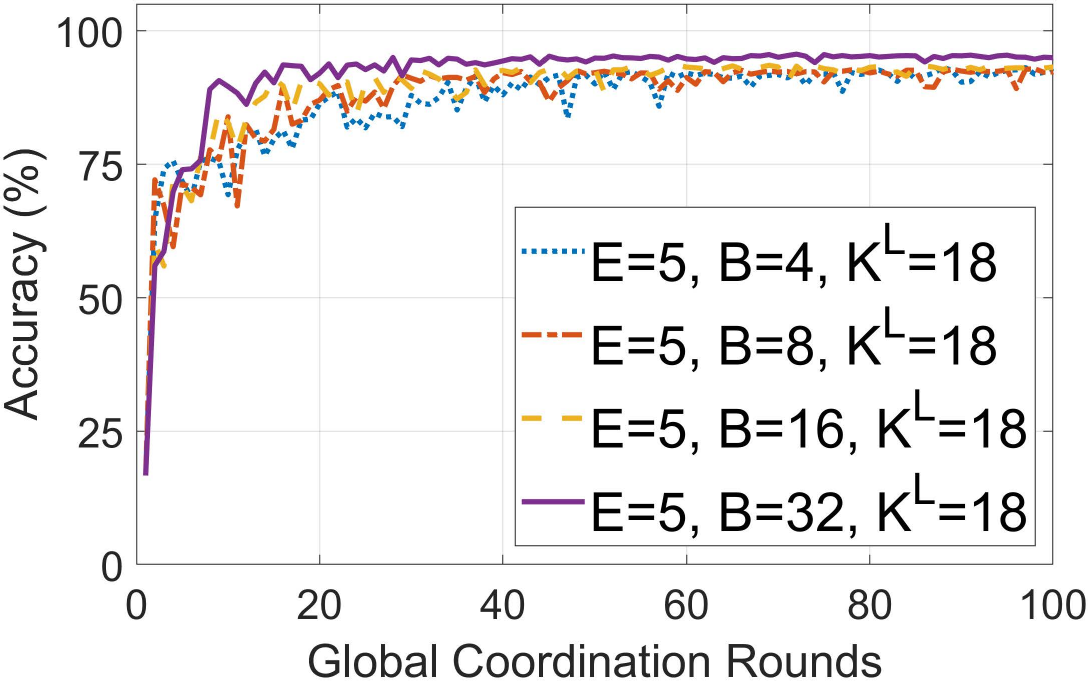}
         \caption{}         \label{accuracy_convergence_different_B}
     \end{subfigure}
     \hfill
     \begin{subfigure}[b]{0.24\textwidth}
         \centering
           \includegraphics[width=1\textwidth]{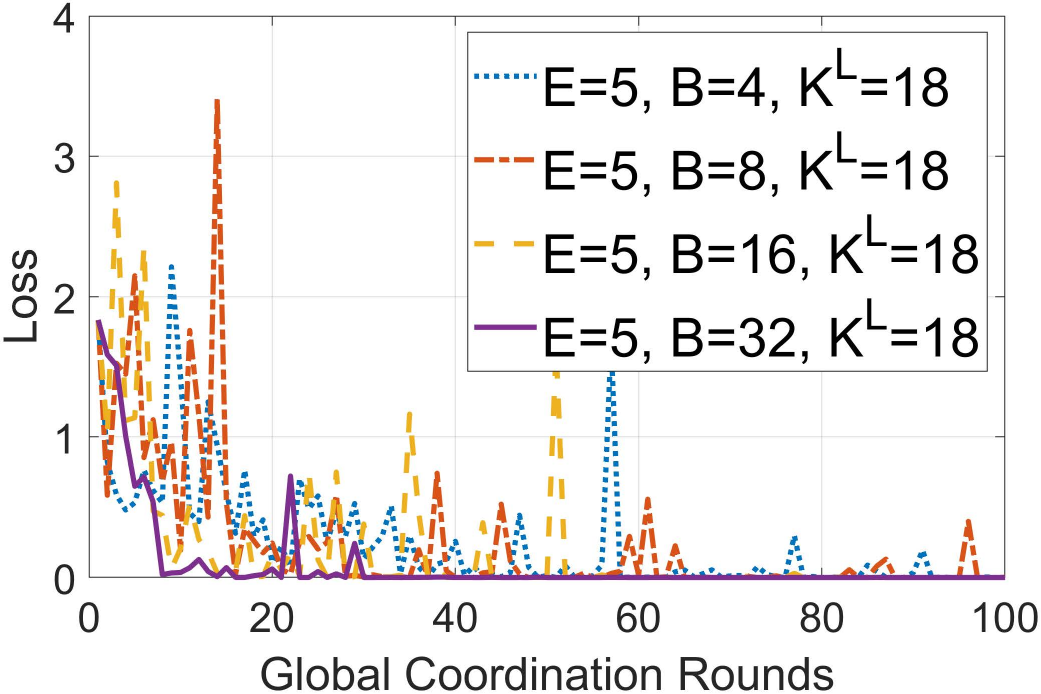}
         \caption{}
         \label{loss_convergence_different_B}
     \end{subfigure}
    \caption{Comparison of convergence rates of model training under different batch-sizes and local epoch numbers.}     \label{Fig_convergence_comparison_across_different_setups_BE}
\end{figure*}

To verify the theoretical results derived in Section \ref{Section_TheorecticalBoundModelTraining}, we evaluate the convergence performance of the model training process at receivers with labelled data under different number of coordination rounds and combinations of key model parameters including $\lambda$, $\sigma_R$, $K^L$, $B$ and $E$ in Fig. \ref{Fig_convergence_comparison_across_different_setups_lambda} and \ref{Fig_convergence_comparison_across_different_setups_BE}.

\higl{Recall that $\lambda$ is the collaboration parameter that controls the weights of the attention-inducing regulation function in the local objective function of each receiver. Increasing $\lambda$ accelerates collaboration between receivers with highly correlated models. We can observe in Fig. 8(a) that when the value of $\lambda$ increases from zero to one, the model convergence speed also increases. However, when $\lambda$ continues to increase from one to 10, the model accuracy will be degraded. This is because $\lambda$ can only control the weight of the regularization term in the local objective function, and when this weight becomes too high, the regularization term will overwhelm the overall local objective function, resulting in high distortion on the original local objective as well as resulting models. Therefore, there is an optimal $\lambda$ for the target problem, which can not only accelerate the model convergence and avoid overfitting, but also prevent the regularization term from overwhelming the effect of the cross-entropy loss term because the large penalty of increasing the weight modulus from 0 distorts the shape of the loss surface.}

Similarly, in Fig. \ref{Fig_convergence_comparison_across_different_setups_lambda}(b), we can observe that another key parameter $\sigma_R$ in the negative exponential regularization function to control the weights of aggregation of correlated model also needs to be carefully chosen to improve the model accuracy level with maximized convergence speed, e.g., as observed in Fig. \ref{Fig_convergence_comparison_across_different_setups_lambda}(b), the highest convergence performance is achieved when $\sigma_R=1$. In the rest of this section, we set both values of $\lambda$ and $\sigma_R$ into 1. In Fig. \ref{Fig_convergence_comparison_across_different_setups_lambda}(c), we present the convergence rate under different numbers of receivers participating in the model training. It is known that for most traditional federated learning solutions, if datasets at different receivers are non-iid, allowing more receivers to participate in the model training generally results in reduced convergence rates. We can observe in Fig. \ref{Fig_convergence_comparison_across_different_setups_lambda}(c),
\higl{however, that the convergence performances of SANSee do not change much even when the number of receivers participating in the model training increases from 2 to 18.}

In Fig. \ref{Fig_convergence_comparison_across_different_setups_BE}, we compare the average model accuracy and the loss values under different coordination rounds and combinations of mini-batch sizes $B$ and local iteration (epoch) numbers between consecutive coordination rounds $E$. We can observe in Theorem 1 that the convergence rate is in the order of ${\cal O} ({\frac{1}{E}})$ when all the other parameters are fixed, which is aligned with Fig. \ref{Fig_convergence_comparison_across_different_setups_BE}(a) and (b), in which we can observe that, as $E$ increases from 1 to 10, the number of global coordination rounds also increases.  Similarly, in Fig. \ref{Fig_convergence_comparison_across_different_setups_BE}(c) and (d), we fix $E=5$ and compare the convergence performance of SANSee under different $B$. We can observe that increasing $B$ results in almost linear reduction of the required number of coordination rounds $T$ to convergence.

\begin{figure}[!ht]
    \hfill
    \begin{subfigure}[b]{0.23\textwidth}
        \includegraphics[width=1\textwidth]{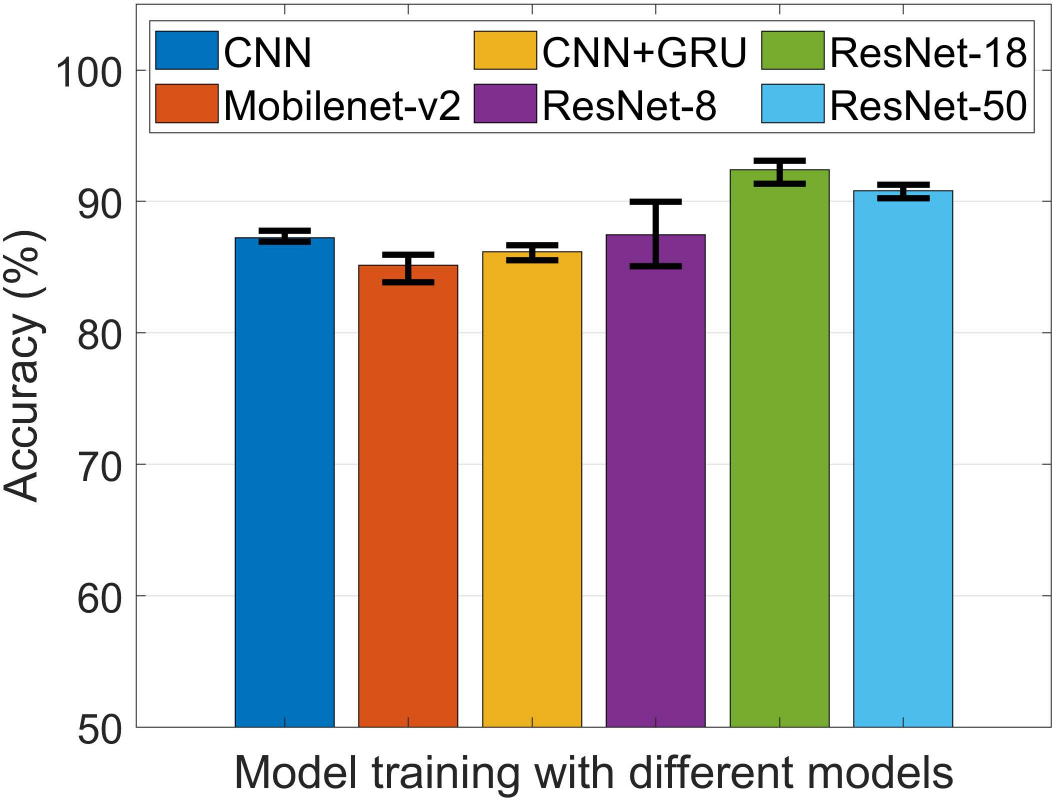}
        \label{Test_accuracy_pSAN_different_models}
        \caption{}
     \end{subfigure}
    \hfill
    \begin{subfigure}[b]{0.22\textwidth}
        \includegraphics[width=1\textwidth]{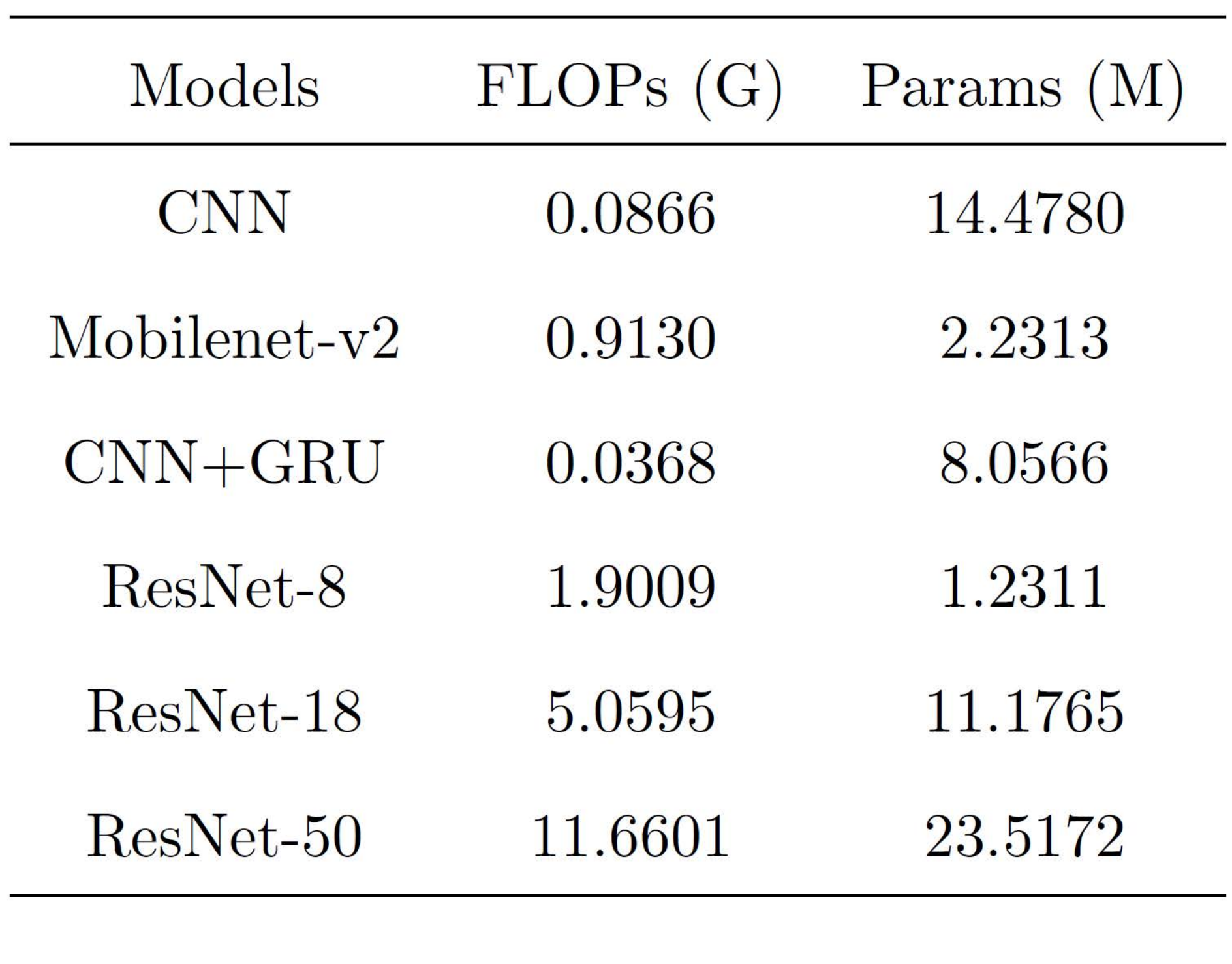}
        \label{model_FLOPs}
         \caption{}
    \end{subfigure}
    \hfill
    \caption{(a): Model accuracies achieved by different models, each with computational complexity (in FLOPs) and size of parameters listed in Figure (b), including a 2-layer CNN network ({CNN}), a lightweight CNN network ({Mobilenet-v2}) \cite{Sandler_2018_MobileNetV2}, a hybrid neural network composed of CNN and GRU ({CNN+GRU}) \cite{zheng2019Widar}, ResNet-8, ResNet-18, and ResNet-50.}
    \label{different_model_FLOPs}
\end{figure}

\higl{In Fig. 11(a), we compare the performance of wireless sensing when adopting different models in SANSee, including a 2-layer CNN, a lightweight CNN called Mobilenet-v2 \cite{Sandler_2018_MobileNetV2}, a hybrid neural network (CNN+GRU) consisting of a CNN for spatial feature extraction and a GRU for temporal modeling \cite{zheng2019Widar}, as well as ResNet-8, ResNet-18, and ResNet-50. We also compare the computational complexity (in FLOPs) and sizes of parameters of these models in Fig. 11(b). We observe that, in terms of average accuracy, ResNet-18 outperforms the other models, achieving 2.31\% to 7.45\% improvements on average. When considering the variance of wireless sensing, however, adopting more complex models (with a higher number of parameters) can always reduce the variance. This is due to the fact that for a given number of training samples, models with small or large numbers of parameters tend to cause underfitting or overfitting issues, resulting in higher bias with lower variance or lower bias with higher variance in performance. Furthermore, we can observe that lightweight models such as Mobilenet-v2 and ResNet-8 can still achieve relatively good wireless sensing accuracy. Furthermore, choosing complex models such as ResNet-50, i.e., models with large numbers of parameters, may not always result in improved performance. This is because large models may result in overfitting. }

\begin{figure}[!ht]
    \hfill
    \begin{subfigure}[b]{0.23\textwidth}
        \centering
        \includegraphics[width=1\textwidth]{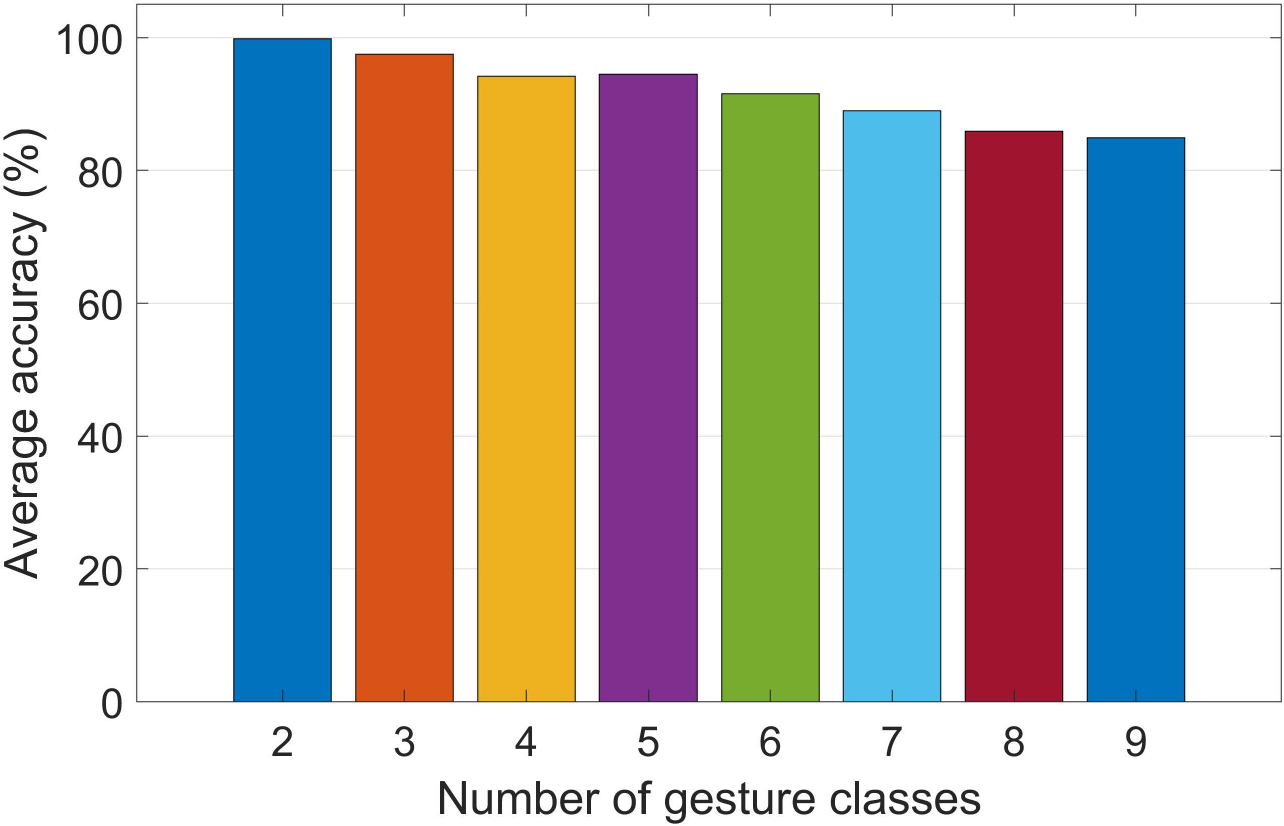}
        \caption{}
         \label{different_classes_pSAN_resnet1}
     \end{subfigure}
    \hfill
    \begin{subfigure}[b]{0.23\textwidth}
        \centering
        \includegraphics[width=1\textwidth]{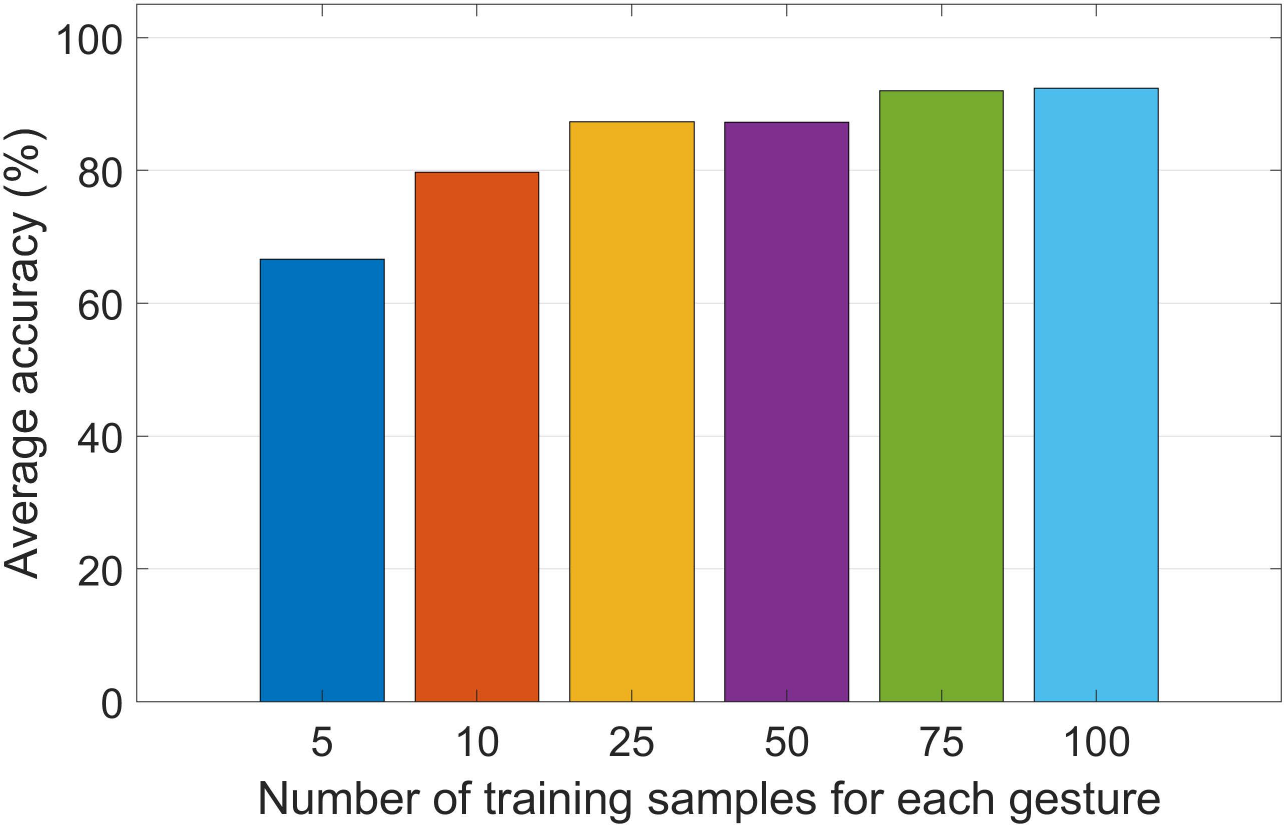}
         \caption{}
         \label{different_samples_pSAN_resnet18}
    \end{subfigure}
    \hfill
    \caption{Average model accuracy with (a) different numbers of gesture classes, and (b) different training dataset sizes per gesture (with six gestures in total) at each receiver.}
    \label{Test_accuracy_pSAN_resnet18}
\end{figure}

\higl{In Fig. \ref{Test_accuracy_pSAN_resnet18}, we compare the average model accuracy of SANSee under different numbers of gesture classes (Fig. \ref{Test_accuracy_pSAN_resnet18}(a)) and different training dataset sizes per gesture (with six gestures in total) (Fig. \ref{Test_accuracy_pSAN_resnet18}(b)) at each receiver. From Fig. \ref{Test_accuracy_pSAN_resnet18}(a), we observe that when the number of gesture classes increases from 2 to
9, the average model accuracy decreases from 99.809$\%$ to 84.915$\%$. This is because, as the number of gesture classes increases, the output dimension of the model also increases, resulting in underfitting issues for each class of gestures. In Fig. \ref{Test_accuracy_pSAN_resnet18}(b), we can observe that when the training dataset size per gesture at each receiver increases, the increasing speed of the average model accuracy decreases. For example, as the number of samples per gesture at each receiver increases from 1 to 25, the model accuracy increases from 49.05\% to 87.28\%, resulting in 38.23$\%$ improvement. However, if the training sample size continues to increase from 75 to 100, the model accuracy improves by only 0.417$\%$.}

\subsection{Model Transfer}
To evaluate the performance of SANSee for transferring the already trained models, i.e., base models, to receivers without labeled dataset, we consider two model transfer scenarios: {\em in-environment model transfer} in which all the receivers with and without labelled data are  in the same environment and {\em cross-environment model transfer} in which models trained by receivers in one environment are transferred to receivers located in a new environment.

We evaluate the in-environment model transfer performance in Fig. \ref{Test_accuracy_at_each_node_different_receivers_transfer}, in which we compare the gesture recognition accuracy of models obtained by  a receiver without labeled data 
when its model are transferred based on different numbers, 1-4, of available models constructed by receivers with labeled data. We can observe that, as the number of available models increases, the accuracy of the transferred model also improves. The increasing speed of the model accuracy however decreases as the number of available models becomes large. This means that SANSee is able to transfer a relatively small number of trained models, e.g., trained by two to three receivers, to any number of location-specific models with sufficiently ``good" accuracy, e.g., above 70\% gesture recognition accuracy.

\begin{figure}[!ht]
    \centering
       \includegraphics[width=0.5\textwidth]{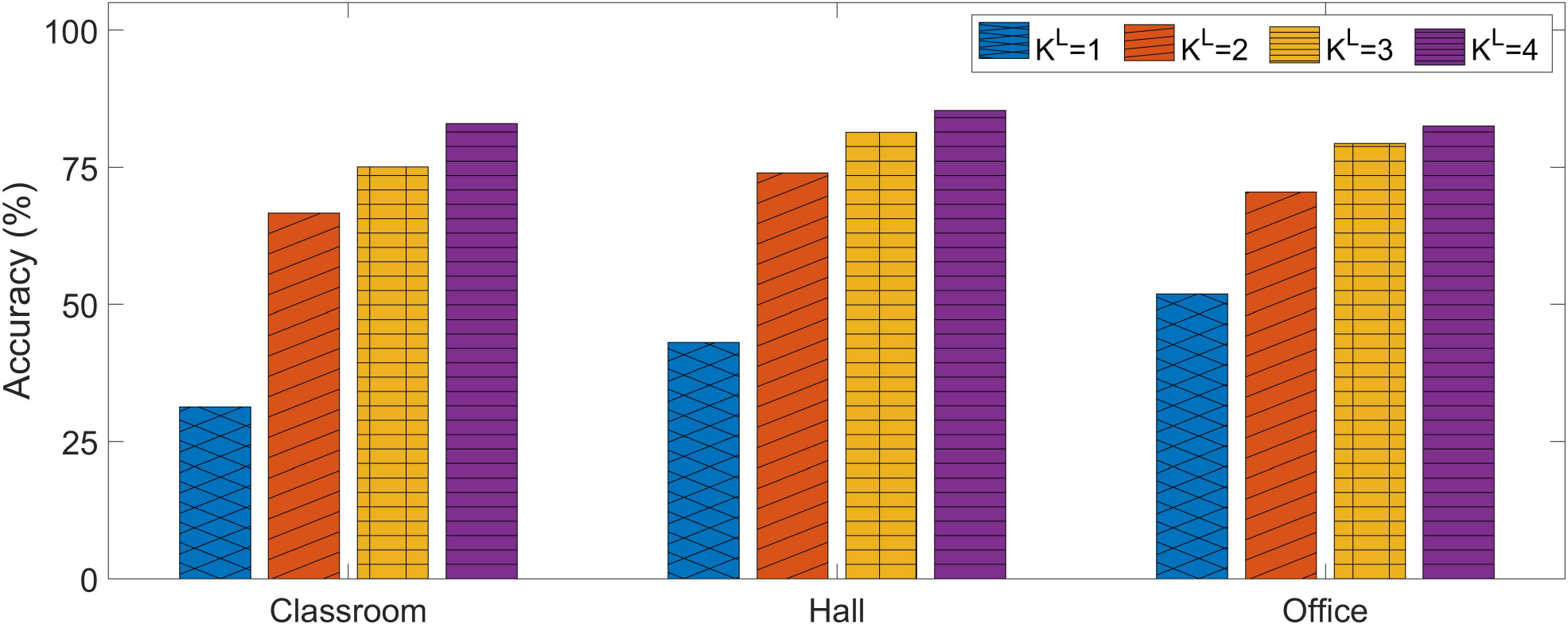}
     \caption{Comparison of in-environment model transfer performance at three environment under different numbers of base models $K^L$.}         \label{Test_accuracy_at_each_node_different_algorithm_transfer}
\end{figure}
\begin{figure}[!ht]
    \centering
       \includegraphics[width=0.5\textwidth]{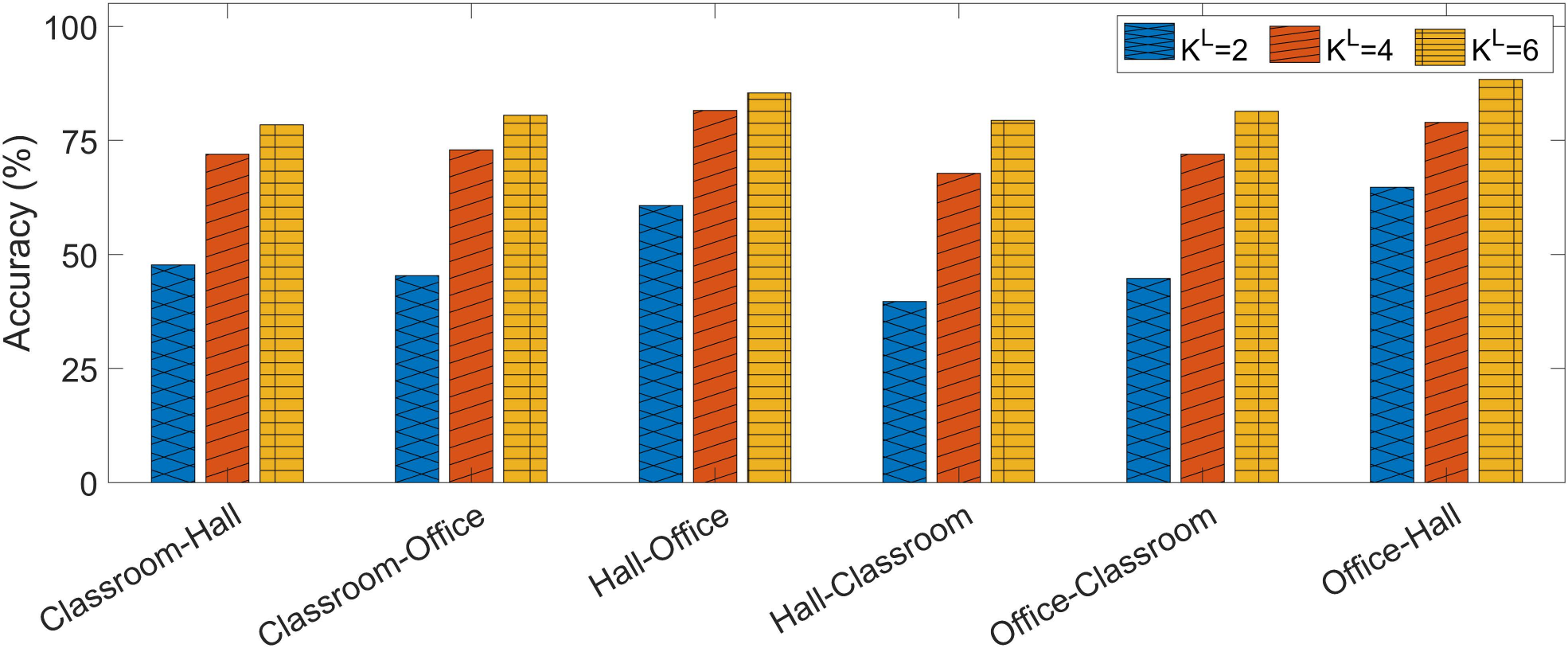}
     \caption{Comparison of cross-environment model transfer performance under different numbers of base models $K^L$.}         \label{Test_accuracy_at_each_node_different_receivers_transfer}
\end{figure}

We evaluate the cross-environment model transfer performance in Fig. \ref{Test_accuracy_at_each_node_different_algorithm_transfer} where models trained by receivers in one environment are transferred to the receivers in another environment. We can observe that, generally speaking, the accuracy of the cross-environment model transfer is slightly worse than that of the in-environment model transfer under the same number of available models. The performance of the transferred model is again affected by the number of models that have already been trained. For example, in hall-classroom cross-environment model transfer scenario, as the number of models increases from 2 to 6, the accuracy of the transferred model improves almost 50\% from accuracies 41\% to 83\%. The increasing speed of the model performance again approaches a stationary level when the number of available models increases. In other words, SANSee is a useful solution for achieving sustainable network AI in a large networking system, in which an almost infinite number of novel models tailored for a wide range of applications and scenarios can be transferred based on a very limited number of base models using their semantic correlations.

\section{Conclusion}\label{Conclusion}
This paper proposed a semantic-aware networking-based framework for distributed wireless sensing, called SANSee, that allowed models constructed in a limited number of locations to be directly transferred to other locations without any training efforts. In particular, a physical-layer semantic-aware network, called pSAN, has been developed to characterize the similarity between physical-layer semantic features and the correlations of wireless sensing data distributions across different locations. We have then proposed a pSAN-based zero-shot transfer learning solution for receivers without labeled data to construct its location-specific model based on the correlated model trained by receivers with labeled data. Finally, extensive experiments have been conducted based on real-world datasets to evaluate the performance of SANSee, and numerical results showed the accuracy of transferred models obtained by SANSee matched that of the models trained by the locally labeled data based on supervised learning approaches.

\vspace{-0.05cm}
\section*{Acknowledgment}
\vspace{-0.1cm}
The work of Y. Xiao was supported in part by the National Natural Science Foundation of China (NSFC) under grant 62071193. The work of Y. Li was supported
in part by the "CUG Scholar" Scientific Research Funds at China University of Geosciences (Wuhan) (Project No.2021164), in part by the International Science and Technology Cooperation Program of Hubei Province under Grant 2023EHA009, and in part by the NSFC under grant 62301516. The work of G. Shi was supported in part by the NSFC under grant 62293483.  Y. Xiao, Y. Li, and G. Shi were supported in part by the Major Key Project of Peng Cheng Laboratory under grant PCL2023AS1-2. 

\bibliographystyle{ieeetr}
\bibliography{myref,DeepLearningRef}

\begin{IEEEbiography}[{\includegraphics[width=1.1in,height=1.3in,clip,keepaspectratio]{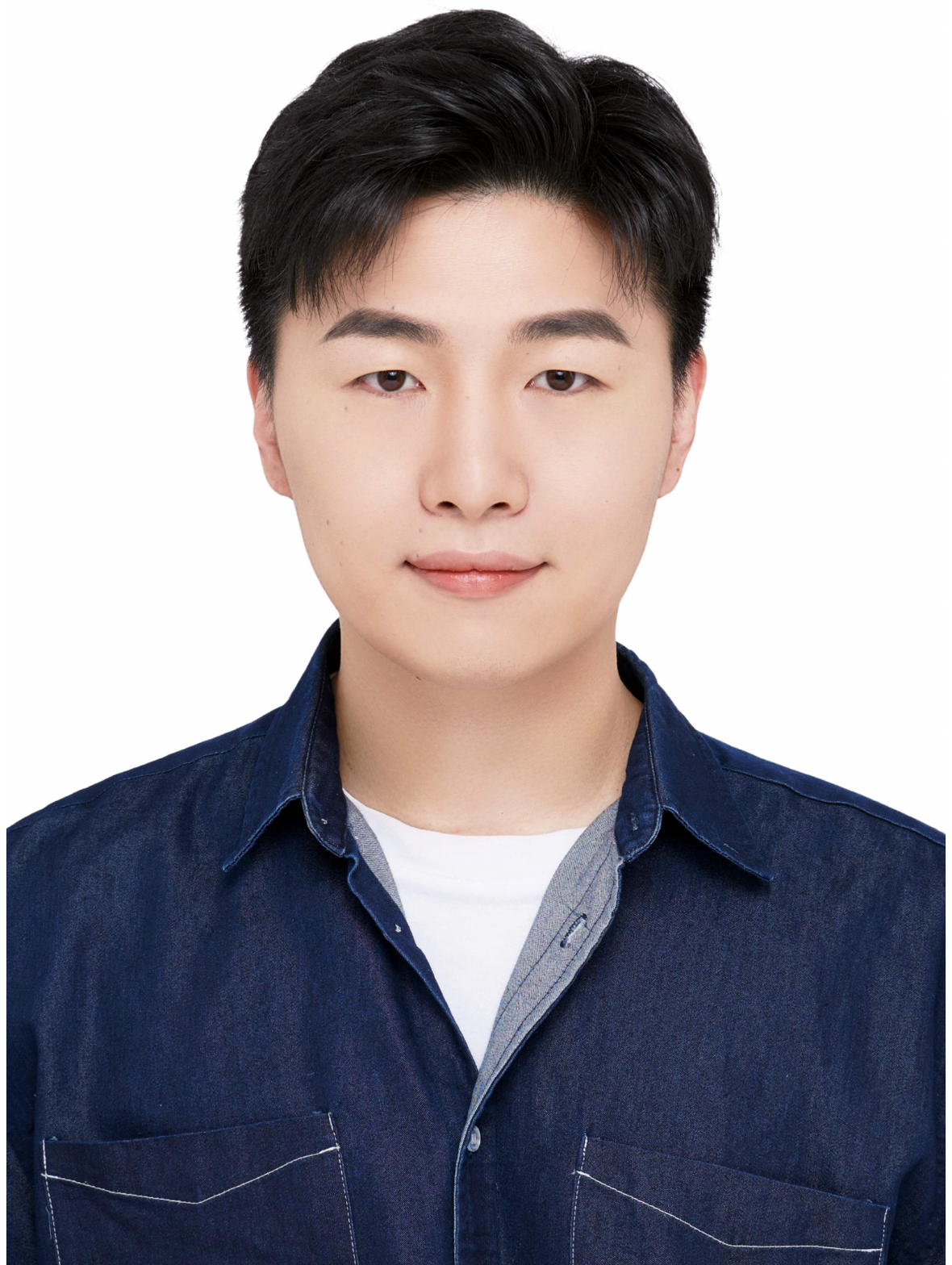}}]{Huixiang Zhu} received his B.S. degree in Wuhan University of Technology, Wuhan, China in 2019. He is currently pursuing his PhD degree in the School of Electronic Information and Communications at the Huazhong University of Science and Technology, Wuhan, China. His research interests include semantic-aware communications and network AI.
\end{IEEEbiography}

\begin{IEEEbiography}[{\includegraphics[width=1.1in,height=1.3in,clip,keepaspectratio]{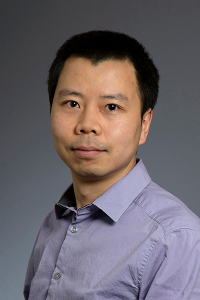}}]{Yong Xiao} (Senior Member, IEEE) received his B.S. degree in electrical engineering from China University of Geosciences, Wuhan, China in 2002, M.Sc. degree in telecommunication from Hong Kong University of Science and Technology in 2006, and his Ph. D degree in electrical and electronic engineering from Nanyang Technological University, Singapore in 2012. He is now a professor in the School of Electronic Information and Communications at the Huazhong University of Science and Technology (HUST), Wuhan, China. He is also with Peng Cheng Laboratory, Shenzhen, China and Pazhou Laboratory (Huangpu), Guangzhou, China. He is the associate group leader of the network intelligence group of IMT-2030 (6G promoting group) and the vice director of 5G Verticals Innovation Laboratory at HUST. Before he joins HUST, he was a  research assistant professor in the Department of Electrical and Computer Engineering at the University of Arizona where he was also the center manager of the Broadband Wireless Access and Applications Center (BWAC), an NSF Industry/University Cooperative Research Center (I/UCRC) led by the University of Arizona.
His research interests include machine learning, game theory, distributed optimization, and their applications in semantic communications and semantic-aware networks, cloud/fog/mobile edge computing, green communication systems, wireless communication networks, and Internet-of-Things (IoT).
\end{IEEEbiography}

\begin{IEEEbiography}[{\includegraphics[width=1.1in,height=1.3in,clip,keepaspectratio]{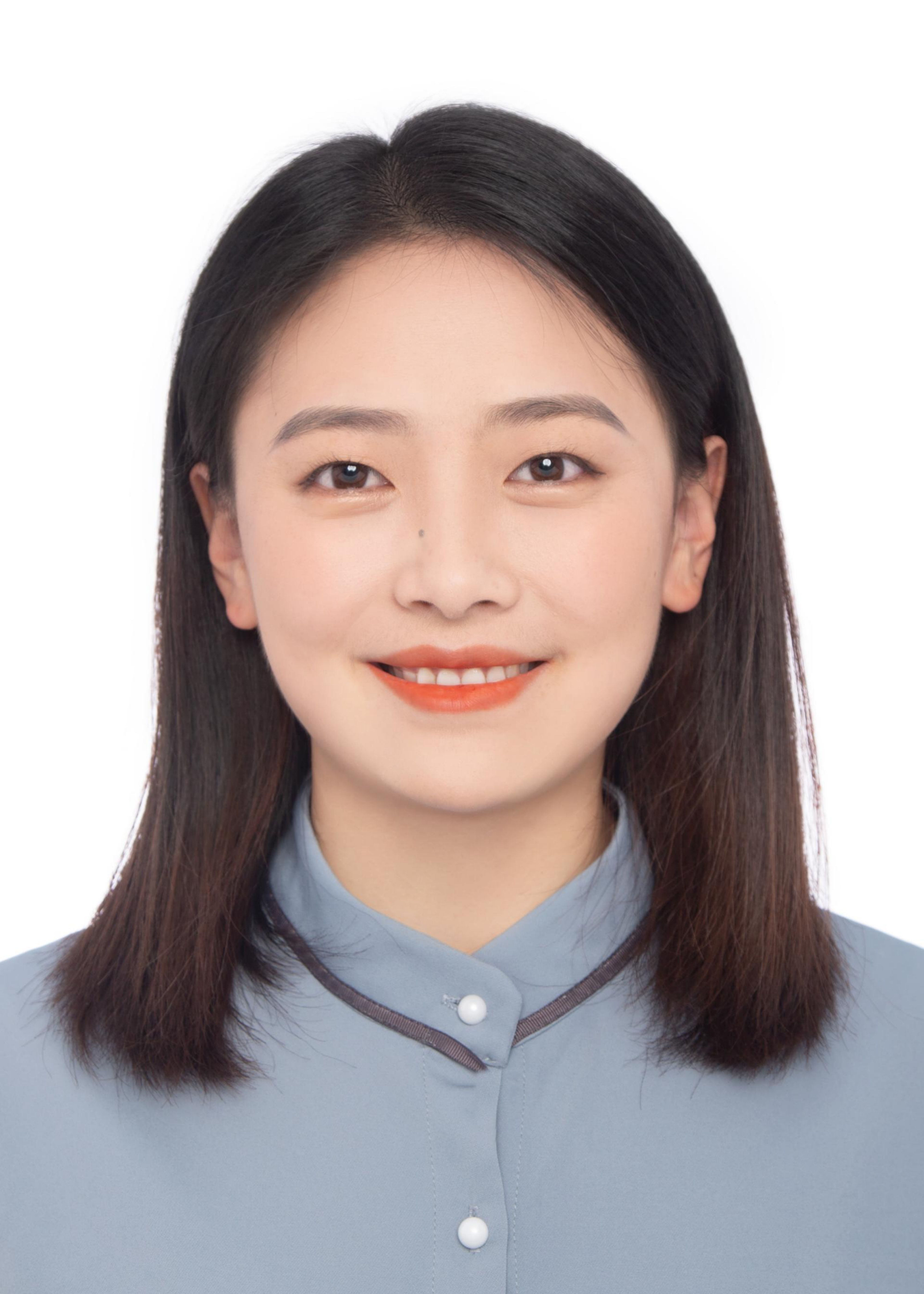}}]{Yingyu Li} (Member, IEEE) received the B.Eng. degree in electronic information engineering and the Ph.D. degree in circuits and systems from the Xidian University, Xi'an, China, in 2012 and 2018, respectively. From 2014 to 2016, she was a Research Scholar with the Department of Electronic Computer Engineering at the University of Houston, TX, USA. She was a postdoctoral researcher in the School of Electronic Information and Communications at Huazhong University of Science and Technology from 2018 to 2021. She is now an associate professor at the School of Mechanical Engineering and Electronic Information, China University of Geosciences (Wuhan). Her research interests include semantic communications, edge intelligence, green communication networks, and IoT.
\end{IEEEbiography}

\begin{IEEEbiography}[{\includegraphics[width=1.1in,height=1.3in,clip,keepaspectratio]{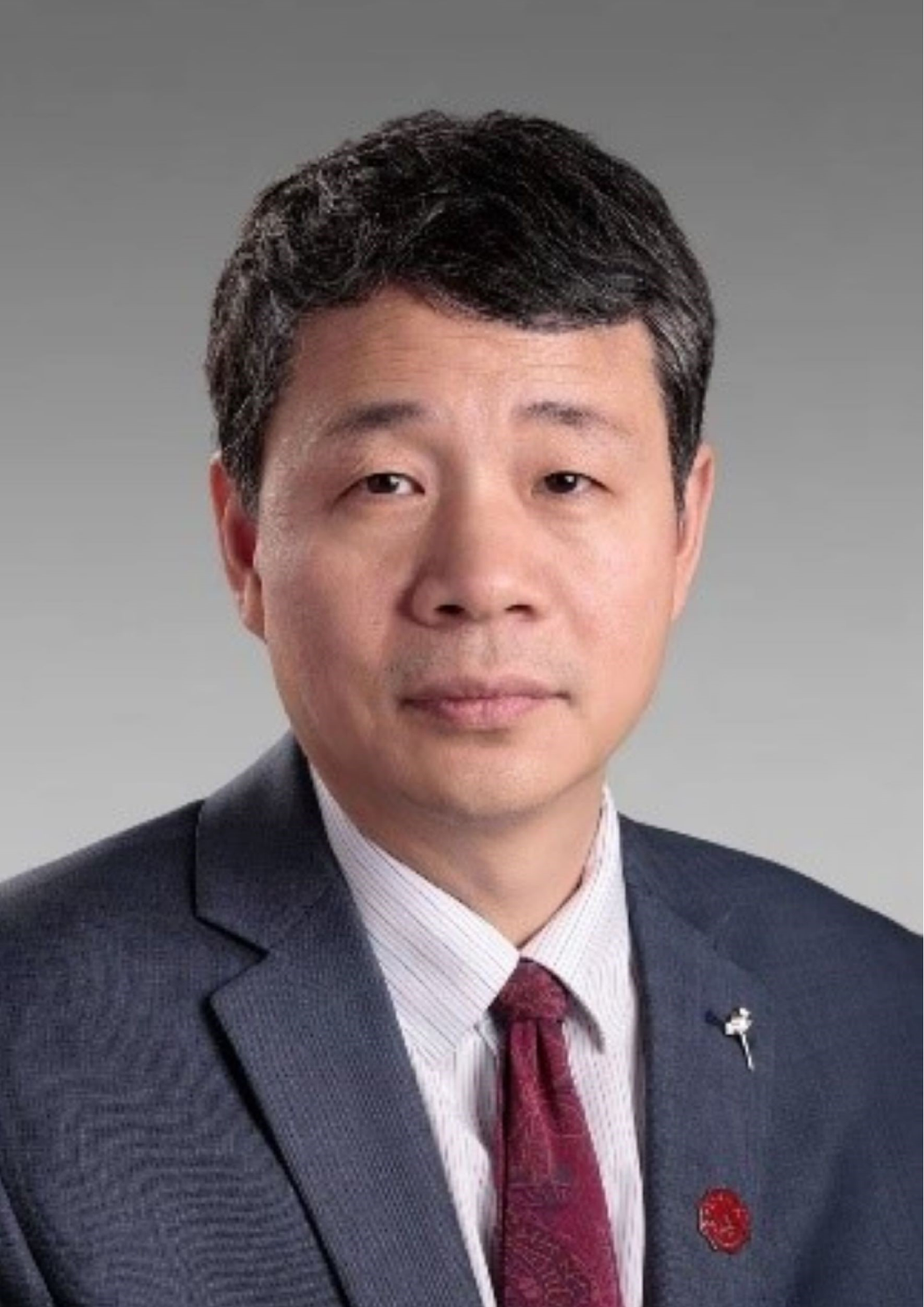}}]{Guangming Shi} (Fellow, IEEE) received the M.S. degree in computer control and the Ph.D. degree in electronic information technology from Xidian University, Xi’an, China, in 1988, and 2002, respectively. He was the vice president of Xidian University from 2018 to 2022. Currently, he is  the Vice Dean of Peng Cheng Laboratory and a Professor with the School of Artificial Intelligence, Xidian University. He is an IEEE Fellow, the chair of IEEE CASS Xi’an Chapter, senior member of ACM and CCF, Fellow of Chinese Institute of Electronics, and Fellow of IET. He was awarded Cheung Kong scholar Chair Professor by the ministry of education in 2012. He won the second prize of the National Natural Science Award in 2017. His research interests include Artificial Intelligence, Semantic Communications, and Human-Computer Interaction.
\end{IEEEbiography}

\begin{IEEEbiography}[{\includegraphics[width=1.1in,height=1.3in,clip,keepaspectratio]{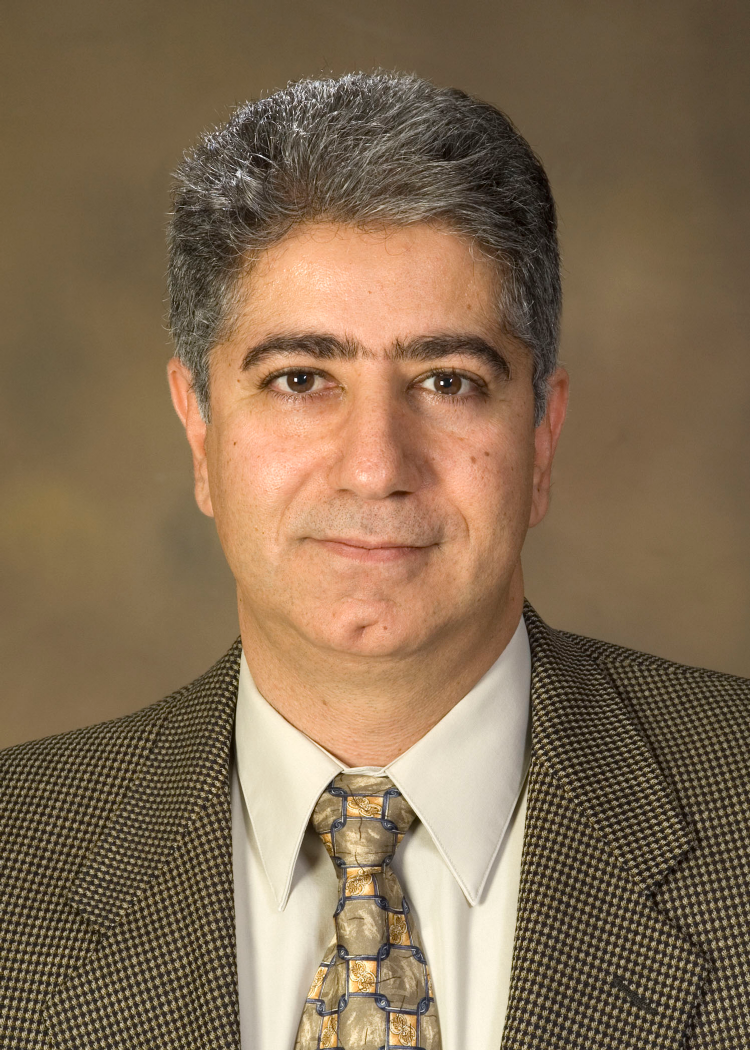}}]{Marwan Krunz}(Fellow, IEEE) is a Regents Professor at the University of Arizona. He holds the Kenneth VonBehren Endowed Professorship in ECE and is also a professor of computer science. He directs the Broadband Wireless Access and Applications Center (BWAC), a multi-university NSF/industry center that focuses on next-generation wireless technologies. He also holds a courtesy appointment as a professor at University Technology Sydney. Previously, he served as the site director for  Connection One, an NSF/industry-funded center of five universities and 20+ industry affiliates. Dr. Krunz’s research is in the fields of wireless communications, networking, and security, with recent focus on applying AI and machine learning techniques for protocol adaptation, resource management, and signal intelligence. He has published more than 320 journal articles and peer-reviewed conference papers, and is a named inventor on 12 patents. His latest h-index is 60. He is an IEEE Fellow, an Arizona Engineering Faculty Fellow, and an IEEE Communications Society Distinguished Lecturer (2013-2015). He received the NSF CAREER award. He served as the Editor-in-Chief for the IEEE Transactions on Mobile Computing. He also served as editor for numerous IEEE journals. He was the TPC chair for INFOCOM’04, SECON’05, WoWMoM’06, and Hot Interconnects 9. He was the general vice-chair for WiOpt 2016 and general co-chair for WiSec’12. Dr. Krunz served as chief scientist/technologist for two startup companies that focus on 5G and beyond wireless systems. 
\end{IEEEbiography}

\newpage
\appendices

\section{Proof of Theorem \ref{Theorem_TrainError}}
\label{Proof_Theorem_TrainError}
In this section, we present detailed derivation of the (local) model training error $\mathbb{E}[\mathcal{J}(\pmb{\Omega})-\mathcal{J}(\pmb{\Omega}^{*})]$ in Theorem \ref{Theorem_TrainError}, where $\mathcal{J}(\pmb{\Omega})$ is defined previously in (\ref{global_obj}), can converge to near the minimum as the number of training rounds $T$ increases. Before we present the detailed proofs, let us first introduce the following lemmas which will be useful for our proofs of Theorem 1.

\subsection{Key Lemmas of Theorem \ref{Theorem_TrainError}}
Recall the definition of ${\cal J}\left({\boldsymbol \Omega}\right)={\cal F}\left({\boldsymbol \Omega}\right)+\lambda {\cal R}\left({\boldsymbol \Omega}\right)$ in (\ref{global_obj}), where $\mathcal{R}({\boldsymbol \Omega})=\sum_{{k'}\in \mathcal{K}^L} \sum_{{k'} \neq {j'}} R\left(\| \pmb{\omega}_{k'}- \pmb{\omega}_{{j'}}\|^2\right)$ and $\mathcal{F}({\boldsymbol \Omega})=\sum_{k'\in \mathcal{K}^L} F_{k'}({\boldsymbol \omega}_{k'})$, respectively.

\begin{lemma}
Suppose that Assumptions 1-3 in Theorem 1 hold. If  $L_J:=L+\lambda \kappa_R$, then for any $\pmb{\Omega}$, $\pmb{\Omega}^{\prime} \in \mathbb{R}^{d K}$, we can derive the following results:

(a) $\left\|\nabla \mathcal{F}(\pmb{\Omega})-\nabla \mathcal{F}\left(\pmb{\Omega}^{\prime}\right)\right\| \leq L\left\|\pmb{\Omega}-\pmb{\Omega}^{\prime}\right\|$;

(b) $\left\|\nabla \mathcal{J}(\pmb{\Omega})-\nabla \mathcal{J}\left(\pmb{\Omega}^{\prime}\right)\right\| \leq L_J\left\|\pmb{\Omega}-\pmb{\Omega}^{\prime}\right\|$;

(c) $\|\nabla \mathcal{F}(\pmb{\Omega})\|^2 \leq 2 L^2\left\|\pmb{\Omega}-\pmb{\Omega}^{\prime}\right\|^2+2\left\|\nabla \mathcal{F}\left(\pmb{\Omega}^{\prime}\right)\right\|^2$;

(d) $\|\nabla \mathcal{J}(\pmb{\Omega})\|^2 \leq 2 L_J^2\left\|\pmb{\Omega}-\pmb{\Omega}^{\prime}\right\|^2+2\left\|\nabla \mathcal{J}\left(\pmb{\Omega}^{\prime}\right)\right\|^2$;

(e) $\mathcal{J}(\pmb{\Omega})-\mathcal{J}\left(\pmb{\Omega}^{\prime}\right) \leq\left\langle\nabla \mathcal{J}\left(\pmb{\Omega}^{\prime}\right), \pmb{\Omega}-\pmb{\Omega}^{\prime}\right\rangle+\frac{L_J}{2}\left\|\pmb{\Omega}-\pmb{\Omega}^{\prime}\right\|^2$.

(f) $\mathbb{E}\|\nabla \widetilde{\mathcal{J}}(\pmb{\Omega}, \zeta)-\nabla \mathcal{J}(\pmb{\Omega})\|^2 \leq \sigma_F^2$
\end{lemma}

\begin{lemma}\label{lemma11}
 Suppose that Assumption 1-2 hold and $\lambda >2L$. Then there exists a positive value $\Gamma^L$ such that, for any $\pmb{\Omega}$, $\pmb{\Omega} \in \mathbb{R}^{d K}$, we have
\begin{equation}
\begin{aligned}
 \|\nabla\mathcal{F}({\boldsymbol \Omega}^{*})\|^2 \le \Gamma^L .
 \label{lemma1_2}
\end{aligned}
\end{equation}
 $\Gamma^L$ is measured only at unique solution ${\boldsymbol \Omega}^{*}$, and thus ${\Gamma^L}$ is finite. The bound is tight in the sense that $\Gamma^L=\lambda^2\|\nabla{\cal R}\left({\boldsymbol \Omega}^*\right)\|^2=0$ for the i.i.d. cases, where $\pmb{\omega}_1=\cdots=\pmb{\omega}_{K^L}$. In the considered wireless scenarios, there is always $\Gamma^L>0$ due to the heterogeneity of wireless data samples.
\end{lemma}

\begin{lemma}
Suppose that Assumptions 4 holds. The gradient of server update in (\ref{Server_gradient}) at the $t$th coordination round, denoted as $g^t$, can be be represented as follows:
\begin{eqnarray}
    g^t&=&\sum_{e=0}^{E-1} \nabla \widetilde{J}\left(\pmb{\Omega}^{t, e}\right)+\lambda \kappa_R \mathcal{L}\left( \pmb{\Omega}^{t,e}- \pmb{\Omega}^{t}\right)\nonumber\\
    &&+\frac{\eta \lambda \kappa_R  \mathcal{L}}{E} \sum_{e=0}^{E-1}   \nabla \widetilde{F}\left(\pmb{\Omega}^{t, e}\right).
\end{eqnarray}
\begin{IEEEproof}
In the $t$th coordination round, the $e$th local update of the model vector $\pmb{\Omega}$ in (\ref{Local_SGD}) can be represented as follows:
\begin{eqnarray}
\pmb{\Omega}^{t, e+1}=\pmb{\Omega}^{t, e}-\eta   \nabla \widetilde{F}\left(\pmb{\Omega}^{t, e}\right)
\end{eqnarray}
implies that after $E$ local update steps, we have
\begin{eqnarray}
\eta   \sum_{e=0}^{E-1} \nabla \widetilde{F}\left(\pmb{\Omega}^{t, e}\right)&=&\sum_{e=0}^{E-1}\left(\pmb{\Omega}^{t, e}-\pmb{\Omega}^{t, e+1}\right)\nonumber\\
&=&\pmb{\Omega}^{t, 0}-\pmb{\Omega}^{t, e}=\pmb{\Omega}^{t}-\pmb{\Omega}^{t, e} .
\label{con_update}
\end{eqnarray}
we then rewrite the server update in (\ref{Server_gradient}) as follows
\begin{eqnarray}
\pmb{\Omega}^{t+1}&=&\left(I-\eta \lambda \kappa_R   \mathcal{L}\right)\left[\pmb{\Omega}^{t}- \eta\sum_{e=0}^{E-1}   \nabla \widetilde{F}\left(\pmb{\Omega}^{t, e}\right)\right] \\
 &=&\pmb{\Omega}^{t}-\eta \sum_{e=0}^{E-1} \nabla \widetilde{F}\left(\pmb{\Omega}^{t, e}\right)-\eta \lambda \kappa_R \mathcal{L} \pmb{\Omega}^{t}\nonumber\\
 &&+\frac{\eta^2 \lambda \kappa_R  \mathcal{L}}{E} \sum_{e=0}^{E-1}   \nabla \widetilde{F}\left(\pmb{\Omega}^{t, e}\right) \nonumber\\
&=&\pmb{\Omega}^{t}-\eta \left(\sum_{e=0}^{E-1} \nabla \widetilde{F}\left(\pmb{\Omega}^{t, e}\right)-\lambda \kappa_R \mathcal{L} \pmb{\Omega}^{t}\right.\nonumber\\
&&\left.+\frac{\eta \lambda \kappa_R  \mathcal{L}}{E} \sum_{e=0}^{E-1}   \nabla \widetilde{F}\left(\pmb{\Omega}^{t, e}\right)\right) \nonumber\\
 &=&\pmb{\Omega}^{t}-\eta \left(\sum_{e=0}^{E-1} \nabla \widetilde{J}\left(\pmb{\Omega}^{t, e}\right)+\frac{\lambda \kappa_R  \mathcal{L}}{E}\sum_{e=0}^{E-1}\left( \pmb{\Omega}^{t,e}- \pmb{\Omega}^{t}\right)\right. \nonumber\\
&&\left.+\frac{\eta \lambda \kappa_R  \mathcal{L}}{E} \sum_{e=0}^{E-1}   \nabla \widetilde{F}\left(\pmb{\Omega}^{t, e}\right)\right), \nonumber
\end{eqnarray}
which finishes the proof.
\end{IEEEproof}
\label{lemma2}
\end{lemma}

\begin{lemma}
Suppose that Assumptions 1 to 3 hold. We can derive the gradient bound of the server update in Lemma \ref{lemma2} as follows:
\begin{eqnarray}
\mathbb{E}\left\|Z^t\right\|^2 &\leq& (1+\eta^2 \lambda^2 \kappa^2_R  \mathcal{L}^2)\left(6 L_J^2 \mathcal{E}^{(t)}+6 \mathbb{E}\left\|\nabla J\left(\pmb{\Omega}^{t}\right)\right\|^2\right. \nonumber\\
&&\left.+\frac{3 \sigma_F^2}{E}\right) +3 \lambda^2 \kappa^2_R \|\mathcal{L}\|^2 \mathcal{E}^{(t)}
\label{30}
\end{eqnarray}

\begin{IEEEproof}
Using Jensen’s inequality, we have that
\begin{eqnarray}
\mathbb{E}\left\|Z^t\right\|^2 &\leq& 3 \mathbb{E}\left\|\frac{1}{E} \sum_{e=0}^{E-1} \nabla \widetilde{J}\left(\pmb{\Omega}^{t, e}\right)\right\|^2\\
&&+3 \mathbb{E}\left\|\frac{\lambda \kappa_R  \mathcal{L} }{E} \sum_{e=0}^{E-1} \mathcal{L}\left(\pmb{\Omega}^{t, e}-\pmb{\Omega}^{t}\right)\right\|^2 \nonumber\\
&& +3 \mathbb{E}\left\|\frac{\eta \lambda \kappa_R  \mathcal{L}}{E} \sum_{e=0}^{E-1}   \nabla \widetilde{F}\left(\pmb{\Omega}^{t, e}\right)\right\|^2. \nonumber
\label{34}
\end{eqnarray}

First, according to \cite[Lemma 9]{dinh2022new} the definition of $\mathcal{E}^{(t)}$, we can bound the the first term in (\ref{34}) as follows:
\begin{eqnarray}
\lefteqn{3 \mathbb{E}\left\|\frac{1}{E} \sum_{e=0}^{E-1}   \nabla \widetilde{J}\left(\pmb{\Omega}^{t, e}\right)\right\|^2} \\
&\leq  &\frac{3}{E} \sum_{e=0}^{E-1} \mathbb{E}\left\|\nabla J\left(\pmb{\Omega}^{t, e}\right)\right\|^2+\frac{3 \sigma_F^2}{E}\nonumber \\
&\leq&  \frac{3}{E} \sum_{e=0}^{E-1}\left(2 L_J^2 \mathbb{E}\left\|\pmb{\Omega}^{t, e}-\pmb{\Omega}^{t}\right\|^2\right. \nonumber\\
&&\left.\;\;+2 \mathbb{E}\left\|\nabla J\left(\pmb{\Omega}^{t}\right)\right\|^2\right)+\frac{3 \sigma_F^2}{E}\nonumber \\
 &=& 6 L_J^2 \mathcal{E}^{(t)}+6 \mathbb{E}\left\|\nabla J\left(\pmb{\Omega}^{t}\right)\right\|^2 +\frac{3 \sigma_F^2}{E}.\nonumber
\label{35}
\end{eqnarray}

Next, we can bound the second term in (\ref{34}) as follows:
\begin{equation}
 \begin{aligned}
\lefteqn{\;\;\;\;\;3 \mathbb{E}\left\|\frac{\lambda \kappa_R  \mathcal{L} }{E} \sum_{e=0}^{E-1} \mathcal{L}\left(\pmb{\Omega}^{t, e}-\pmb{\Omega}^{t}\right)\right\|^2} \\
& =3 \lambda^2 \kappa^2_R \mathbb{E}\left\|\frac{1}{E} \sum_{e=0}^{E-1} \mathcal{L}\left(\pmb{\Omega}^{t, e}-\pmb{\Omega}^{t}\right)\right\|^2 \\
& \leq \frac{3 \lambda^2 \kappa^2_R}{E} \sum_{e=0}^{E-1} \mathbb{E}\left\|\mathcal{L}\left(\pmb{\Omega}^{t, e}-\pmb{\Omega}^{t}\right)\right\|^2 \\
& \leq \frac{3 \lambda^2 \kappa^2_R}{E}\|\mathcal{L}\|^2 \sum_{e=0}^{E-1} \mathbb{E}\left\|\left(\pmb{\Omega}^{t, e}-\pmb{\Omega}^{t}\right)\right\|^2=3 \lambda^2 \kappa^2_R \|\mathcal{L}\|^2 \mathcal{E}^{(t)}.
\label{36}
\end{aligned}
\end{equation}

Then, proceeding as in (\ref{35}), the third term in (\ref{34}) can be bounded as follows:
\begin{equation}
 \begin{aligned}
&3 \mathbb{E}\left\|\frac{\eta \lambda \kappa_R  \mathcal{L}}{E} \sum_{e=0}^{E-1}   \nabla \widetilde{F}\left(\pmb{\Omega}^{t, e}\right)\right\|^2\\
& \leq 3 \eta^2 \lambda^2 \kappa^2_R  \mathcal{L}^2 \mathbb{E}\left\|\frac{1}{E} \sum_{e=0}^{E-1}   \nabla \widetilde{F}\left(\pmb{\Omega}^{t, e}\right)\right\|^2 \\
& \leq \eta^2 \lambda^2 \kappa^2_R  \mathcal{L}^2\left(6 L_J^2 \mathcal{E}^{(t)}+6 \mathbb{E}\left\|\nabla J\left(\pmb{\Omega}^{t}\right)\right\|^2+\frac{3 \sigma_F^2}{E}\right) .
\label{37}
\end{aligned}
\end{equation}

Substituting (\ref{35}), (\ref{36}), and (\ref{37}) into (\ref{34}), the result of this Lemma in (\ref{30}) can be obtained. This concludes the proof.

\end{IEEEproof}
\label{lemma3}
\end{lemma}

\begin{lemma}
Let $\mathcal{E}^{t}\coloneqq\frac{1}{E} \sum_{e=0}^{E-1} \mathbb{E}\left\|\pmb{\Omega}^{t, e}-\pmb{\Omega}^{t}\right\|^2$ be the drift caused by $E$ local update steps at clients, where $\mathbb{E}$ is the expectation taken over all random sources and $\pmb{\Omega}^{t}=\pmb{\Omega}^{t,0}$. Suppose that Assumption 3 holds, we have
\begin{eqnarray*}
\mathcal{E}^{t} \leq 4 \eta^2  \mathbb{E}\left\|\nabla F\left(\pmb{\Omega}^{t}\right)\right\|^2+2\eta ^2 \sigma_F^2 E
\end{eqnarray*}
\begin{IEEEproof}
By Assumption 3, using Lemmas 3(a) and 3, we derive that
\begin{eqnarray}
\mathbb{E}\left\|\pmb{\Omega}^{t, e}-\pmb{\Omega}^{t}\right\|^2&=&\mathbb{E}\left\|\pmb{\Omega}^{t, e-1}-\pmb{\Omega}^{t}-\eta   \nabla \widetilde{F}\left(\pmb{\Omega}^{t, e-1}\right)\right\|^2 \nonumber\\
& \leq &\mathbb{E}\left\|\pmb{\Omega}^{t, e-1}-\pmb{\Omega}^{t}-\eta \nabla F\left(\pmb{\Omega}^{t, e-1}\right)\right\|^2\nonumber\\
&&+\eta^2 \sigma_F^2 \nonumber\\
& \leq & \left(1+\frac{1}{E}\right) \mathbb{E}\left\|\pmb{\Omega}^{t, e-1}-\pmb{\Omega}^{t}\right\|^2\nonumber\\
&&+(1+E) \eta ^2 \mathbb{E}\left\|\nabla F\left(\pmb{\Omega}^{t}\right)\right\|^2+\eta ^2 \sigma_F^2\nonumber \\
& \leq &\left(1+\frac{1}{E}\right) \mathbb{E}\left\|\pmb{\Omega}^{t, e-1}-\pmb{\Omega}^{t}\right\|^2\nonumber\\
&&+\frac{2 \eta^2}{E} \mathbb{E}\left\|\nabla F\left(\pmb{\Omega}^{t}\right)\right\|^2+\eta ^2 \sigma_F^2,
\end{eqnarray}
where the last inequality is due to the fact that $1+R \tau \leq R+R=2 R$ since $R \geq 1$ and $\tau \leq 1$. Telescoping the last inequality yields
\begin{eqnarray}
\lefteqn{\mathbb{E}\left\|\pmb{\Omega}^{t, e}-\pmb{\Omega}^{t}\right\|^2} \\
&&\leq\left(\frac{2 \tilde{\eta}^2}{E} \mathbb{E}\left\|\nabla F\left(\pmb{\Omega}^{t}\right)\right\|^2+\frac{\tilde{\eta}^2 \sigma_F^2}{R^2}\right) \sum_{r=1}^{R-1}\left(1+\frac{1}{E}\right)^r \nonumber.
\end{eqnarray}
Since $\sum_{j=0}^{m-1} x_j=\frac{x^m-1}{x-1}$ and $\left(1+\frac{x}{n}\right)^n \leq e^x, \forall x \in \mathbb{R}, n \in \mathbb{N}$, we have $\sum_{e=0}^{E-1}\left(1+\frac{1}{E}\right)^r=\frac{\left(1+\frac{1}{E}\right)^E-1}{\left(1+\frac{1}{E}\right)-1} \leq(e-1) E \leq 2 E$, and thus
\begin{equation}
 \begin{aligned}
\mathbb{E}\left\|\pmb{\Omega}^{t, e}-\pmb{\Omega}^{t}\right\|^2 \leq 4 \eta^2 \mathbb{E}\left\|\nabla F\left(\pmb{\Omega}^{t}\right)\right\|^2+2\eta ^2 \sigma_F^2 E .
\end{aligned}
\end{equation}
Averaging it over $r$, we get the conclusion.
\end{IEEEproof}
\label{lemma4}
\end{lemma}

\subsection{Proof of Theorem 1}
First, according to the result of Lemma \ref{lemma2}, we can have
\begin{eqnarray}
    \mathbb{E}\left\|\pmb{\Omega}^{t+1}-\pmb{\Omega}^*\right\|^2&=&\mathbb{E}\left\|\pmb{\Omega}^{t}-\pmb{\Omega}^*\right\|^2+\tilde{\eta}^2 \mathbb{E}\left\|Z^t\right\|^2\nonumber\\
    &&-2 \tilde{\eta} \mathbb{E}\left\langle Z^t, \pmb{\Omega}^{t}-\pmb{\Omega}^*\right\rangle.
    \label{42}
\end{eqnarray}
For the third term, we can obtain its bound from the Lemma \ref{lemma2} as follows:
\begin{eqnarray}
\lefteqn{-2 \tilde{\eta} \mathbb{E}\left\langle Z^t, \pmb{\Omega}^{t}-\pmb{\Omega}^*\right\rangle}\\
&=&  \frac{2 \tilde{\eta}}{E} \sum_{e=0}^{E-1} \mathbb{E}\left\langle\nabla J\left(\pmb{\Omega}^{t, e}\right), \pmb{\Omega}^*-\pmb{\Omega}^{t}\right\rangle \nonumber\\
&&+\frac{2 \tilde{\eta} \lambda \kappa_R}{E} \sum_{e=0}^{E-1} \mathbb{E}\left\langle\mathcal{L}\left(\pmb{\Omega}^{t, e}-\pmb{\Omega}^{t}\right), \pmb{\Omega}^{t}-\pmb{\Omega}^*\right\rangle \nonumber\\
&& +\frac{2 \tilde{\eta}^2 \lambda \kappa_R}{E} \sum_{e=0}^{E-1} \mathbb{E}\left\langle\mathcal{L}   \nabla F\left(\pmb{\Omega}^{t, e}\right),  \pmb{\Omega}^{t}-\pmb{\Omega}^* \right\rangle \nonumber\\
&\leq & \frac{2 \tilde{\eta} }{E} \sum_{e=0}^{E-1}\left(\mathbb{E}\left[J\left(\pmb{\Omega}^*\right)-J\left(\pmb{\Omega}^{t}\right)\right]-\frac{\mu}{4} \mathbb{E}\left\|\pmb{\Omega}^{t}-\pmb{\Omega}^*\right\|^2\right.\nonumber\\
&&\left.+L_J \mathbb{E}\left\|\pmb{\Omega}^{t, e}-\pmb{\Omega}^{t}\right\|^2\right) \nonumber\\
&& +\frac{2 \tilde{\eta} \lambda \kappa_R}{E} \sum_{e=0}^{E-1}\left(\frac{m}{2}\|\mathcal{L}\|^2 \mathbb{E}\left\|\pmb{\Omega}^{t, e}-\pmb{\Omega}^{t}\right\|^2\right.\nonumber\\
&&\left.+\frac{1}{2 m} \mathbb{E}\left\|\pmb{\Omega}^{t}-\pmb{\Omega}^*\right\|^2\right) \nonumber\\
&& +\frac{2 \tilde{\eta}^2 \lambda \kappa_R}{E} \sum_{e=0}^{E-1}\left(\frac{n}{2}\|\mathcal{L}\|^2 \mathbb{E}\left\|  \nabla F\left(\pmb{\Omega}^{t, e}\right)\right\|^2\right.\nonumber\\
&&\left.+\frac{1}{2 n} \mathbb{E}\|\pmb{\Omega}^{t}-\pmb{\Omega}^* \|^2\right),\nonumber
\label{43}
\end{eqnarray}
where $m, n>0$ will be chosen later. Due to the smoothness of $F(\cdot)$, we have
\begin{eqnarray}
\mathbb{E}\left\|\nabla F\left(\pmb{\Omega}^{t, e}\right)\right\|^2 & \leq& 2 L^2 \mathbb{E}\left\|\pmb{\Omega}^{t, e}-\pmb{\Omega}^{t}\right\|^2+2 \mathbb{E}\left\|\nabla F\left(\pmb{\Omega}^{t}\right)\right\|^2 \nonumber\\
 &\leq& 2 L^2 \mathbb{E}\left\|\pmb{\Omega}^{t, e}-\pmb{\Omega}^{t}\right\|^2\nonumber\\
&&+\frac{8 L^2}{\mu} \mathbb{E}\left[J\left(\pmb{\Omega}^{t}\right)-J\left(\pmb{\Omega}^*\right)\right]+ 4 \Gamma^L .
\label{44}
\end{eqnarray}

Substituting (\ref{43}) into (\ref{44}) and setting $m=\frac{8 \lambda}{\mu}$ and $n=\frac{8 \tilde{\eta} \lambda}{\mu}$, we have
\begin{eqnarray}
\lefteqn{-2 \tilde{\eta} \mathbb{E}\left\langle Z^t, \pmb{\Omega}^{t}-\pmb{\Omega}^*\right\rangle} \\
&\leq & -\left(2 \tilde{\eta}-\frac{64 \tilde{\eta}^3 \lambda^2 \kappa^2_R \|\mathcal{L}\|^2 L^2}{\mu^2}\right) \mathbb{E}\left[J\left(\pmb{\Omega}^{t}\right)-J\left(\pmb{\Omega}^*\right)\right]\nonumber\\
&&-\frac{\tilde{\eta}  \mu}{4} \mathbb{E}\left\|\pmb{\Omega}^{t}-\pmb{\Omega}^*\right\|^2 +\frac{32 \tilde{\eta}^3 \lambda^2 \kappa^2_R \|\mathcal{L}\|^2 \Gamma^L}{\mu}\nonumber\\
&& +\tilde{\eta} \left(2 L_J+\frac{\lambda^2 \kappa^2_R \|\mathcal{L}\|^2}{\mu}+\frac{16 \tilde{\eta}^2 \lambda^2 \kappa^2_R \|\mathcal{L}\|^2 L^2}{\mu}\right) \mathcal{E}^{(t)} .\nonumber
\label{45}
\end{eqnarray}
Combining this with (\ref{45}) and Lemma \ref{lemma2}, we rewrite (\ref{42}) as follow:
\begin{eqnarray}
\lefteqn{\mathbb{E} \left\|\pmb{\Omega}^{t+1}-\pmb{\Omega}^*\right\|^2} \\
&\leq & \left(1-\frac{\tilde{\eta} \mu}{4}\right) \mathbb{E}\left\|\pmb{\Omega}^{t}-\pmb{\Omega}^*\right\|^2+\tilde{\eta} p \mathcal{E}^{(t)}+6 \tilde{\eta}^2 \mathbb{E}\left\|\nabla J\left(\pmb{\Omega}^{t}\right)\right\|^2\nonumber\\
&&-\left(2 \tilde{\eta}-\frac{64 \tilde{\eta}^3 \lambda^2 \kappa^2_R \|\mathcal{L}\|^2 L^2}{\mu^2}\right) \mathbb{E}\left[J\left(\pmb{\Omega}^{t}\right)-J\left(\pmb{\Omega}^*\right)\right] \nonumber\\
&& +6 \tilde{\eta}^4 \lambda^2 \kappa^2_R \|\mathcal{L}\|^2 \mathbb{E}\left\|\nabla F\left(\pmb{\Omega}^{t}\right)\right\|^2+\frac{32 \tilde{\eta}^3 \lambda^2 \kappa^2_R \|\mathcal{L}\|^2 \Gamma^L}{\mu}\nonumber\\
&&+\frac{3 \tilde{\eta}^2 \left(1+\tilde{\eta}^2 \lambda^2 \kappa^2_R \|\mathcal{L}\|^2\right) \sigma_F^2}{E},\nonumber
\label{46}
\end{eqnarray}
where $p=2 L_J+\frac{8 \lambda^2 \kappa^2_R \|\mathcal{L}\|^2}{\mu}+\frac{64 \beta^2}{\mu}+\frac{12 L_J^2}{\lambda \kappa_R \|\mathcal{L}\|}+6 \lambda \kappa_R \|\mathcal{L}\|+\frac{48 L^2}{\lambda \kappa_R \|\mathcal{L}\|}$.

Using Lemma \ref{lemma4}, we have
\begin{eqnarray}
\lefteqn{\mathbb{E}\left\|\pmb{\Omega}^{t+1}-\pmb{\Omega}^*\right\|^2 } \\
& \leq&\left(1-\frac{\tilde{\eta}  \mu}{4}\right) \mathbb{E}\left\|\pmb{\Omega}^{t}-\pmb{\Omega}^*\right\|^2\nonumber\\
&&-\left(2 \tilde{\eta}-\frac{64 \tilde{\eta}^3 \lambda^2 \kappa^2_R \|\mathcal{L}\|^2 L^2}{\mu^2}\right) \mathbb{E}\left[J\left(\pmb{\Omega}^{t}\right)-J\left(\pmb{\Omega}^*\right)\right] \nonumber\\
&& +\left(4 p \tilde{\eta}^3 +6 \tilde{\eta}^4 \lambda^2 \kappa^2_R \|\mathcal{L}\|^2\right)\left(4 \frac{L^2}{\mu} \mathbb{E}\left[J(\pmb{\Omega})-J\left(\pmb{\Omega}^*\right)\right]\right.\nonumber\\
&&\left.+2 \Gamma^L\right)+\frac{2 p \tilde{\eta}^3 \tau^2 \sigma_F^2/B}{E}  +12 \tilde{\eta}^2 L_J \mathbb{E}\left[J(\pmb{\Omega})-J\left(\pmb{\Omega}^*\right)\right]\nonumber\\
&&+\frac{32 \tilde{\eta}^3 \lambda^2 \kappa^2_R \|\mathcal{L}\|^2\Gamma^L}{\mu}+\frac{3 \tilde{\eta}^2\left(1+\tilde{\eta}^2\lambda^2 \kappa^2_R \|\mathcal{L}\|^2\right) \sigma_F^2/B}{E} \nonumber\\
& \leq&\left(1-\frac{\tilde{\eta}  \mu}{4}\right) \mathbb{E}\left\|\pmb{\Omega}^{t}-\pmb{\Omega}^*\right\|^2+\tilde{\eta}^3 \underbrace{\frac{p\left(8 E \Gamma^L+2 \sigma_F^2/B\right)}{E}}_{C_2}\nonumber\\
&&-\left[2 \tilde{\eta}-\tilde{\eta}^2 q\right] \mathbb{E}\left[J\left(\pmb{\Omega}^{t}\right)-J\left(\pmb{\Omega}^*\right)\right]  \nonumber\\
&& +\tilde{\eta}^2 \underbrace{\frac{(64 \lambda \kappa_R \|\mathcal{L}\| E+48) \Gamma^L+15 \sigma_F^2/B}{\mu E}}_{C_1},\nonumber
\label{47}
\end{eqnarray}
where $q=\left(\frac{128 \lambda \kappa_R \|\mathcal{L}\| L^2}{\mu^2}+12 L_J+\frac{96 L^2}{\mu}+\frac{32 p L^2}{\mu \lambda \kappa_R \|\mathcal{L}\|}\right)$.
Let $\mu \leq \frac{\tilde{\eta}_1}{E}$. Then $\tilde{\eta} \leq \tilde{\eta}_1=\min \left\{\frac{1}{q}, \frac{2}{\lambda \kappa_R \|\mathcal{L}\|}\right\} \leq \frac{1}{q}$, which implies that $2 \tilde{\eta}-\tilde{\eta}^2 q \geq \tilde{\eta}$, and so
\begin{equation}
 \begin{aligned}
\mathbb{E}\left\|\pmb{\Omega}^{t+1}-\pmb{\Omega}^*\right\|^2 \leq & \left(1-\frac{\tilde{\eta}  \mu}{4}\right) \mathbb{E}\left\|\pmb{\Omega}^{t}-\pmb{\Omega}^*\right\|^2 \\
& -\tilde{\eta} \mathbb{E}\left[J\left(\pmb{\Omega}^{t}\right)-J\left(\pmb{\Omega}^*\right)\right]+\tilde{\eta}^3 C_2+\tilde{\eta}^2 C_1
\end{aligned}
\end{equation}

Recalling that $\Delta^{(t)}=\left\|\pmb{\Omega}^{t}-\pmb{\Omega}^*\right\|^2$, rearranging the terms, and multiplying both sides of (66) with $\frac{\theta^{(t)}}{\tilde{\eta} \tau \Theta_T}$, where $\Theta_T=\sum_{t=0}^{T-1} \theta^{(t)}$, we obtain that

\begin{equation}
 \begin{aligned}
\sum_{t=0}^{T-1} & \frac{\theta^{(t)} \mathbb{E}\left[J\left(\pmb{\Omega}^{t}\right)\right]}{\Theta_T}-J\left(\pmb{\Omega}^*\right) \\ & \leq \sum_{t=0}^{T-1} \mathbb{E}\left[\left(1-\frac{\tilde{\eta}  \mu}{4}\right) \frac{\theta^{(t)} \Delta^{(t)}}{\tilde{\eta} \tau \Theta_T}-\frac{\theta^{(t)} \Delta^{(t+1)}}{\tilde{\eta} \tau \Theta_T}\right] \\
& \quad +\mu^2 \tau C_2+\tilde{\eta} C_1 \\
& =\sum_{t=0}^{T-1} \mathbb{E}\left[\frac{\theta^{(t-1)} \Delta^{(t)}-\theta^{(t)} \Delta^{(t+1)}}{\tilde{\eta} \tau \Theta_T}\right]+\mu^2 \tau C_2+\tilde{\eta} C_1 \\
& =\frac{1}{\tilde{\eta} \tau \Theta_T} \Delta^{(0)}-\frac{\theta^{(T-1)}}{\tilde{\eta} \tau \Theta_T} \mathbb{E} \Delta^{(T)}+\tilde{\eta}^2 C_2+\tilde{\eta} C_1 \\
& \leq \frac{1}{\tilde{\eta} \tau \Theta_T} \Delta^{(0)}+\tilde{\eta}^2 C_2+\tilde{\eta} C_1 .
\label{49}
\end{aligned}
\end{equation}

Here, (\ref{49}) follows from the fact that $\left(1-\frac{\tilde{\eta}  \mu}{4}\right) \theta^{(t)}=\theta^{(t-1)}$ due to $\theta^{(t)}=\left(1-\frac{\tilde{\eta}  \mu}{4}\right)^{-(t+1)}$. Now, let $T \geq \frac{4 E}{\tilde{\eta}_1 \mu S}$. There is $\left(1-\frac{\tilde{\eta}  \mu}{4}\right)^T \leq \exp \left(-\frac{\tilde{\eta} \mu T}{4}\right) \leq \exp (-1) \leq \frac{3}{4}$, and thus
\begin{equation}
 \begin{aligned}
\Theta_T \geq\left(1-\frac{\tilde{\eta}  \mu}{4}\right)^{-T} \frac{1}{\tilde{\eta} \mu}=\frac{\theta^{(T-1)}}{\tilde{\eta} \mu}.
\end{aligned}
\end{equation}
which yields $\frac{1}{\tilde{\eta} \Theta_T} \leq \frac{\mu}{\theta^{(T-1)}} \leq \mu e^{-\frac{\tilde{\eta} \mu T}{4}}$. Therefore, (\ref{49}) can be rewritten as follows:
\begin{equation}
 \begin{aligned}
\sum_{t=0}^{T-1} \frac{\theta^{(t)} \mathbb{E}\left[J\left(\pmb{\Omega}^{t}\right)\right]}{\Theta_T}-J\left(\pmb{\Omega}^*\right) \leq \mu \Delta^{(0)} e^{-\frac{\tilde{\eta} \mu T}{4}}+\tilde{\eta}^2 C_2+\tilde{\eta} C_1,
\label{51}
\end{aligned}
\end{equation}
which together with the convexity of $J$ implies that
\begin{equation}
 \begin{aligned}
\mathbb{E}\left[J\left(\widetilde{\pmb{\Omega}}^{T}\right)-J\left(\pmb{\Omega}^*\right)\right]&=\mathbb{E}\left[J\left(\sum_{t=0}^{T-1} \frac{\theta^{(t)}}{\Theta_T} \pmb{\Omega}^{t}\right)\right]-J\left(\pmb{\Omega}^*\right) \\
&\leq \mu \Delta^{(0)} e^{-\frac{\tilde{\eta} \mu T}{4}}+\tilde{\eta}^2 C_2+\tilde{\eta} C_1 .
\label{52}
\end{aligned}
\end{equation}
Using (\ref{51}) -(\ref{52}) and by the L-smoothness of $J(\cdot)$, we can easily obtain ,
\begin{equation}
 \begin{aligned}
\mathbb{E}\left[J\left(\pmb{\Omega}^{T}\right)-J\left(\pmb{\Omega}^*\right)\right] &\le \frac{L}{2} \left(\mathbb{E}\left[\|\pmb{\Omega}^{T}-\pmb{\Omega}^*\|^2\right]\right)\\
&\leq \frac{L}{\mu}\left( \mu \Delta^{(0)} e^{-\frac{\tilde{\eta} \mu T}{4}}+\tilde{\eta}^2 C_2+\tilde{\eta} C_1 \right)\\
&=\frac{L}{\mu} \mathcal{O}\left(\mathbb{E}\left[J\left(\widetilde{\pmb{\Omega}}^{T}\right)-J\left(\pmb{\Omega}^*\right)\right]\right).
\label{56}
\end{aligned}
\end{equation}
Following the same approaches in \cite{ t2020personalized}, we consider the following cases.
\begin{itemize}
    \item If $\tilde{\eta}_1 \geq \widehat{\mu}:=\max \left\{\frac{4}{\mu T}, \frac{4}{\mu T} \log \left(\frac{\mu^2 \Delta^{(0)} T}{C_1}\right)\right\}$, then we choose $\mu=\widehat{\mu}$ and have
    \begin{equation}
    \begin{aligned}
    \mathbb{E}\left[J\left(\widetilde{\pmb{\Omega}}^{T}\right)-J\left(\pmb{\Omega}^*\right)\right] \leq \widetilde{\mathcal{O}}\left(\frac{C_2}{\mu^2 T^2}\right)+\widetilde{\mathcal{O}}\left(\frac{C_1}{\mu T}\right)
    \end{aligned}
    \end{equation}

    \item If $\frac{4}{\mu T} \leq \tilde{\eta}_1 \leq \widehat{\mu}$, then we choose $\mu=\tilde{\eta}_1$ and have
    \begin{equation}
    \begin{aligned}
    \mathbb{E}\left[J\left(\widetilde{\pmb{\Omega}}^{T}\right)-J\left(\pmb{\Omega}^*\right)\right] \leq & \mathcal{O}\left(\alpha \Delta^{(0)} e^{-\frac{\tilde{\eta}_1 \mu T}{4}}\right) \\ &+\widetilde{\mathcal{O}}\left(\frac{C_2}{\mu^2 T^2}\right)+\widetilde{\mathcal{O}}\left(\frac{C_1}{\mu T}\right)
    \end{aligned}
    \end{equation}
\end{itemize}
By combining (\ref{56}) and the above two cases,
\begin{eqnarray}
\lefteqn{\mathbb{E}\left[J\left(\pmb{\Omega}^{T}\right)-J\left(\pmb{\Omega}^*\right)\right]} \\
&&\leq \mathcal{O}\left( \Delta^{(0)} e^{-\frac{\tilde{\eta}_1 \mu^2 T}{4}} +\frac{(1+\mu T)(E \Gamma^L+\sigma_F^2/B)}{\mu^3 T^2 E K^L} \right) ,\nonumber
\end{eqnarray}
where $\Delta^{(0)}=\|\pmb{\Omega}^{0}-\pmb{\Omega}^*\|^2$. This concludes the proof.

\section{Proof of Theorem \ref{Theorem_TransferError}}
\label{Proof_Theorem_TransferError}

In this section, we would like to bound the transfer objective $\mathbb{E}[F_{k''}(\pmb{\omega}_{k''})-F_{k''}(\pmb{\omega}^{*}_{k''})]$ in (\ref{main_objective}) to capture the above-performance gap between the transfer model obtained by our proposed transfer solution and the local optimal model. Note that the expected error bound of $\mathbb{E}[F_{k''}(\pmb{\omega}_{k''})-F_{k''}(\pmb{\omega}^{*}_{k''})]$ can be decomposed into the following form:
\begin{equation}
\begin{aligned}
&\underbrace{\mathbb{E}[F_{k''}(\pmb{\omega}_{k''})-F_{k''}(\pmb{\omega}^{*}_{k''})]}_{\text{transfer error}} \\
&\le
\underbrace{\mathbb{E}[F_{k''}(\pmb{\omega}_{k''})-\frac{1}{K^L}\sum_{k'\in {\cal K}^L}F_{k'}(\pmb{\omega}_{k'})]}_{\text{generalization error}} \\
&+\underbrace{\mathbb{E}[\frac{1}{K^L}\sum_{k'\in {\cal K}^L}F_{k'}(\pmb{\omega}_{k'})]-\frac{1}{K^L}\sum_{k'\in {\cal K}^L}F_{k'}(\pmb{\omega}^{*}_{k'})}_{\text{training error}}\\
&+\underbrace{\frac{1}{K^L}\sum_{k'\in {\cal K}^L}F_{k'}(\pmb{\omega}^{*}_{k'})-F_{k''}(\pmb{\omega}^{*}_{k''})}_{\text{transfer gap}}.
\label{theorem0}
\end{aligned}
\end{equation}
Hence, to bound the expected transfer error, we should bound the expectation of training and generalization errors. To begin with, we first introduce the following lemmas.

\subsection{Key Lemmas of Theorem \ref{Theorem_TransferError}}

\begin{lemma}\label{lemma6}
Suppose Assumptions 5 holds. According to the proposed model transfer solution, we can bound the generalization error term in (\ref{theorem0}) as follows:
\begin{eqnarray}
  \lefteqn{\mathbb{E}[F_{k''}(\pmb{\omega}_{k''})-\frac{1}{K^L}\sum_{k'\in {\cal K}^L}F_{k'}(\pmb{\omega}_{k'})] }  \nonumber \\
     &&\le M \sum_{k'\in \cal{K}^L} \| \xi_{k'', k'}\mathcal{P}_{k''}-\frac{\mathcal{P}_{k'}}{K^L} \|_{TV}.
\end{eqnarray}
\begin{IEEEproof}
 Due to $\pmb{\omega}_{k''}=\sum_{k'\in {\cal K}^L}\xi_{k'', k'}\pmb{\omega}_{k'}$, using Jensen's inequality, we have $F_{k''}(\pmb{\omega}_{k''}) \le \sum_{k'\in {\cal K}^L}\xi_{k'', k'}F_{k''}(\pmb{\omega}_{k'})$. Therefore, we can derive the following result:
\begin{eqnarray}
     \lefteqn{\mathbb{E}[F_{k''}(\pmb{\omega}_{k''})-\frac{1}{K^L}\sum_{k'\in {\cal K}^L}F_{k'}(\pmb{\omega}_{k'})]}   \nonumber \\
    &&\le \sum_{k'\in {\cal K}^L}  \mathbb{E}[\xi_{k'', k'}F_{k''}(\pmb{\omega}_{k'})-\frac{1}{K^L}F_{k'}(\pmb{\omega}_{k'})]   \nonumber\\
  &&\le M \sum_{k'\in \cal{K}^L}\| \xi_{k'', k'}\mathcal{P}_{k''}-\frac{\mathcal{P}_{k'}}{K^L} \|_{TV}.
    \label{62}
\end{eqnarray}
The last inequality in (\ref{62}) can be obtained by using Definition 2. This concludes the proof.
\end{IEEEproof}
\end{lemma}

\begin{lemma}\label{lemma7}
Let $\mathbb{E}[\mathcal{J}(\pmb{\Omega}^{T})-\mathcal{J}(\pmb{\Omega}^{*})] \le \varepsilon$ hold. When choosing $R(\|\pmb{\omega}_{k'}- \pmb{\omega}_{{j'}}\|^2)=1-e^{-\|\pmb{\omega}_{k'}- \pmb{\omega}_{{j'}}\|^2/{\sigma_R}}$ with parameter ${\sigma_R}$, the training error term in (\ref{theorem0}) can be bounded as follows:
\begin{equation}
\begin{aligned}
\mathbb{E}[\frac{1}{K^L}\sum_{k'\in {\cal K}^L}F_{k'}(\pmb{\omega}_{k'})]-\frac{1}{K^L}\sum_{k'\in {\cal K}^L}F_{k'}(\pmb{\omega}^{*}_{k'})\le \frac{\varepsilon}{K^L}+\lambda.
\label{theorem1_b}
\end{aligned}
\end{equation}

\begin{IEEEproof}
According to the Theorem 1, we have $\mathbb{E}\left[J\left(\pmb{\Omega}^{T}\right)-J\left(\pmb{\Omega}^*\right)\right] =  \mathbb{E}\left[F\left(\pmb{\Omega}^{T}\right)-F\left(\pmb{\Omega}^*\right)\right] + \lambda \left( \mathcal{R}({\boldsymbol \Omega})-\mathcal{R}({\boldsymbol \Omega}^*)\right) \le \varepsilon$. When choosing $R(\|\pmb{\omega}_{k'}- \pmb{\omega}_{{j'}}\|^2)=1-e^{-\|\pmb{\omega}_{k'}- \pmb{\omega}_{{j'}}\|^2/{\sigma_R}}$, we can derive that
\begin{eqnarray}
\lefteqn{\frac{1}{K^L}\mathbb{E}\left[\mathcal{F}\left(\pmb{\Omega}^{T}\right)-\mathcal{F}\left(\pmb{\Omega}^*\right)\right]}\\
&=&\frac{1}{K^L}\mathbb{E}\left[\mathcal{J}\left(\pmb{\Omega}^{T}\right)-\mathcal{J}\left(\pmb{\Omega}^*\right)\right]  \nonumber + \frac{\lambda}{K^L} \left( \mathcal{R}({\boldsymbol \Omega})-\mathcal{R}({\boldsymbol \Omega}^*)\right) \nonumber \\
& <&  \frac{\varepsilon}{K^L} + \lambda .\nonumber
\label{60}
\end{eqnarray}
The inequality in (\ref{60}) holds due to $0< R(\|\pmb{\omega}_{k'}- \pmb{\omega}_{{j'}}\|^2) < 1$ in this case. This concludes the proof.
\end{IEEEproof}
\end{lemma}

\subsection{Proof of Theorem \ref{Theorem_TransferError}}
\begin{IEEEproof}
    By combining Lemma \ref{lemma6} and Lemma \ref{lemma7}, we can rewrite (\ref{theorem0}) as follows:
\begin{eqnarray}
\lefteqn{\mathbb{E}[F_{k''}(\boldsymbol{\omega}_{k''}) - F_{k''}(\boldsymbol{\omega}^{*}_{k''})] }  \\
&&\leq M\sum_{k'\in \mathcal{K}^L} \left\| \xi_{k'', k'}\mathcal{P}_{k''} - \frac{\mathcal{P}_{k'}}{K^L} \right\|_{TV} + \frac{\epsilon + (\lambda + \Gamma^N)K^L}{K^L},\nonumber
\end{eqnarray}
where $\Gamma^N=\frac{1}{K^L}\sum_{k'\in {\cal K}^L}F_{k'}(\pmb{\omega}^{*}_{k'})-F_{k''}(\pmb{\omega}^{*}_{k''})$. This concludes the proof.
\end{IEEEproof}

\end{document}